\documentclass[
aip,
jcp,
reprint,%
]{revtex4-1}

\usepackage{savesym}% Package for solving redefinition conflict

\usepackage{graphicx}% Include figure files
\usepackage{dcolumn}% Align table columns on decimal point
\usepackage{bm}% bold math
\usepackage[mathlines]{lineno}% Enable numbering of text and display math
\usepackage{tabularx}% Include figure files
\usepackage[utf8]{inputenc}
\usepackage[T1]{fontenc}
\usepackage{mathptmx}
\usepackage{etoolbox}

\usepackage[english]{babel}
\usepackage{array}
\usepackage{multirow}
\usepackage{appendix}
\usepackage{grffile}
\usepackage{amssymb}
\usepackage{amsmath}

\usepackage[caption=false]{subfig}

\usepackage[colorlinks=true, allcolors=blue]{hyperref}% for creating links
\usepackage[nonumberlist,acronym]{glossaries}
\usepackage[nolist]{acronym}

\usepackage{enumitem}
\usepackage{txfonts}
\usepackage{cleveref}
\usepackage{comment}
\usepackage{soul} % Load the soul package for highlighting

\newcommand{\pd}{\partial}

\newcommand{\ie}{\textit{i.e.}}

\newcommand{\etal}{\textit{et al. }}

\makenoidxglossaries 
\newacronym{SI}{SI}{Supporting Information}
\newacronym{DFT}{DFT}{density functional theory}
\newacronym{MD}{MD}{molecular dynamics}
\newacronym{MC}{MC}{Monte Carlo}
\newacronym{HR}{HR}{hard-rod}
\newacronym{LJ}{LJ}{Lennard–Jones}
\newacronym{QM}{QM}{quantum-mechanical}
\newacronym{SM}{SM}{statistical mechanics}
\newacronym{KS-DFT}{KS-DFT}{Kohn-Sham density functional theory}
\newacronym{FMT}{FMT}{fundamental measure theory} 
\newacronym{AIMD}{AIMD}{\emph{ab initio} molecular dynamics}
\newacronym{cDFT}{cDFT}{classical density functional theory}
\newacronym{OZ}{OZ}{Ornstein-Zernike} 
\newacronym{IET}{IET}{integral-equation theory}
\newacronym{SFT}{SFT}{statistical field theory}
\newacronym{EL}{EL}{Euler-Lagrange}

\newacronym{NS}{NS}{Numerical Solver}
\newacronym{L-BFGS}{L-BFGS}{Limited-memory Broyden, Fletcher, Goldfarb, and Shanno}
\newacronym{ML}{ML}{machine learning}
\newacronym{LLM}{LLM}{Large Language Models}
\newacronym{GPR}{GPR}{Gaussian process regression}
\newacronym{ALEC}{ALEC}{Active Learning by Error Control}
\newacronym{DeepONet}{DeepONet}{Deep Operator Networks}
\newacronym{FNO}{FNO}{Fourier neural operators}
\newacronym{PDE}{PDE}{partial differential equation}
\newacronym{FFT}{FFT}{Fast Fourier Transform}
\newacronym{DNN}{DNN}{Dense Neural Network}
\newacronym{CNN}{CNN}{Convolution Neural Network}
\newacronym{ResNet}{ResNet}{Residual neural network}
\newacronym{RMSCNN}{RMSCNN}{Residual Multi Scale CNN}
\newacronym{sRelu}{sRelu}{squared Relu}
\newacronym{Relu}{Relu}{Rectified linear unit} 
\newacronym{Selu}{Selu}{Scaled Exponential linear unit}
\newacronym{NLL}{NLL}{negative log-likelihood}
\newacronym{GNLL}{GNLL}{Gaussian negative log-likelihood}
\newacronym{FLOPs}{FLOPs}{Floating point operations}
\newacronym{GK}{GK}{Gaussian Kernel}
\newacronym{DK}{DK}{Discrete Kernel}

\newacronym{DISCO}{DISCO}{Discrete Scale Convolutions}
\newacronym{TW-functions}{TW-functions}{Tauber-Wiener functions}
\newacronym{MSE}{MSE}{Mean Squared Error}
\newacronym{GNN}{GNN}{message passing graph neural network}
\newacronym{RDF}{RDF}{radial distribution functions}
\newacronym{PP-GP}{PP-GP}{parallel partial Gaussian process}

\def \disxvec{\mathbf{x}}
\def \xin{\disxvec_{I}}
\def \xout{\disxvec_{O}}
\def \HRdia{d_{hr}}
\def \rcorr {\xi}
\def \rcutoff {x_{cf}}
\def \Dset {\mathbb{D}}
\def \nsample {N_{sample}}
\def \nset {N_{set}}
\def \ngrid {N_{G}}
\def \nin {N_{GI}}
\def \nout {N_{GO}}
\def \msize {D_{model}}
\def \comcost {C_{t}}
\def \corrf {c_{1}}
\def \FFT {\mathcal{F}}
\def \Goperator {\mathcal{G}}
\def \acti {\mathcal{\sigma}}
\def \bnetout {\mathbf{b}}
\def \tnetout {\bm{\tau}}
\def \ufunction {\textit{u}}
\def \vfunction {\textit{v}}
\def \umeasures {\mathbf{u}}
\def \vmeasures {\mathbf{v}}
\def \zmatrix {\hat{\mathbf{z}}}
\def \nfl {n_{l}}
\def \Fex {F_{ex}}
\newglossary*{symbol}{Symbols}
\newglossary*{term}{Terms}
\newglossaryentry{kB}{
name={\ensuremath{k_B}},
symbol={\ensuremath{k_B}},
sort={The Boltzmann constant},
type={symbol},
description={The Boltzmann constant}
}
\newglossaryentry{beta}{
name={\ensuremath{\beta}},
symbol={\ensuremath{\beta}},
sort={Thermodynamic beta},
type={symbol},
description={Thermodynamic beta $\beta=1/(k_BT)$}
}
\newglossaryentry{Lambda}{
name={\ensuremath{\Lambda}},
symbol={\ensuremath{\Lambda}},
sort={the thermal wavelength},
type={symbol},
description={The thermal wavelength, which is used as the unit length $\Lambda=1$ in this paper}
}
\newglossaryentry{rho}{
name={\ensuremath{\rho(x)}},
symbol={\ensuremath{\rho(x)}},
sort={The Density Profile},
type={symbol},
description={The one-body density profile, which is a function of position $x$}
}
\newglossaryentry{Vext}{
name={\ensuremath{V_{ext}(x)}},
symbol={\ensuremath{V_{ext}(x)}},
sort={The external potential},
type={symbol},
description={The external potential changing with position $x$}
}
\newglossaryentry{mu}{
name={\ensuremath{\mu}},
symbol={\ensuremath{\mu}},
sort={The chemical potential},
type={symbol},
description={The chemical potential of a specific species in bulk}
}
\newglossaryentry{Vloc}{
name={\ensuremath{V_{loc}(x)}},
symbol={\ensuremath{V_{loc}(x)}},
sort={Density Profile},
type={symbol},
description={The background potential, which is the sum of chemical potential and external potential, $V_{loc}=V_{ext}(x)-\mu$}
}
\newglossaryentry{Fex}{
name={\ensuremath{F_{ex}}},
symbol={\ensuremath{F_{ex}}},
sort={Excess Free energy},
type={symbol},
description={The excess Helmholtz energy}
}
\newglossaryentry{Omega}{
name={\ensuremath{\Omega}},
symbol={\ensuremath{\Omega}},
sort={The grand potential},
type={symbol},
description={The grand potential}
}
\newglossaryentry{corrf}{
name={\ensuremath{c_{1}(x)}},
symbol={\ensuremath{c_{1}(x)}},
sort={Direct correlation function},
type={symbol},
description={The one-body direct correlation function, defined as $c_1(x)=-\frac{\pd\beta \Fex}{\pd \rho(x)}$}
}
\newglossaryentry{disxvec}{
name={\ensuremath{\mathbf{x}}},
symbol={\ensuremath{\mathbf{x}}},
sort={x vector 0},
type={symbol},
description={The sensor positions as a vector}
}
\newglossaryentry{xin}{
name={\ensuremath{\mathbf{x}_{I}}},
symbol={\ensuremath{\mathbf{x}_{I}}},
sort={x vector 1 in},
type={symbol},
description={The sensor positions for input function}
}
\newglossaryentry{xout}{
name={\ensuremath{\mathbf{x}_{O}}},
symbol={\ensuremath{\mathbf{x}_{O}}},
sort={x vector 2 out},
type={symbol},
description={The sensor positions for output function}
}
\newglossaryentry{HRdia}{
name={\ensuremath{d_{HR}}},
symbol={\ensuremath{d_{HR}}},
sort={hard rod diameter},
type={symbol},
description={The diameter of the hard rod. For convenience, $d_{hr}=\Lambda=1$ is used in this paper}
}
\newglossaryentry{dx}{
name={\ensuremath{\Delta x}},
symbol={\ensuremath{\Delta x}},
sort={grid resolution},
type={symbol},
description={The resolution of sensor grids. $\Delta x=x_{i+1}-x_i$. Usually $0.01$}
}
\newglossaryentry{rcorr}{
name={\ensuremath{\xi}},
symbol={\ensuremath{\xi}},
sort={correlation length},
type={symbol},
description={correlation length of the physical system in thermodynamics}
}
\newglossaryentry{rcutoff}{
name={\ensuremath{x_{cf}}},
symbol={\ensuremath{x_{cf}}},
sort={cutoff length},
type={symbol},
description={the cutoff distance used for a quasi-local mapping between different physical quantities }
}
\newglossaryentry{Dset}{
name={\ensuremath{\mathbb{D}}},
symbol={\ensuremath{\mathbb{D}}},
sort={Dset},
type={symbol},
description={The space which the function is defined in. In this paper, it means the physical spatial domain between the walls, where the density profile and other functions of cDFT are valid.}
}
\newglossaryentry{L}{
name={\ensuremath{L}},
symbol={\ensuremath{L}},
sort={box size},
type={symbol},
description={The length of $\Dset$ for functions including $V_{ext}(x)$,$\corrf(x)$, and $\rho(x)$ has a valid non-zero value. $L+1$ is the distances between the walls. }
}
\newglossaryentry{nsample}{
name={\ensuremath{N_{sample}}},
symbol={\ensuremath{N_{sample}}},
sort={number of samples},
type={symbol},
description={The number of samples used for machine learning.}
}
\newglossaryentry{msize}{
name={\ensuremath{D_{model}}},
symbol={\ensuremath{D_{model}}},
sort={model size},
type={symbol},
description={Number of trainable parameters of a machine learning model, representing the size of that model}
}
\newglossaryentry{Loss_model}{
name={\ensuremath{\mathcal{L}}},
symbol={\ensuremath{\mathcal{L}}},
sort={model Loss},
type={symbol},
description={Loss of machine learning method measured by $\log({\textbf{MSE}})$}
}
\newglossaryentry{ideal_Loss_model}{
name={\ensuremath{\mathcal{L}_{\infty}}},
symbol={\ensuremath{\mathcal{L}_{\infty}}},
sort={model Loss},
type={symbol},
description={Fitted parameter of neural scaling law when $\msize\rightarrow\infty$ and $\nsample\rightarrow\infty$, showing the irreducible loss of the given method}
}
\newglossaryentry{nset}{
name={\ensuremath{N_{set}}},
symbol={\ensuremath{N_{set}}},
sort={number of samples},
type={symbol},
description={The number of samples in the related dataset, used for loss comparison on different datasets.}
}
\newglossaryentry{ngrid}{
name={\ensuremath{N_{G}}},
symbol={\ensuremath{N_{G}}},
sort={number of grids},
type={symbol},
description={The number of sensor grids used to measure functional values, $N_G=\frac{L}{\Delta x}+1$}
}
\newglossaryentry{nin}{
name={\ensuremath{N_{GI}}},
symbol={\ensuremath{N_{GI}}},
sort={number of input grids},
type={symbol},
description={$\ngrid$ for the input function (usually $\rho(x)$).}
}
\newglossaryentry{nout}{
name={\ensuremath{N_{GO}}},
symbol={\ensuremath{N_{GO}}},
sort={number of input grids},
type={symbol},
description={$\ngrid$ for the output function (usually $c_1(x)$)}
}
\newglossaryentry{tFFT}{
name={\ensuremath{\mathcal{F}}},
symbol={\ensuremath{\mathcal{F}}},
sort={truncated Fourier transform},
type={symbol},
description={truncated Fourier transformation, with high frequency modes filled with zeros.}
}
\newglossaryentry{itFFT}{
name={\ensuremath{\mathcal{F}^{-1}}},
symbol={\ensuremath{\mathcal{F}^{-1}}},
sort={inverse truncated Fourier transform},
type={symbol},
description={inverse truncated Fourier transformation.  }
}
\newglossaryentry{ufunction}{
name={\ensuremath{\textit{u}}},
symbol={\ensuremath{\textit{u}}},
sort={input function},
type={symbol},
description={The input function of an operator, which is usually $\rho(x)$ in this paper}
}
\newglossaryentry{vfunction}{
name={\ensuremath{\textit{v}}},
symbol={\ensuremath{\textit{v}}},
sort={output function},
type={symbol},
description={The output function of an operator, which is usually $c_1(x)$ in this paper.}
}
\newglossaryentry{umeasures}{
name={\ensuremath{\mathbf{u}}},
symbol={\ensuremath{\mathbf{u}}},
sort={input vector},
type={symbol},
description={The vector of measured values of input function $\ufunction$ on the input sensor grids $\xin$ }
}
\newglossaryentry{vmeasures}{
name={\ensuremath{\mathbf{v}}},
symbol={\ensuremath{\mathbf{v}}},
sort={output vector},
type={symbol},
description={The vector of measured values of output function $\vfunction$ on the input sensor grids $\xout$}
}
\newglossaryentry{Goperator}{
name={\ensuremath{\mathcal{G}}},
symbol={\ensuremath{\mathcal{G}}},
sort={The Operator},
type={symbol},
description={General Expression of an operator showing the mapping between input function $\ufunction$ and output function $\vfunction$, expressed as $\Goperator:\ufunction\mapsto\vfunction$. So $v(x)=\Goperator(u)(x)$.}
}
\newglossaryentry{corrfoperator}{
name={\ensuremath{\mathcal{G}_{c}}},
symbol={\ensuremath{\mathcal{G}_{c}}},
sort={The correlation Operator},
type={symbol},
description={The operator of mapping $\rho(x)\mapsto\corrf(x)$.}
}
\newglossaryentry{rhooperator}{
name={\ensuremath{\mathcal{G}_{\rho}}},
symbol={\ensuremath{\mathcal{G}_{\rho}}},
sort={The density Operator},
type={symbol},
description={The operator of mapping $V_{ext}(x)\mapsto\rho(x)$.}
}
\newglossaryentry{acti}{
name={\ensuremath{\mathcal{\sigma}}},
symbol={\ensuremath{\mathcal{\sigma}}},
sort={The activation function},
type={symbol},
description={The activation function for neural networks.}
}
\newglossaryentry{bnet}{
name={the Branch Net},
type={term},
description={the Branch subnet of Deep Operator Networks, takes the value of input function. }
}
\newglossaryentry{tnet}{
name={the Trunk Net},
type={term},
description={The Trunk subnet of Deep Operator Networks, takes the value of output sensor positions.}
}
\newglossaryentry{bnetout}{
name={\ensuremath{\mathbf{b}}},
symbol={\ensuremath{\mathbf{b}}},
sort={Branch Net output},
type={symbol},
description={The output of the Branch Net in Deep Operator Networks, in the form of a vector for each sample}
}
\newglossaryentry{tnetout}{
name={\ensuremath{\bm{\tau}}},
symbol={\ensuremath{\bm{\tau}}},
sort={Trunk Net Output},
type={symbol},
description={The output of the Trunk net in Deep Operator Networks, in the form of a vector for each sample}
}
\newglossaryentry{Player}{
name={\ensuremath{\mathcal{P}}},
symbol={\ensuremath{\mathcal{P}}},
sort={FNO Input layer},
type={term},
description={}
}
\newglossaryentry{Qlayer}{
name={\ensuremath{\mathcal{Q}}},
symbol={\ensuremath{\mathcal{Q}}},
sort={FNO Output layer},
type={term},
description={}
}
\newglossaryentry{Flayer}{
name={\ensuremath{\mathcal{F}}},
symbol={\ensuremath{\mathcal{F}}},
sort={FNO Fourier layer},
type={term},
description={}
}
\newglossaryentry{N_FL}{
name={\ensuremath{n_l}},
symbol={\ensuremath{n_l}},
sort={Number of Fourier layer},
type={symbol},
description={The number of Fourier layer in FNO}
}
\newglossaryentry{d_v}{
name={\ensuremath{d_v}},
symbol={\ensuremath{d_v}},
sort={Channels of Fourier layer},
type={symbol},
description={The number of hidden channels in each Fourier Layer in FNO}
}
\newglossaryentry{zmatrix}{
name={\ensuremath{\hat{\mathbf{z}}}},
symbol={\ensuremath{\hat{\mathbf{z}}}},
sort={zmatrix output of Fourier layer},
type={symbol},
description={The output matrix of Fourier layers in FNO, with $\zmatrix_{j}$ is the output of j-th Fourier layer}
}
\numberwithin{equation}{section}
\makeatletter
\def\@email#1#2{%
 \endgroup
 \patchcmd{\titleblock@produce}
  {\frontmatter@RRAPformat}
  {\frontmatter@RRAPformat{\produce@RRAP{*#1\href{mailto:#2}{#2}}}\frontmatter@RRAPformat}
  {}{}
}%
\makeatother

\newcommand{\centercells}[2]{%
  \begingroup\sbox0{\begin{minipage}{3cm}\raggedright#1\end{minipage}}%
  \sbox2{\begin{minipage}{3cm}\raggedright#2\end{minipage}}%
  \xdef\finalheight{\the\dimexpr\ht0+\dp0+\smallskipamount\relax}%
  \xdef\finalheightB{\the\dimexpr\ht2+\dp2+\smallskipamount\relax}%
  \ifdim\finalheightB>\finalheight
    \global\let\finalheight\finalheightB
  \fi\endgroup
  \begin{minipage}[t][\finalheight][t]{3cm}\raggedright#1\end{minipage}&
  \begin{minipage}[t][\finalheight][t]{3cm}\raggedright#2\end{minipage}}

\begin{document}
%\begin{frontmatter}
\preprint{AIP/123-QED}

\title[]{Neural Operators for Forward and Inverse Potential–Density Mappings in Classical Density Functional Theory}

\author{Runtong Pan}
\affiliation{\mbox{Department of Chemical and Environmental Engineering, University of California, Riverside, CA 92521, USA}}
\author{Xinyi Fang}
\affiliation{\mbox{Department of Statistics and Applied Probability, University of California, Santa Barbara, CA 93106, USA}}
\author{Kamyar Azizzadenesheli}
\affiliation{Nvidia Corporation, 2788 San Tomas Expressway, Santa Clara, CA 95051, USA}
\author{Miguel Liu-Schiaffini}
\affiliation{\mbox{Department of Computing and Mathematical Sciences, California Institute of Technology, Pasadena, CA 91125, USA}}
\author{Mengyang Gu}
\affiliation{\mbox{Department of Statistics and Applied Probability, University of California, Santa Barbara, CA 93106, USA}}
\author{Jianzhong Wu$^*$}
\affiliation{\mbox{Department of Chemical and Environmental Engineering, University of California, Riverside, CA 92521, USA}}
\email{jianzhong.wu@ucr.edu}

% Define affiliations
% Use affiliation or adress instead of affil 
%\affil[1]{Department of Chemical and Environmental Engineering, University of California, Riverside, CA 92521, USA}
%\affil[2]{Department of Statistics and Applied Probability, University of California, Santa Barbara, CA 93106, USA}
\date{Submitted: May 7, 2025}

%\maketitle  

%% Abstract
\begin{abstract}
Neural operators are capable of capturing nonlinear mappings between infinite-dimensional functional spaces, offering a data-driven approach to modeling complex functional relationships  in classical density functional theory (cDFT). In this work, we evaluate the performance of several neural operator architectures in learning the functional relationships  between the one-body density profile $\rho(x)$, the one-body direct correlation function ${\corrf}(x)$, and the external potential $V_{ext}(x)$ of inhomogeneous one-dimensional (1D) hard-rod fluids, using training data generated from analytical solutions of the underlying statistical-mechanical model. Several variants of the Deep Operator Network (DeepONet) and the Fourier Neural Operator (FNO) were considered, each incorporating different machine-learning (ML) architectures, activation functions, and training strategies. These operator learning methods are benchmarked against a fully connected dense neural network (DNN), which serves as a baseline. We compared their performance in terms of the Mean Squared Error (MSE) loss in establishing the functional relationships  as well as in predicting the excess free energy across two test sets: (1) a group test set generated via random cross-validation (CV) to assess interpolation capability, and (2) a newly constructed dataset for leave-one-group CV to evaluate extrapolation performance. Our results show that FNO achieves the most accurate predictions of the excess free energy, with the squared ReLU activation function outperforming other activation choices. Among the DeepONet variants, the Residual Multiscale Convolutional Neural Network (RMSCNN) combined with a trainable Gaussian derivative kernel (GK-RMSCNN-DeepONet) demonstrates the best performance. Additionally, we applied the trained models to solve for the density profiles at various external potentials and compared the results with those obtained from the direct mapping $V_{ext} \mapsto \rho$ with neural operators, as well as with Gaussian Process Regression (GPR) combined with Active Learning by Error Control (ALEC), which has shown strong performance in previous studies. While the direct mapping from $V_{ext} \mapsto \rho$ suffers from high extrapolation error and proves inefficient for out-of-distribution predictions, the neural-operator mapping $\rho \mapsto {\corrf}$ can effectively be used to solve the density profile via the Euler-Lagrange equation or be integrated with other surrogate methods. Moreover, neural operators offer additional flexibility through specialized operations, such as significance-based predictions on uneven grids (as in GK-CNN-DeepONet) and adaptive grid resolution adjustment (as in FNO), both of which can enhance prediction accuracy.
\end{abstract}

\maketitle

\clearpage

%\printglossary[type=\acronymtype]
%\printnoidxglossary[sort=use,
%type=\acronymtype, title=Abbreviations,
%nonumberlist]
%\printnoidxglossary[sort=use,type=symbol, title=symbols,nonumberlist]

%\end{frontmatter}

\section{Introduction\label{sec:introduction}}
\Acrfull{cDFT} is a powerful tool in statistical mechanics to establish structure-property relationships  for inhomogeneous thermodynamic systems.\cite{Evans79,doi:10.1021/ed500049m,RN8386} The variational approach relies on formulation of the grand potential as a functional of the one-body density profile(s). For an equilibrium state specified by temperature, system volume, the external potentials as well as chemical potentials of individual species,  minimization of the grand potential leads to a set of self-consistent equations to predict both structural and thermodynamic properties, enabling a seamless connection between microscopic variables and the macroscopic behavior. Although the theoretical procedure is formally rigorous, the practical applications of \acrshort{cDFT} are often hindered by challenges in deriving a reliable free-energy functional and in numerical complexity for solving the density profiles.\cite{wu2006density,wu2007,SERMOUD2024114177} These limitations have sparked considerable interest in using \acrfull{ML} methods to develop reliable free-energy functionals and/or to solve the computationally demanding \acrfull{EL} equation.\cite{Sammüller_2024,simon2024,doi:10.1021/acs.jpclett.3c02804} 

\acrshort{ML} methods can be utilized to establish functional relationships  in \acrshort{cDFT} for the free energy (or the excess free energy \glssymbol{Fex}) of an entire thermodynamic system, the local thermodynamic potential, or the external potential. For example, Oettel and coworkers explored the feasibility of using neural networks to construct free-energy functionals for one-dimensional (1D) \acrfull{HR} and \acrfull{LJ} fluids.\cite{10.21468/SciPostPhys.6.2.025, 10.1063/1.5135919,2017arXiv171110604D} With the free-energy functionals expressed in terms of artificial neural networks, they found that the \acrshort{ML} functionals can reproduce the density profiles at conditions beyond the training region. Kelley \etal formulated a universal \acrshort{ML} framework and training protocol to learn nonlocal functionals, combining a \acrfull{CNN} with weighted-density approximation for the free energy and a smooth convolution kernel. The \acrshort{ML} functionals achieve excellent accuracy for a diverse set of systems, including the 1D \acrshort{HR} fluids, liquid water, and the Ising model, as well as the exchange energy and the kinetic energy of inhomogeneous electrons.\cite{10.1063/5.0223792} Yatsyshin \etal  introduced a Bayesian method for predicting the external potential of classical particles from the density profiles generated through \acrfull{MC} simulation.\cite{10.1063/5.0146920,10.1063/5.0071629} The \acrshort{ML} regression predicts the grand potential and density profile of 1D \acrshort{HR} systems in good agreement with exact results. Cats \etal combined the \acrshort{cDFT} formalism with a stochastic optimization method to regress the free-energy functional of \acrshort{LJ} fluids.\cite{10.1063/5.0042558} They expressed the non-mean-field component of the free energy in terms of the density profile in quadratic and cubic forms with the kernels obtained by fitting with numerical results obtained from grand-canonical \acrshort{MC} simulation. The \acrshort{ML} functional  provides good predictions of both the thermodynamic properties of inhomogeneous systems and the pair correlation functions of bulk \acrshort{LJ} fluids. Dijkman \etal proposed training the free-energy functional of \acrshort{LJ} fluids exclusively on the \acrfull{RDF} of bulk systems, circumventing the need to sample costly heterogeneous density profiles in a wide variety of external potentials.\cite{PhysRevLett.134.056103,PhysRevE.110.L032601} Samm\"uller \etal applied a fully connected \acrfull{DNN} to map the local value of the one-body direct correlation function as a functional of the one-body density profile.\cite{RN141} In comparison to alternative methods, learning the one-body direct correlation function has advantages because it is typically short-ranged, allowing for quasi-local approximation to reduce the length scale of input space. The neural functional outperforms the state-of-the-art \acrshort{cDFT} methods for hard-sphere fluids and is able to capture a broad spectrum of physical behavior associated with liquid-gas phase coexistence in bulk and at interfaces.\cite{PhysRevX.15.011013} Bui and Cox incorporated the neural functional approach for one-body direct correlation functions with the local molecular field theory to achieve an accurate description of the structure and thermodynamics of electrolyte solutions and ionic liquids near various charged surfaces.\cite{PhysRevLett.134.148001} Yang \etal demonstrated that an optimized operator learning method can be established to represent the functional relation between the density profile and the one-body direct correlation function for polyatomic systems.\cite{doi:10.1021/acs.jctc.5c00484} 

In the realm of \acrshort{cDFT}, artificial neural networks can also be used as surrogates for the iterative solution of the  \acrfull{EL} equation.\cite{PhysRevE.89.053316} Furthermore, \acrshort{ML} methods can be used to establish operator relationships  among the one-body external potential, the one-body density profile, and the one-body direct correlation function. For example, Fang \etal utilized a \acrfull{PP-GP} emulator to predict the one-body density profile from the one-body external potential for 1D \acrshort{HR} fluids.\cite{RN139} The infinite-dimensional mapping is applicable to high-fidelity predictions with a controlled predictive error and is more computationally scalable than direct \acrshort{cDFT} calculations. 

While neural networks can achieve high accuracy in learning functional relationships, they typically require extensive training data and do not provide reliable error estimates when applied outside the training distribution.\cite{li2020fourier} Moreover, neural functionals trained on a specific class of external potentials may not generalize well to other classes due to distributional shifts between the training and testing inputs, limiting their applicability to realistic molecular systems.\cite{wang2023scientific} In contrast, operator learning offers a more fundamental and flexible approach by learning relationships  between functions—i.e., mathematical operators—directly in function spaces. Compared to traditional functional learning methods, which map functions to scalar outputs, operator learning naturally accommodates functional inputs, is invariant to discretization, and provides improved generalization and computational efficiency for inference at variable resolutions.\cite{JMLR2023}

In this work, we investigate the performance of several neural operators in learning the mappings from the one-body density profile \glssymbol{rho} to the one-body direct correlation function \glssymbol{corrf} and that from the external potential \glssymbol{Vext} to \glssymbol{rho}. Reference data for training and testing the \acrshort{ML} models are generated from the analytical equations for 1D \acrshort{HR} fluids.\cite{RN183, RN139} This prototypical model enables fast generation of the training data and easy construction of interpolation and extrapolation data sets for calibrating various \acrshort{ML} methods.\cite{RN144} Several variants of the \acrfull{DeepONet} and \acrfull{FNO} models are considered, which incorporate different network architectures, activation functions, and training strategies. These operator learning methods are evaluated against \acrshort{DNN} as the baseline. We demonstrate that operator learning methods can be used to effectively learn the functional relationship between the density profile and the one-body direct correlation function, and that between the external potential and the density profile. Analysis based on the Chinchilla scaling law,\cite{ kaplan2020scalinglawsneurallanguage} which describes how model performance scales with compute, model size, and data size, shows that operator learning methods generalize better than traditional discrete \acrshort{ML} approaches. 

\section{Data Preparation and ML Methods} \label{section:data methods}
\subsection{cDFT}
To illustrate the basic features of \acrshort{cDFT}, consider a thermodynamic system of classical particles specified by absolute temperature $T$ and chemical potential \glssymbol{mu}. The training data for establishing the functional relation between the one-body density profile \glssymbol{rho} and the one-body direct correlation function \glssymbol{corrf} can be generated from molecular simulation (or analytical methods) by systematically varying the external potential \glssymbol{Vext}. These real-valued functions are defined in the same spatial domain \glssymbol{Dset} that defines the system volume, and they are interconnected through the \acrshort{EL} equation:
\begin{equation} \label{eqn:cdft:EL-eqn}
\rho(x)\Lambda^D=\exp\left[ -\beta V_{loc}(x) + {\corrf}(x) \right]
\end{equation}   
where $\glssymbol{Vloc}\equiv \glssymbol{Vext} - \glssymbol{mu}$ represents the background potential, \glssymbol{Lambda} denotes the thermal wavelength, and $D$ stands for the dimensionality of spatial domain \glssymbol{Dset}. As usual, $\glssymbol{beta}=1/(k_BT)$ and \glssymbol{kB} is the Boltzmann constant. The exact form of \glssymbol{Dset} is system-specific, depending on the space accessible to the classical particles and their configuration. 

\acrshort{cDFT} asserts that both the external potential \glssymbol{Vext} and the correlation function \glssymbol{corrf} are uniquely determined by the density profile \glssymbol{rho}, and that all thermodynamic properties can be derived from these functional relationships.\cite{RN40, RN71, Evans79} By learning the functional relationship between \glssymbol{corrf} and \glssymbol{rho} using \acrshort{ML} methods, it is, in principle, possible to predict the microscopic structure and thermodynamic properties under arbitrary external potentials—without explicitly sampling the system’s microstates. This data-driven approach eliminates the need for approximations often required in constructing the free-energy functional. Furthermore, learning the mapping from \glssymbol{Vloc} to \glssymbol{rho} allows us to bypass the computationally intensive task of minimizing the grand potential or solving the \acrshort{EL} equation.

\begin{figure}[ht!]
\centering
\includegraphics[width=1.0\linewidth]{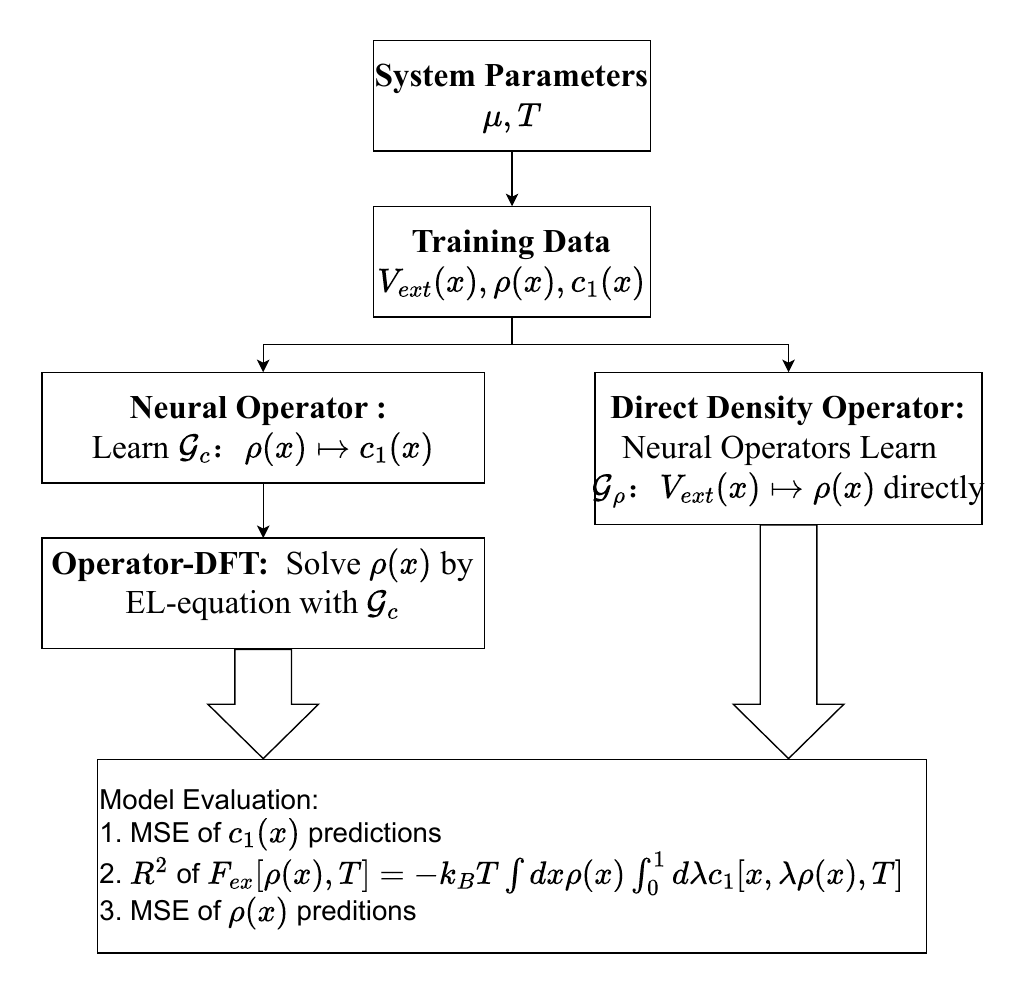}
\caption{\label{fig:flowchart} Conceptual diagram illustrating application of neural operators in cDFT.}
\end{figure}

Figure \ref{fig:flowchart} illustrates the overall workflow for applying the operator learning methods in \acrshort{cDFT} calculations. Our goal is to develop a neural operator $\glssymbol{Goperator}_C$ that maps the one-body direct correlation function \glssymbol{corrf} as a functional of the density profile \glssymbol{rho}. For a given external potential \glssymbol{Vext}, \glssymbol{rho} can then be obtained by solving the \acrshort{EL} equation. To overcome the numerical difficulties, we may develop another operator $\glssymbol{Goperator}_{\rho}$ mapping \glssymbol{rho} as a functional \glssymbol{Vext}. The performance of various neural operators  for predicting ${\corrf}(x)$ and $\rho(x)$ can be measured in terms of the \acrfull{MSE}.  Given absolute temperature $T$ and bulk chemical potential $\mu$, the excess Helmholtz energy $\glssymbol{Fex}[\rho(x)]$ and other thermodynamic properties can be subsequently calculated through functional integration of ${\corrf}[x,\rho(x)]$. 
 
To investigate the efficiency of different \acrshort{ML} methods, we select the one-dimensional \acrshort{HR} model as a benchmark due to its efficiency in data generation. This statistical-mechanical model provides analytical solutions for the density profile $\rho(x)$ and the grand potential \glssymbol{Omega} as functions of the background potential $V_{loc}(x)$.\cite{RN183,RN158} The corresponding analytical expressions are presented in Appendix~\ref{appendix:c1_HR}. The \acrshort{HR} system is athermal, meaning that temperature effects are implicitly included by expressing all energetic quantities in units of $k_B T$. For the \acrshort{HR} system, the local value of the correlation function \glssymbol{corrf} is strictly determined by the one-body density profile within a neighborhood of size equal to the particle diameter \glssymbol{HRdia}. More generally, the local \glssymbol{corrf} is influenced by the density profile within a range of a few times the correlation length \glssymbol{rcorr}. In this work, learning the functional relation between \glssymbol{corrf} and \glssymbol{rho} within a cutoff distance of $\rcutoff = \rcorr$ is referred to as the quasi-local approximation, while learning the mapping over the entire domain \glssymbol{Dset} is termed the "full-range" approach. This quasi-local concept parallels the use of periodic boundary conditions in molecular simulations. Notably, the correlation length diverges at the vapor–liquid critical point, rendering the quasi-local approximation invalid in the critical region and posing a significant theoretical challenge for methods relying on finite-range assumptions.

For the preparation of training data, we use dimensionless units, with length expressed in terms of the \acrshort{HR} diameter $\HRdia = 1$ and energy in units of $k_B T$. The one-body external potential $V_{ext}(x)$, the one-body density profile $\rho(x)$, and the one-body direct correlation function ${\corrf}(x)$ are all one-dimensional functions defined over the domain $x \in \Dset = [0, L]$, where \glssymbol{L} denotes the system size in units of the \acrshort{HR} diameter. All \acrshort{ML} models were trained and tested using five groups of $V_{loc}(x)$, each defined over the same domain $\Dset$ with $\glssymbol{L} = 8$. Each group contains 2000 samples. The corresponding values of the dimensionless chemical potential $\beta \mu$ span a range from 0 to 4, with the thermal wavelength set to $\Lambda = 1$. The various functional forms of the external potential $V_{ext}(x)$ are provided in Appendix~\ref{appendix:V_ext}. For training all \acrshort{ML} models, data points (i.e., sensors) are taken at $x_{sen} = [x_i]$, where $x_i \in [0, L]$, are uniformly spaced with a grid resolution of $\glssymbol{dx} = x_{i+1} - x_i = 0.01$. Data of $\glssymbol{L} = 4, 5, 6, 7, 10, 12$ with same data structure is used for box system size impact analysis for the best performance model. 1D \acrshort{LJ} fluid with $\glssymbol{L} = 8$, the same  $V_{ext}(x)$ and $\rho_b\in (0.01,1)$, generated using cDFT with weighted density approximation, is  used to show the model performances on systems with intermolecular interactions. 

\begin{figure}[ht!]
\centering
\vspace{-0.75\baselineskip}
%\begin{subfigure}{0.45\linewidth}
%\vspace{0\baselineskip}
%\caption{}
\subfloat[]{
\includegraphics[width=0.45\linewidth,trim= {0em 1em 0em 1em},clip]{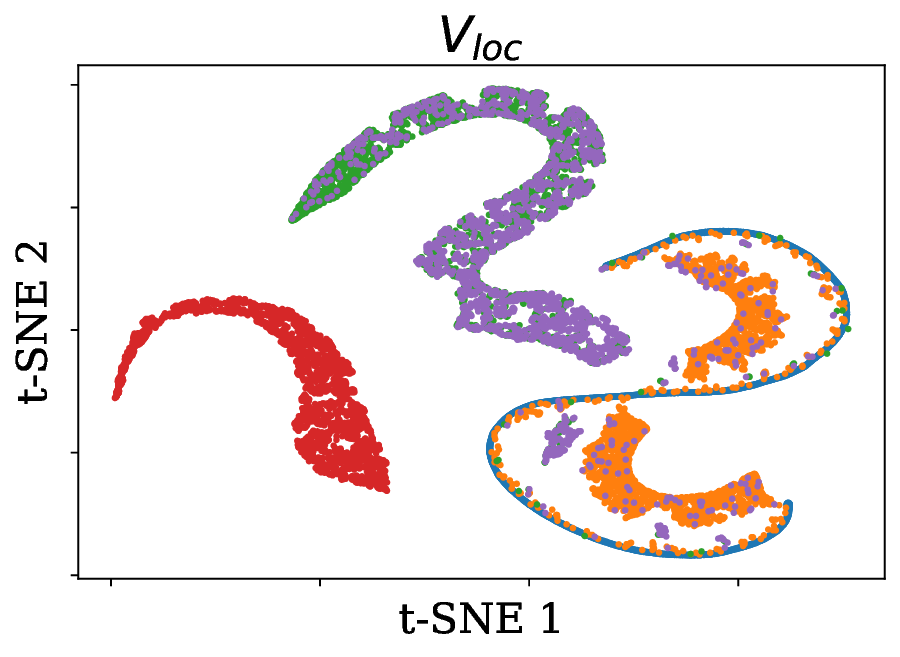} }
%\caption{t-SNE visualization of backgroud potential $\mu-V_{ext}(\disxvec)$}
%\end{subfigure}
%\begin{subfigure}{0.45\linewidth}
%\vspace{0\baselineskip}
%\caption{}
\vspace{-0.75\baselineskip}
\subfloat[]{
\includegraphics[width=0.45\linewidth,trim= {0em 1em 0em 1em},clip]{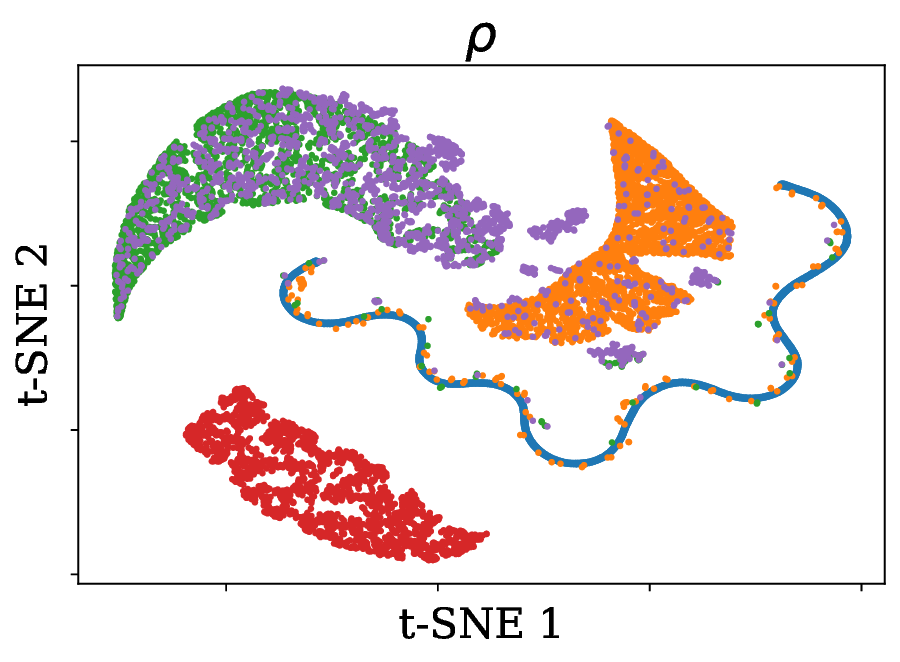} 
}

%\caption{t-SNE visualization of input $\rho(\disxvec)$}
%\end{subfigure}
%\begin{subfigure}{0.45\linewidth}
%\vspace{0\baselineskip}
%\caption{}
\vspace{-0.4\baselineskip}
\subfloat[]{
\includegraphics[width=0.45\linewidth,trim= {0em 1em 0em 1em},clip]{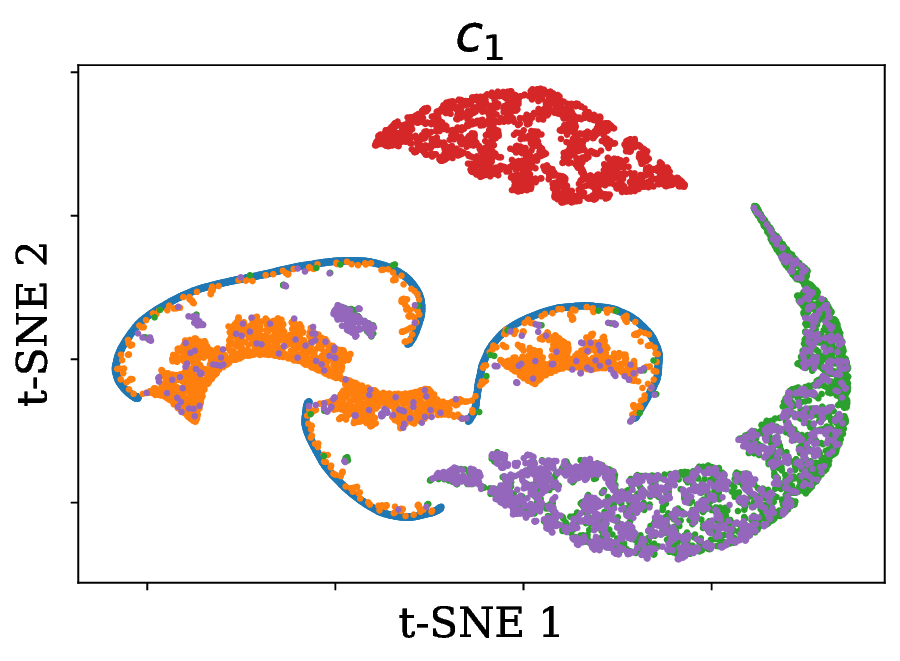} 
}
%\caption{t-SNE visualization of output ${\corrf}(\disxvec)$}
%\end{subfigure}
%\vspace{-0.75\baselineskip}
%\begin{subfigure}{0.45\linewidth}
%\vspace{0\baselineskip}
%\caption{}
%\vspace{-0.4\baselineskip}
\vspace{-0.4\baselineskip}
\subfloat[]{
\includegraphics[width=0.45\linewidth,trim= {2em 0em 2em 0em},clip]{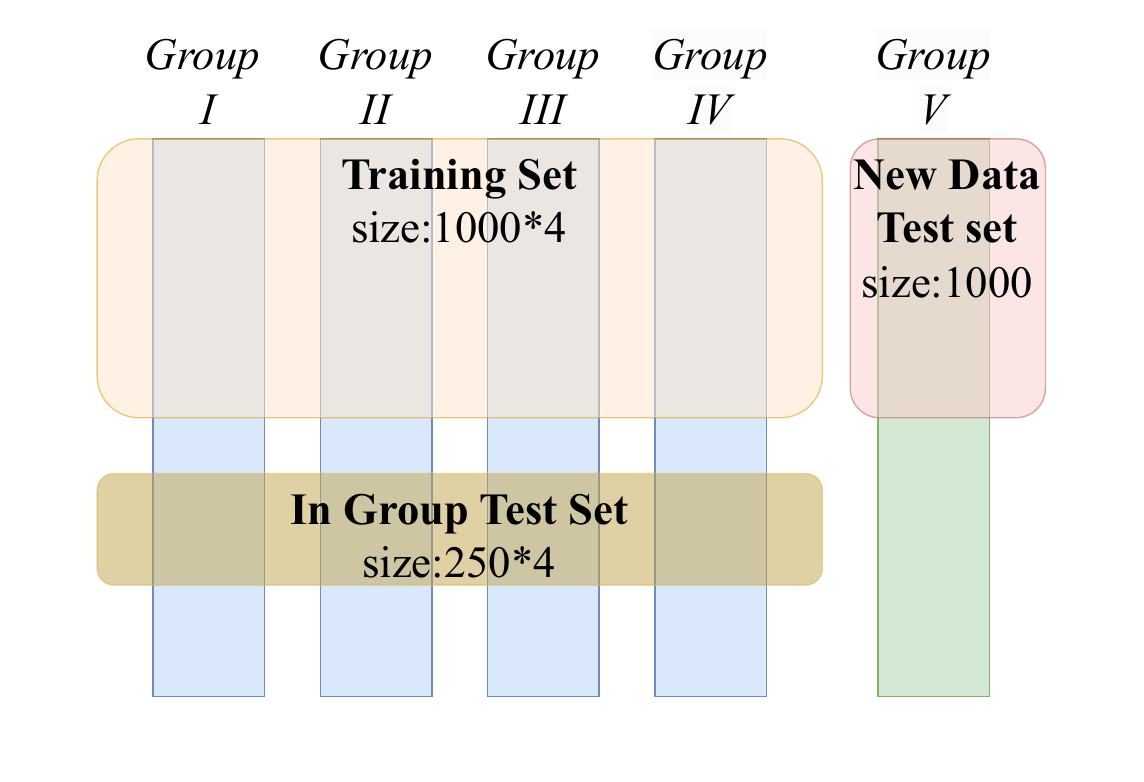} }
%\caption{Data split of training set, In-group test set and New Data test set}
%\end{subfigure}
%\begin{subfigure}{0.8\linewidth}

\includegraphics[width=0.8\linewidth,trim= {2em 0em 0em 0em},clip]{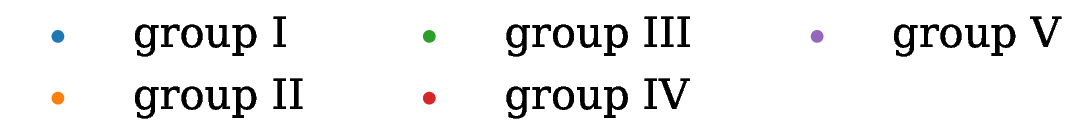} 
%\caption{Data split of training set, In-group test set and New Data test set}
%\end{subfigure}
\vspace{-0.75\baselineskip}
\caption{\label{fig:data} t-SNE visualization of (a) the reduced background potential $\beta V_{loc}(x)$, (b) the one-body density profile $\rho(x)$, and (c) the one-body direct correlation function ${\corrf}(x)$. The t-SNE plots were generated using the Barnes-Hut algorithm with the following parameters: perplexity = 25, learning rate = 100, and maximum number of iterations = 1000.(d) Data split diagram illustrating the training set, in-group test set, and new data test set. 
Legends for the points shown in panels a-c are: Group I: Cyan; Group II: Orange; Group III: Green; Group IV: Red; Group V: Violet. }
\end{figure}

Figure~\ref{fig:data}(a--c) shows the high-dimensional data for different groups of background potential $V_{loc}(x)$, density profile $\rho(x)$, and one-body direct correlation function ${\corrf}(x)$ using t-distributed stochastic neighbor embedding (t-SNE), a statistical technique that projects high-dimensional data into two or three dimensions for visualization. These t-SNE plots reveal latent relationships among the functional datasets, emphasizing emergent clustering behavior that reflects underlying phase states or variations in external potentials. The functions in Group I are shown as curves because they are generated from a spatially uniform background potential $V_{loc}(x)$, \ie, the external potential is a constant. For Group V, the external potential $V_{ext}$ is constructed as a combination of potentials from Groups II and III. Consequently, the t-SNE embeddings of the Group V functions overlap with those of Groups II and III, with some data points positioned in between, reflecting their hybrid nature. 

Figure~\ref{fig:data}(d) illustrates the structure of the dataset used for training and testing the \acrshort{ML} models. The training set and in-group test set are sampled from Groups I–IV. The training set consists of 1000 randomly selected samples from each group, while the in-group test set, used for random cross-validation (CV), includes 250 samples per group. The new data test set comprises 1000 samples from Group V and is used for leave-one-group cross-validation (LOG-CV) to evaluate extrapolation performance. All data are normalized using Z-score normalization, with the mean and standard deviation computed across all five groups. Each function has a total of $N_{totsample} = 2000 \times 5 = 10^4$ samples, and each sample contains $\glssymbol{ngrid} = 801$ uniformly spaced sensor grid points. In the following tables and figures, the in-group test set and new data test set are referred to as \textit{in-group} and \textit{new data}, respectively.
 
\subsection{Operator learning models} \label{sec:ML_method}
\acrshort{DeepONet} and \acrshort{FNO} are two powerful \acrshort{ML} models that have been commonly used to solve partial differential equations and learn operator mappings between function spaces.\cite{LU2022114778} While \acrshort{DeepONet} provides a general approach for learning operators across diverse domains, \acrshort{FNO} leverages Fourier transforms to capture global spatial correlations.\cite{li2021} The spectral method is known for efficiency in structured problems, especially in handling smooth functions and structured grids, and works well in the field of potential theory. However, its performance may deteriorate greatly for systems with complex geometries. In both \acrshort{DeepONet} and \acrshort{FNO} models, the input and output functions are discretized and evaluated at sets of distinct positions. For convience, the list of abbreviations and inputs for each machine learning methods are listed in the beginning of the paper. 
All machine learning methods take $\rho(x)$ as the input function and ${\corrf}(x)$ as the output function, so $\rho(x)\mapsto\corrf(x)$ mapping are learned in this project. Besides this, $V_{ext}(x)\mapsto \rho(x)$ mapping is also tested, means $V_{ext}(x)$ is used as input function and $\rho(x)$ is used as the output function for machine learning to fir, All functions, including $\rho(x)$, $\corrf(x)$ and $V_{ext}(x)$  are measured at the sensor  \glssymbol{disxvec} with $\Delta x=0.01$ and 801 points. 

\subsubsection{ \glsentrylong{DeepONet} (\glsentryshort{DeepONet})}
\begin{figure}[ht!]
\centering
\includegraphics[width=\linewidth]{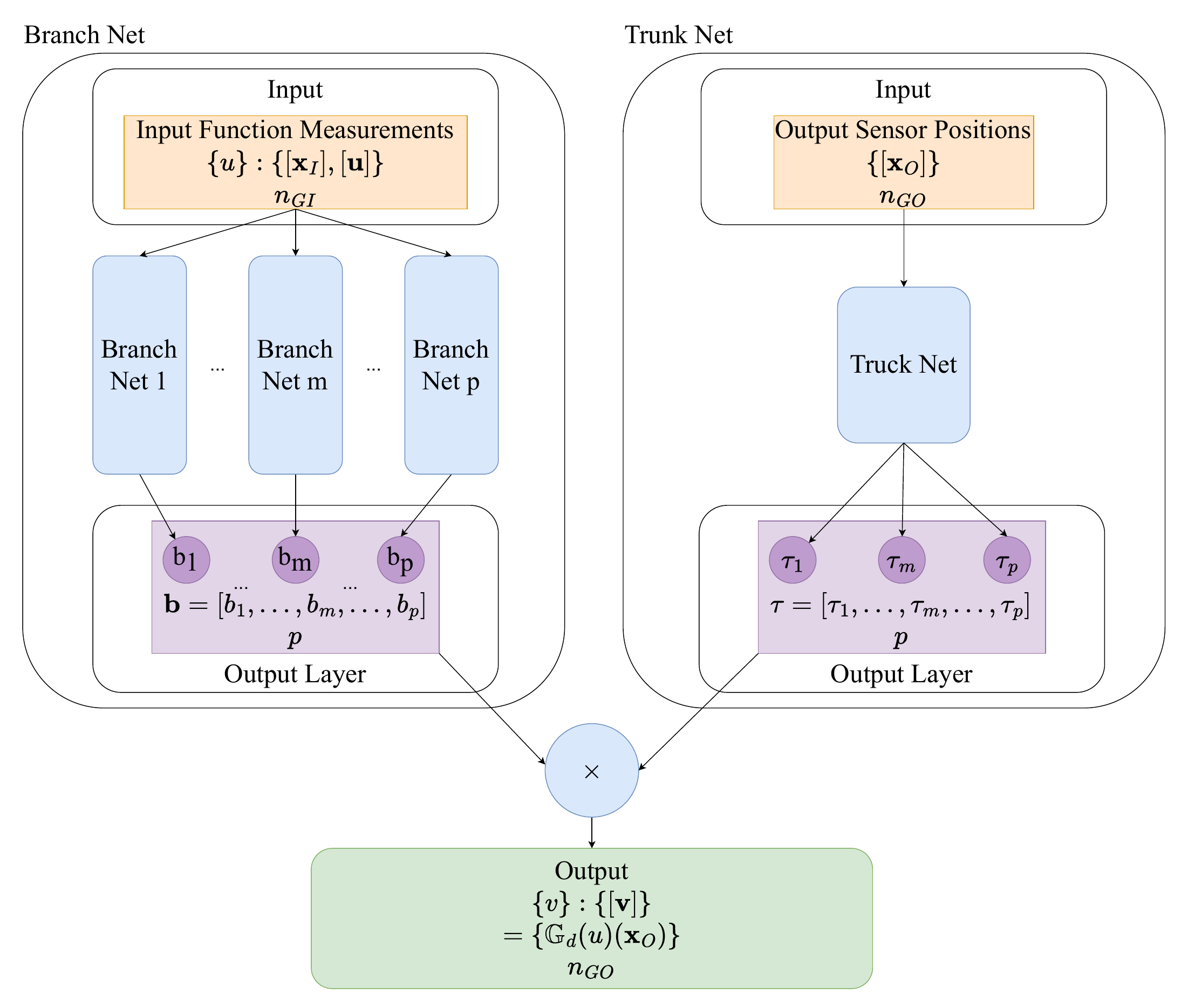}
\caption{\label{fig:DeepONet} Schematic of the general \acrshort{DeepONet} architecture. The branch net consists of $p$ unstacked sub-networks, each processing components of the discretized input function. The trunk net is a single stacked network that produces $p$ outputs corresponding to the sensor locations of the output functions.}
\end{figure}

As illustrated in Figure~\ref{fig:DeepONet}, \acrshort{DeepONet} consists of two subsidiary neural networks, known as the branch net and the trunk net. The branch net processes the discretized input function $\glssymbol{ufunction}(x)$,  which is $\rho(x)$ in this work, as a multidimensional vector by evaluating it at a set of positions $ \glssymbol{xin} = (x_{I,1}, x_{I,2}, \cdots, x_{I,n_{GI}})$:
\begin{equation}
\glssymbol{umeasures} = [u(x_{I,1}), u(x_{I,2}), \cdots, u(x_{I,n_{GI}})],
\end{equation}
where $n_{GI}$ denotes the number of input grid points. The trunk net, in contrast, takes the output sensor positions $ \glssymbol{xout}$ as its input, $ \glssymbol{xout} = (x_{O,1}, x_{O,2}, \cdots, x_{O,n_{GO}})$, where $n_{GO}$ is the number of grid points for the output function. While the grid points for the input and output functions need not be the same, for any given sample, the output layers of the branch net, ${\bnetout} = [b_1, \ldots, b_p]$, and the trunk net, ${\tnetout} = [\tau_1, \ldots, \tau_p]$, are both one-dimensional vectors of the same length $p$. In our implementation of \acrshort{DeepONet}, the output size of both networks is fixed at $p = 512$, commensurate with the domain size $\Dset = [0, L]$. The final output function $\glssymbol{vfunction}(x)$,  which is $\corrf(x)$, is computed as the inner product of \glssymbol{bnetout} and \glssymbol{tnetout}, plus a bias term $b_0$, and is represented as another multidimensional vector:
\begin{equation}
\glssymbol{vmeasures}=[v(x_{O,1}),v(x_{O,2}),\cdots,v(x_{O, n_{GO}})],   
\end{equation}
where
\begin{equation} \label{eqn:DON:Operator}
\vfunction(x)=\Goperator(\ufunction)(x)={\bnetout}(\umeasures)\cdot {\tnetout}(x) + b_0.
\end{equation}

\acrshort{DeepONet} is not tied to any specific architecture for its branch or trunk networks. Instead, it adopts \acrshort{DNN}s as the default implementation. In this work, we refer to this architecture as \acrshort{DNN}-\acrshort{DeepONet}, and use it as one of the benchmarks for comparing various \acrshort{ML} methods. As illustrated in \Cref{appendix:additional} (Figure~\ref{fig:Structures_sub}), \acrshort{DNN}-\acrshort{DeepONet} consists of two hidden layers with 512 nodes each in the branch \acrshort{DNN} net, and a single hidden layer with 512 nodes in the trunk \acrshort{DNN} net.

The performance and training efficiency of \acrshort{DeepONet} can be enhanced by optimizing the structures of both the trunk and branch networks.\cite{U-DeepONet,RN171,RN148} In our case, the input to the trunk net is simply a Cartesian grid. Prior studies have shown that for such structured sensor inputs, the choice of trunk net architecture has minimal impact on training outcomes due to the limited information content.\cite{RN143} Therefore, we adopt the default architecture for the trunk net without further optimization.

In contrast, the branch net encodes rich information from the input function, and its architecture plays a critical role in determining the performance. To optimize the network structure, we explore two advanced architectures for the branch net: a standard \acrshort{CNN} and a \acrfull{RMSCNN} (multi-scale \acrshort{CNN} with \acrfull{ResNet}). These network designs have been widely applied to computer vision tasks such as image detection and restoration.\cite{ResUNet,Res2Net_2021,sosnovik2021disco} The architectural flowcharts of these branch networks are presented in \Cref{appendix:additional}  (Figures~\ref{fig:CNN_Structures} and~\ref{fig:MSCNN_Structures}).

We consider \acrshort{CNN} in our investigation due to its ability to generate translation-equivariant responses, which is especially suitable for image-like functional inputs. In the context of \acrshort{cDFT}, the translational symmetry applies both to the entire system and to any non-interacting subsystems. Pooling, a fundamental operation in \acrshort{CNN}, is used to reduce the spatial resolution of feature maps while retaining essential information. This approach improves computational efficiency and enhances robustness to variations in the input function.

\acrshort{RMSCNN} is also considered due to its deeper architecture and use of smaller convolutional kernels compared to standard \acrshort{CNN}s. This design enables the extraction of long-range dependencies in deeper layers. Additionally, the skip-connections in \acrshort{ResNet} help preserve high-resolution features that might otherwise be lost due to pooling operations.

Kelley \etal\cite{2017arXiv171110604D} demonstrated that Gaussian functions used as convolution kernels yield the best performance among a wide class of smooth functions in \acrshort{CNN}-based models for \acrshort{DFT} problems. Following their findings, we adopt the same Gaussian convolution kernels in this work, defined as
\begin{equation*}
    G_{\mathrm{gau}}(\sigma)(x) = \mathcal{N}(0, \sigma^2)(x) * (C_0 + C_1 x),
\end{equation*}
where $\mathcal{N}(0, \sigma^2)(x)$ denotes a Gaussian distribution with zero mean and variance $\sigma^2$, and $*$ represents convolution. These kernels are referred to as \acrfull{GK} when used within the \acrshort{CNN}-\acrshort{DeepONet} framework. For benchmarking purposes, the performance of the \acrshort{GK} kernels is compared with that of the discrete kernels.

\begin{figure*}[ht!]
\centering
\vspace{-0.75\baselineskip}
%\begin{subfigure}{0.9\linewidth}
%\vspace{0\baselineskip}
%\caption{}
%\vspace{-0.4\baselineskip}
\subfloat[]{
\includegraphics[width=0.9\linewidth,trim= 1em 1em 1em 1em,clip]{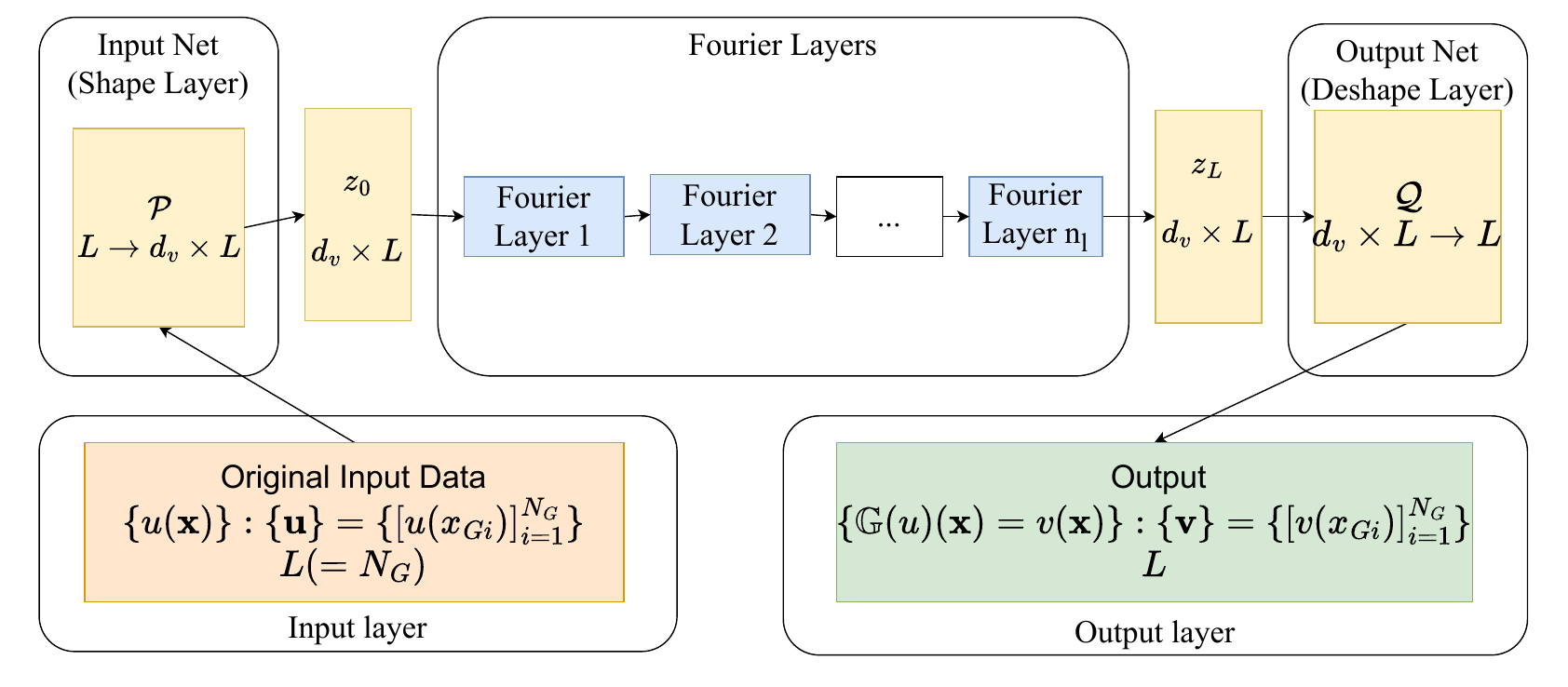}
}

%\end{subfigure}
%\begin{subfigure}{0.9\linewidth}
%\vspace{0\baselineskip}
%\caption{}
\vspace{-0.4\baselineskip}
\subfloat[]{
\includegraphics[width=0.9\linewidth,trim= {2em 3em 2em 1.5em},clip]{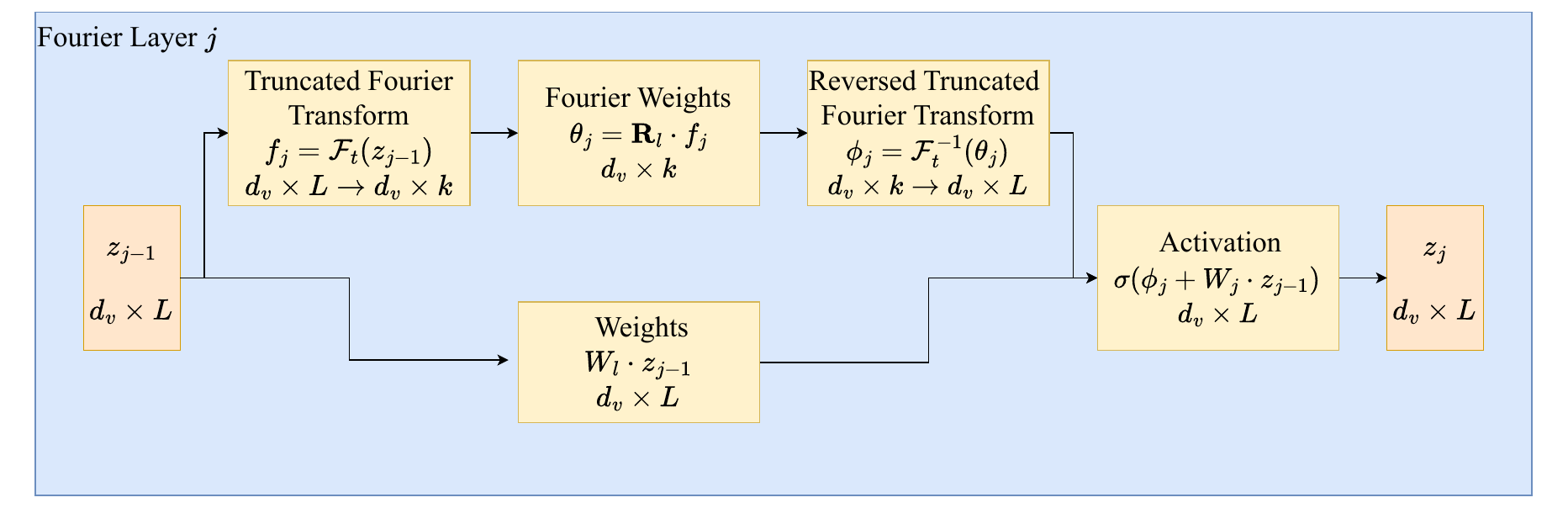}
}
%\end{subfigure}
\caption{\label{fig:FNO} Schematic structures of \acrfull{FNO}. (a) A \acrshort{FNO} model typically consists of three main layers: a lifting layer, iterative Fourier layers (or kernel integration layers), and a projection layer. In the lifting layer, the input network $\mathcal{P}$ `lifts' the original input data $\umeasures$ to a higher-dimensional space represented by the matrix $\zmatrix_0$. This lifted representation is then processed through a sequence of Fourier layers, resulting in matrix $\zmatrix_{\nfl}$, which is subsequently projected back to the target dimension as $\vmeasures$ by the output network $\mathcal{Q}$. (b) The structure of a single Fourier layer. In the $j$-th Fourier layer, the truncated Fourier transform ($\FFT$) of the input matrix $\zmatrix_{j-1}$ is linearly transformed using the Fourier weights matrix $\mathbf{R}_l$. The transformed data is then subjected to an inverse truncated Fourier transform ($\FFT^{-1}$), with the unused modes padded with zeros. The resulting output is added to a bias-weighted version of the input, $\mathbf{W}_i \cdot \zmatrix_{j-1}$, and passed through an activation function $\acti$.}
\end{figure*}

\subsubsection{ \glsentrylong{FNO}(\glsentryshort{FNO})}
\acrshort{FNO} leverages Fourier transforms to capture global spatial correlations between input and output functions. In contrast to \acrshort{DeepONet}, it requires that both input and output functions,  $\glssymbol{ufunction}(x)$ and $\glssymbol{vfunction}(x)$, be defined over the same domain, $\Dset = \Dset_{in} = \Dset_{out}$. For generating the training data, these functions are discretized on a uniform grid:
\begin{equation}
{  \xin} = {  \xout} = { \disxvec} = 
(x_1, x_2, \cdots,x_{N_G})
\end{equation}
where the grid spacing $\Delta x = x_{i+1} - x_i$ is constant. In the application of \acrshort{FNO} to \acrshort{cDFT}, all local properties, including the density profile $\rho(x)$, the reduced background potential $\beta V_{loc}(x)$, and the one-body direct correlation function ${\corrf}(x,[\rho(x)])$, are defined on a real-space domain $x \in \Dset$ with a uniform grid spacing. As a result, these functions naturally satisfy the input format requirements of the \acrshort{FNO} framework. \glssymbol{corrfoperator} is our major aim to learn, with \glssymbol{rho} and \glssymbol{corrf} being the input and output function. 

To establish the operator mapping between functions $\glssymbol{ufunction}(x)$ and $\glssymbol{vfunction}(x)$ within \acrshort{FNO}, the input network \glssymbol{Player} and output network \glssymbol{Qlayer} are realized through linear transformations that expand the data dimensionality:
\begin{equation}
\begin{aligned}
    &\mathcal{P}: [{\zmatrix}_0( \glssymbol{disxvec})]_j = u( \glssymbol{disxvec}) \\
    &\mathcal{Q}: v( \glssymbol{disxvec}) = \sum_{i=1}^{d_v} w_i \cdot [{\zmatrix}_{\nfl}( \glssymbol{disxvec})]_i
\end{aligned}
\end{equation}
where $i \in (1, d_v)$, \glssymbol{d_v} denotes the number of hidden channels, and \glssymbol{N_FL} is the number of Fourier layers. The intermediate representation ${\glssymbol{zmatrix}}_j \in \mathbb{R}^{d_v \times N_G}$ is the output of the $j$-th Fourier layer evaluated on the $N_G$ uniform grid points. As shown in Figure \ref{fig:FNO}, ${\glssymbol{zmatrix}}_0$ is the lifted input matrix, and ${\glssymbol{zmatrix}}_{\glssymbol{N_FL}}$ is the final hidden representation passed to the output net. The $i$-th channel of ${\zmatrix}_0$ and ${\zmatrix}_{\glssymbol{N_FL}}$ is denoted by $[{\zmatrix}_0]_i$ and $[{\zmatrix}_{\glssymbol{N_FL}}]_i$, respectively. The output net \glssymbol{Qlayer} computes the predicted function $v(x)$ by performing a weighted summation across the \glssymbol{d_v} channels of ${\zmatrix}_{\glssymbol{N_FL}}$, using the learned weights $\mathbf{w} = [w_1, \cdots, w_{d_v}]$.

Several Fourier layers are applied between \glssymbol{Player} and \glssymbol{Qlayer}. In each Fourier layer, $\glssymbol{zmatrix}_j$ are transformed into Fourier space, and multiplied with a trainable weight matrix $\mathbf{R}_l$, then transformed back to spatial space undergoes the activation function \glssymbol{acti}.
The convolution in the spatial domain is equal to a point-wise multiplication in the Fourier domain, so linear transformations in the Fourier space can do the same job as a smooth convolution filter in the spatial space, and being much faster than convolution calculations. The truncated \acrshort{FFT} is used to reduce computational cost without causing a significant loss in prediction performance, as the pooling layer in \acrshort{CNN} does. In general, the Fourier layers are surrogate of a multi-layer \acrshort{CNN} for smooth functions.

\subsubsection{ \glsentrylong{DNN}(\glsentryshort{DNN})}
For comparison between conventional functional learning and the operator learning methods, we also trained a  full-scale \acrshort{DNN}, meaning it learns the full-scale mapping, and quasi-local  \acrshort{DNN} as implemented in Neural-DFT \cite{Sammüller_2024} which learns the quasi-local mapping of $\rho\mapsto c_1$. These functional networks take $\rho(x)$ as the input and ${\corrf}(x)$ as the output. Both functions are measured at the sensor \glssymbol{disxvec} with $\Delta x=0.01$ and 801 points.   

\subsection{Hyperparameters}  \label{sec:Hyperparameters}
Different hyperparameters were used in evaluating the effect of model size by neural scaling law analysis, including the number of layers and nodes in each layer of all the networks, the detailed hyperparameters of models chosen are listed in Table \ref{tab:SI:Hyperparameter}.\cite{Bahri_2024}  For full-scale \acrshort{DNN}, quasi-local \acrshort{DNN} and \acrshort{DNN}-\acrshort{DeepONet}, we tested architectures with 2 to 5 layers of neural networks, using layer widths $N_d$ selected from $4,16,64,128\text{ and~} 512$.  For \acrshort{FNO}, the number of Fourier layers \glssymbol{N_FL} varied from 2 to 4. In each layer, the number of hidden channels \glssymbol{d_v} was chosen from $4,8,16,64,128$, and the number of retained modes $k$ was chosen from $8,32,128$. Zero-padding was used for all convolution layers. The padded zeros represent the zero-density regions outside of the hard walls. Model-size effects for \acrshort{CNN}-\acrshort{DeepONet} and \acrshort{RMSCNN}-\acrshort{DeepONet} were not tested due to the prohibitively long training times required by these models.

For all \acrshort{ML} models, training was terminated either when the in-group test loss failed to decrease for more than 1,000 consecutive steps or when the total number of training iterations reached 100{,}000. For neural scaling law analysis, the best-performing model within each order of magnitude in model size was selected.

Models with parameter counts \glssymbol{msize} ranging from \(5 \times 10^5\) to \(1.2 \times 10^6\) were chosen for each \acrshort{ML} model type. This includes the best models from standard architectures, as well as \acrshort{CNN}-\acrshort{DeepONet} and \acrshort{RMSCNN}-\acrshort{DeepONet} within the same model size range. These models were also used for additional comparative analyses, including iteration–loss relationships , excess free energy (\(\Fex\)) calculations, and density profile predictions \(\rho(x)\).

The chosen model parameters are as follows: both \acrshort{DNN} and Neural-DFT consist of three hidden layers, each containing 512 nodes. The FNO model includes three Fourier layers, each with \(d_v = 256\) channels and \(k = 32\) truncated modes. In \acrshort{GK}-\acrshort{CNN}-\acrshort{DeepONet}, the branch network (\texttt{bnet}) is constructed with two Gaussian kernel-weighted convolution layers, followed by a dense hidden layer of width 512. The Gaussian kernel employs a cutoff width of 10 grid points, corresponding to a spatial cutoff of \(x_{\text{conv,cut}} = 10 \times \Delta x = 0.1\), to control computational cost. For comparison, a standard \acrshort{DK}-\acrshort{CNN}-\acrshort{DeepONet} model is also tested, featuring two convolutional layers with width 10 and a dense hidden layer of width 512. In \acrshort{GK}-\acrshort{RMSCNN}-\acrshort{DeepONet}, the branch network begins with a 10-width convolution layer, as used in GK-CNN-DeepONet, followed by four layers with kernel width 5. Additionally, a standard \acrshort{DK}-\acrshort{RMSCNN} architecture with six convolutional layers of width 3—similar to that used in Res-U-Net—is tested for comparison.

\subsection{Data structure and methods}
To establish the functional relation \(\rho \mapsto {\corrf}\), the direct correlation functions \({\corrf}(x, \rho(x)) = \Goperator(\rho)(x)\) are computed from a given set of density profiles \(\{\rho(x)\}\). As explained above, all data are normalized prior to training, and all neural network layers are configured to perform auto-normalization. The derivatives of the activation functions remain non-negative during training. Multiple activation functions are employed across the \acrshort{ML} models, including the \acrfull{Relu}, \acrfull{Selu}, the logistic function (commonly referred to as sigmoid in software packages), and the \acrfull{sRelu}. The \acrshort{Selu} and logistic functions are continuous over the entire real line, i.e., \((-\infty, \infty)\), whereas \acrshort{Relu} and \acrshort{sRelu} are discontinuous at zero. Notably, \acrshort{sRelu} has compact support, and activation functions of this type have shown significantly improved performance for \acrshort{CNN}-based \acrshort{PDE} solvers in previous studies~\cite{CNN_compact_acti}. These functions are described in detail in Appendix~\ref{appendix:activations}.

All machine learning routines were implemented in Python~3 and R, using packages from Anaconda and PyPI. \acrshort{DeepONet}s were implemented using the DeepXDE package~\cite{lu2021deepxde}, built on Keras~2 and TensorFlow~2, with TensorFlow Probability as the backend~\cite{2017arXiv171110604D}.  The Trainable Gaussian Derivative Layer for \acrshort{GK}-\acrshort{CNN}s was adapted from the implementation by Penaud-Polge~\cite{RN142}, while the scalable \acrshort{CNN} was implemented using \acrfull{DISCO} by Sosnovik~\cite{sosnovik2021disco}. \acrshort{FNO} models were implemented using the NeuralOperator package~\cite{RN174, kovachki2021neural}, with PyTorch and the \texttt{torch-harmonics} backend. The L-GBFS optimization algorithm is implemented in both TensorFlow Probability and PyTorch. The \acrshort{ALEC}-\acrshort{GPR} models are developed in the R programming language using the RobustGaSP package, with integration into the Python workflow via the \texttt{rpy2} interface~\cite{2019arXiv191201703P, bonev2023spherical}.

\section{Numerical Results} \label{section:Results}
The following eight \acrshort{ML} models were evaluated to approximate the functional relation \(\glssymbol{corrfoperator}: \rho(x) \mapsto \corrf\) between the one-body correlation function ${\corrf}(x)$ and the one-body density profile $\rho(x)$ for 1D \acrshort{HR} fluids:
\begin{enumerate}[label=(\arabic*), itemsep=0pt, topsep=3pt]
    \item Full-scale \acrshort{DNN};
    \item Quasi-local \acrshort{DNN};
    \item \acrshort{DNN}-\acrshort{DeepONet};
    \item \acrshort{GK}-\acrshort{RMSCNN}-\acrshort{DeepONet}; % Retained \acrfull as per original
    \item \acrshort{GK}-\acrshort{CNN}-\acrshort{DeepONet};
    \item \acrshort{DK}-\acrshort{CNN}-\acrshort{DeepONet};
    \item \acrshort{DK}-\acrshort{RMSCNN}-\acrshort{DeepONet};
    \item \acrshort{FNO}.
\end{enumerate}
Models (1)--(3) served as baseline comparisons. The four \acrshort{DeepONet} variants (4)--(7) were specifically included to investigate the influence of different kernel choices (\acrshort{GK} vs. \acrshort{DK}) and neural network architectures (\acrshort{CNN} vs. \acrshort{RMSCNN}) on the neural operator's performance. 

\subsection{Model evaluation metrics}
These \acrshort{ML} models were compared based on several key metrics: the model size, the number of training iterations (epochs), the converged \acrshort{MSE} values achieved on the training set, the in-group test set, and the `new data' test set, as well as the overall training time. The model size refers to the number of trainable parameters, designated as \glssymbol{msize}.

The training loss function is defined as the total \acrfull{MSE} of the predicted operator output \({\corrf}(x)\) across all sensor locations:
\begin{equation} \label{eqn:Loss:MSE}
\text{MSE} = \frac{1}{\nset} \sum_{j=1}^{\nset} \frac{1}{\ngrid} \sum_{x \in \Dset} \left[{\corrf}_{\text{pred},j}(x) - {\corrf}_j(x)\right]^2,
\end{equation}
where \glssymbol{nset}is the number of samples in the dataset, and \glssymbol{ngrid}is the number of spatial grid points used to discretize the domain, given by \(\ngrid = \nin = \nout = L/\Delta x + 1 = 801\). The loss is evaluated separately for the training set, the in-group test set, and the new data test set, as described earlier. A method with a lower \acrshort{MSE} on in-group test set is better on interpolation prediction, while lower \acrshort{MSE} on the new data test set shows better extrapolation performance. 

In addition to the standard analysis of loss versus iteration number, we examine how the loss varies with the number of training samples (\(\nsample\)) and model size (\glssymbol{msize}). Toward that end, the \acrshort{ML} models are trained using \(\nsample\) values ranging from 8 to 4000, with training samples randomly drawn from data groups I–IV. The two test sets remain fixed for all training runs, with identical termination criteria applied in every case. For quasi-local \acrshort{DNN} training, the number of training samples is $\nsample=N_{training,set}*\ngrid$, since quasi-local mapping generates $\ngrid$ training samples on $\corrf$ pointwise from each different \(V_{\text{loc}}(x)\).

\subsection{MSE loss and training time}
Table \ref{tab:Loss_compare_ONs} summarizes the performance of various \acrshort{ML} models. Additional results for three other \acrshort{DeepONet} variants are provided in \Cref{appendix:additional}. Here, the \acrshort{sRelu} activation function was used for all \acrshort{ML} models. This selection was based on its superior extrapolation capabilities compared to other activation functions for most of the tested models, a point discussed further in~\cref{sec:compare_actis}. Furthermore, this particular architecture achieved the lowest \acrshort{MSE} among the four \acrshort{DeepONet} variants evaluated on the in-group test set and the `new data' test set.

The number of trainable parameters was comparable across the eight \acrshort{ML} models, generally on the order of \(10^6\). Notably, the \acrshort{GK}-\acrshort{RMSCNN}-\acrshort{DeepONet} model utilized slightly fewer parameters (\(\approx 6.9 \times 10^5\)). Notably, the quasi-local \acrshort{DNN} exhibits significantly higher accuracy than the full-scale \acrshort{DNN} across the training, in-group, and extrapolation (`new data') sets. Although the quasi-local \acrshort{DNN} requires far fewer iterations, its training time is considerably longer—exceeding that of the full-scale model by more than an order of magnitude.

Both \acrshort{DNN} models show a marked increase in \acrshort{MSE} when evaluated on the `new data' set, indicating limited generalization ability. The hybrid \acrshort{DNN}-\acrshort{DeepONet} model achieves accuracy comparable to the quasi-local \acrshort{DNN}, but with a much larger number of epochs yet a shorter overall training time. Its performance can be further enhanced through alternative neural network architectures, though at the cost of increased training time. Among all \acrshort{ML}  methods tested in this work, \acrshort{FNO} consistently achieves the best performance, particularly in terms of minimizing the loss on the `new data' set. While the full-scale \acrshort{DNN} requires the lest training time, its generalization to new data is significantly weaker compared to other models.

\begin{table*}[ht!] %\acrshort{DeepONet} is abbreviated as "DON" for brevity.
    \centering
    \renewcommand{\arraystretch}{1.2} % Adjust row height for better readability
 \caption{Comparison of five \acrshort{ML} models evaluated in this work in terms of the number of training parameters (\glssymbol{msize}), the number of iterations/epochs ($N_{\text{Epo}}$), and the terminal \acrshort{MSE} losses on different data sets—$\corrf$-\text{MSE}$_T$ (training set), $\corrf$-\text{MSE}$_{IG}$ (in-group set), and $\corrf$-\text{MSE}$_{ND}$ (new data set)—as well as the total training time.  Training settings: The activation function for all models is \acrshort{sRelu}, as detailed in \cref{appendix:activations}. The learning rate is $10^{-3}$ with an exponential decay factor of 0.95 every 100 iterations.}
    \begin{ruledtabular}\begin{tabular}{l m{7em}<{\centering} m{10em}<{\centering} m{7em}<{\centering} m{15em}<{\centering} m{7em}<{\centering}}
        %\hline
    & Full-scale \acrshort{DNN} & Quasi-local \acrshort{DNN} & \acrshort{DNN}-\acrshort{DeepONet} & \acrshort{GK}-\acrshort{RMSCNN}-\acrshort{DeepONet} & \acrshort{FNO} \\
    \hline
    $\msize$ & \(1.1 \times 10^6\) & \(7.8 \times 10^5\) & \(1.0 \times 10^6\) & \(6.9 \times 10^5\) & \(7.5 \times 10^5\) \\
    $N_{Epo}$ & 20000 & 100 & 100000 & 100000 & 2500 \\
    $\corrf-\text{MSE}_T$ & \(5.79 \times 10^{-4}\) & \(1.25 \times 10^{-7}\) & \(8.66 \times 10^{-5}\) & \(3.4 \times 10^{-5}\) & \(5.28 \times 10^{-7}\) \\
    $\corrf-\text{MSE}_{IG}$ & \(5.84 \times 10^{-4}\) & \(1.5 \times 10^{-7}\) & \(2.20 \times 10^{-4}\) & \(4.4 \times 10^{-5}\) & \(5.10 \times 10^{-7}\) \\
    $\corrf-\text{MSE}_{ND}$ & \(2.21 \times 10^{-2}\) & \(7.8 \times 10^{-3}\) & \(8.60 \times 10^{-3}\) & \(9.55 \times 10^{-4}\) & \(7.76 \times 10^{-6}\) \\
    time (s) & 1923 & 58420 & 2285 & 11145 & 6430 \\
       % \hline
    \end{tabular}\end{ruledtabular}
    \label{tab:Loss_compare_ONs}
\end{table*}

\begin{figure}[ht!]
\centering
\vspace{-0.75\baselineskip}
\subfloat[]{
\includegraphics[width=0.45\linewidth,trim={0em 2pt 0em 2pt},clip]{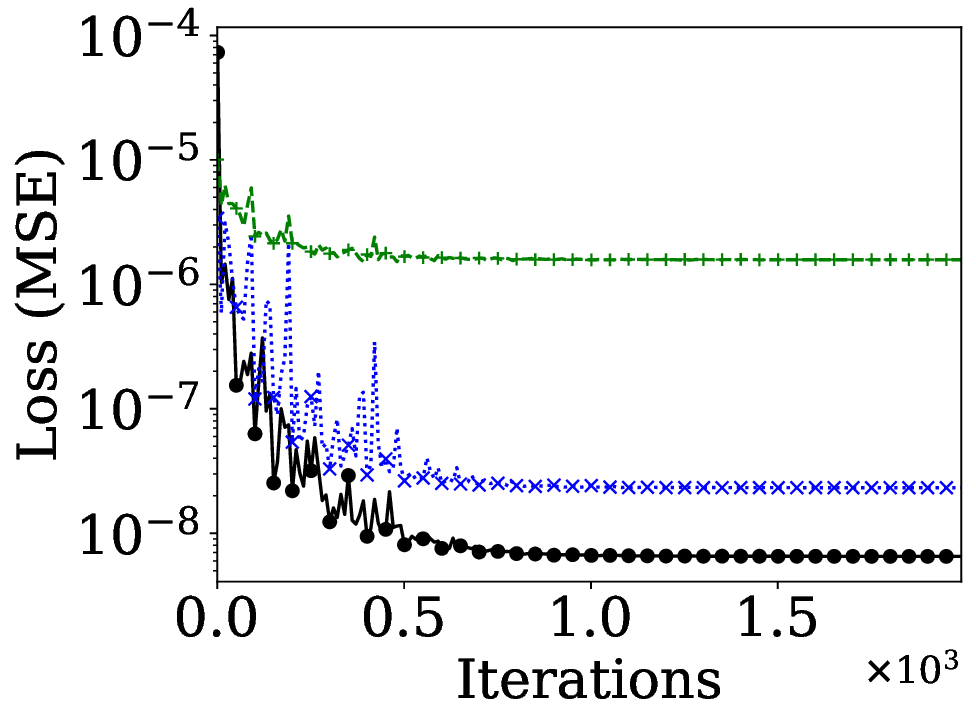}
}
\vspace{-0.75\baselineskip}
\subfloat[]{
\includegraphics[width=0.45\linewidth,trim={0em 5pt 0em 5pt},clip]{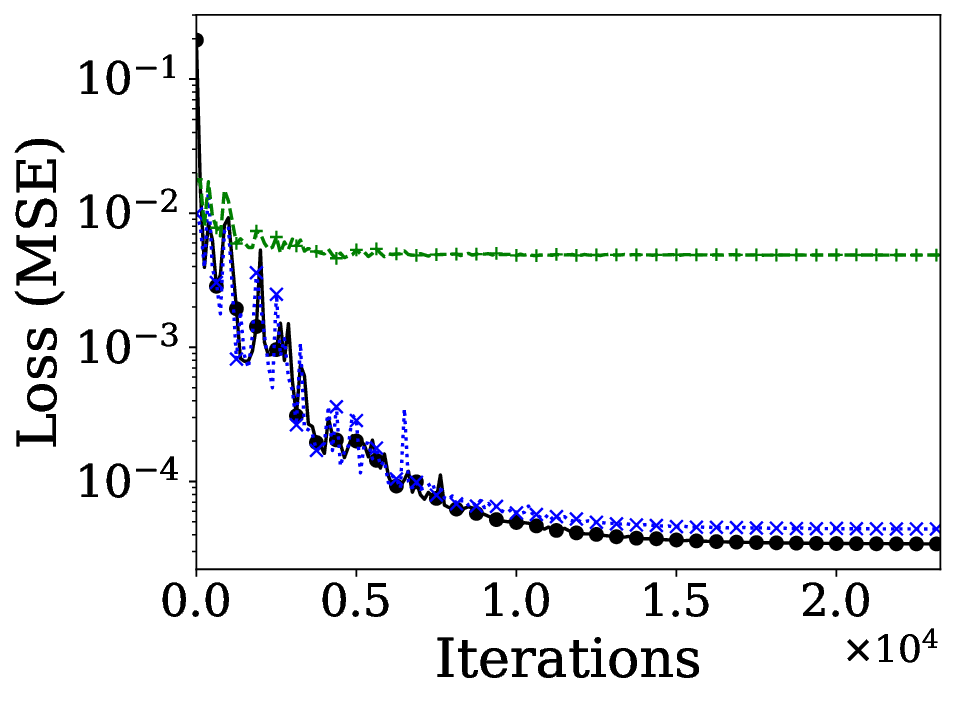}
}

\vspace{-0.75\baselineskip}
\subfloat[]{
\includegraphics[width=0.45\linewidth,trim={0em 5pt 0em 5pt},clip]{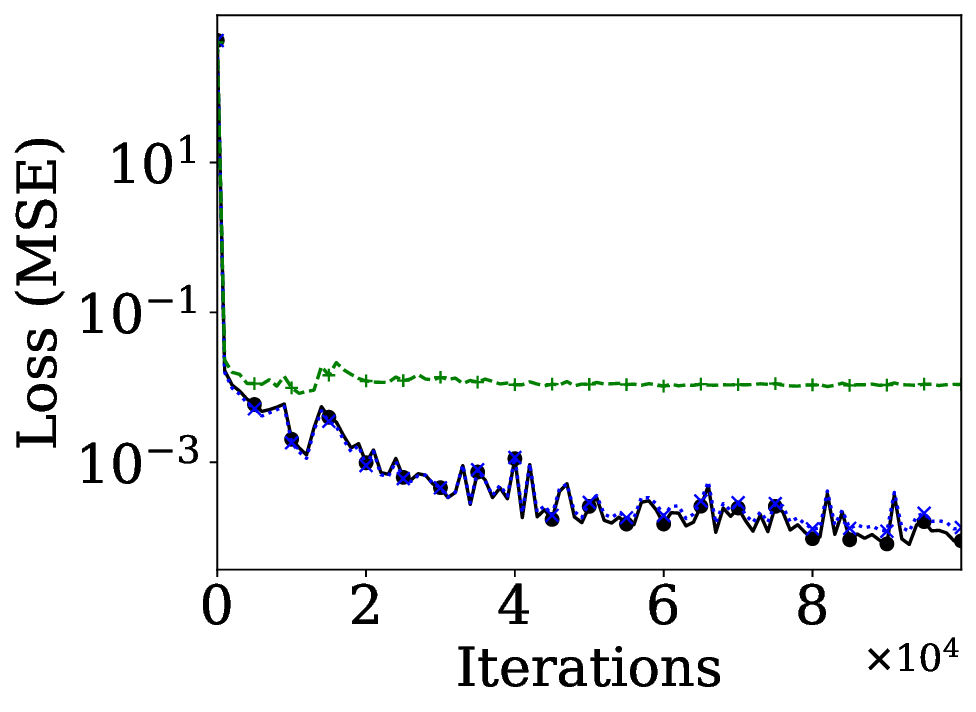}
}
\vspace{-0.75\baselineskip}
\subfloat[]{
\includegraphics[width=0.45\linewidth,trim={0em 2pt 0em 2pt},clip]{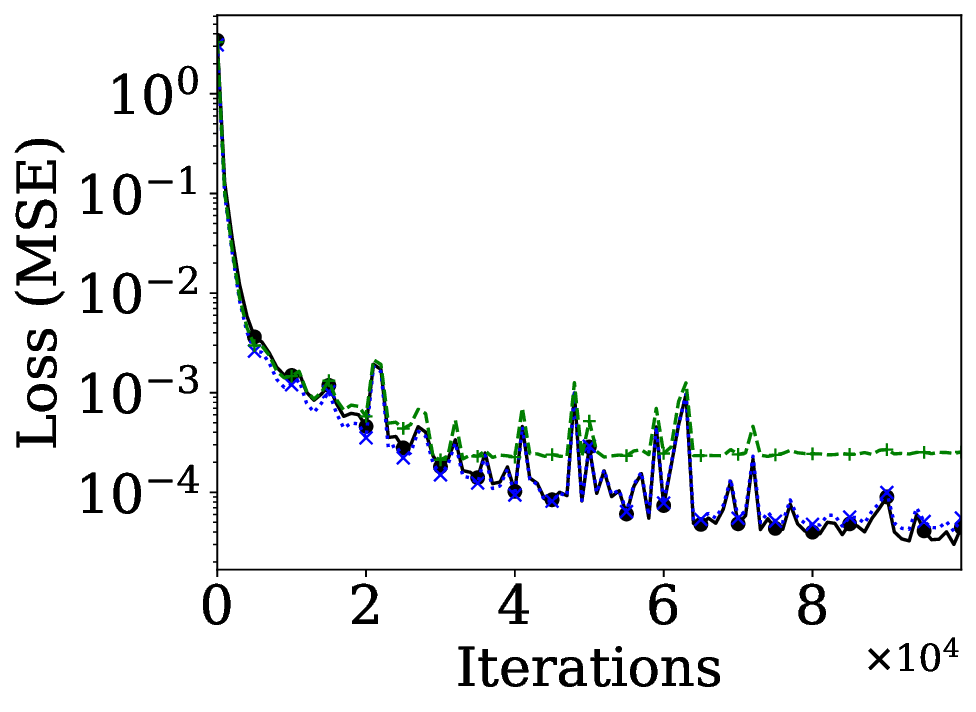}
}

%\begin{subfigure}{0.45\linewidth}
%%\vspace{-2\baselineskip}
%\includegraphics[width=0.4\linewidth]{Figure5leg.eps}
%\end{subfigure}
\subfloat[]{
\includegraphics[width=0.45\linewidth,trim={0em 2pt 0em 2pt},clip]{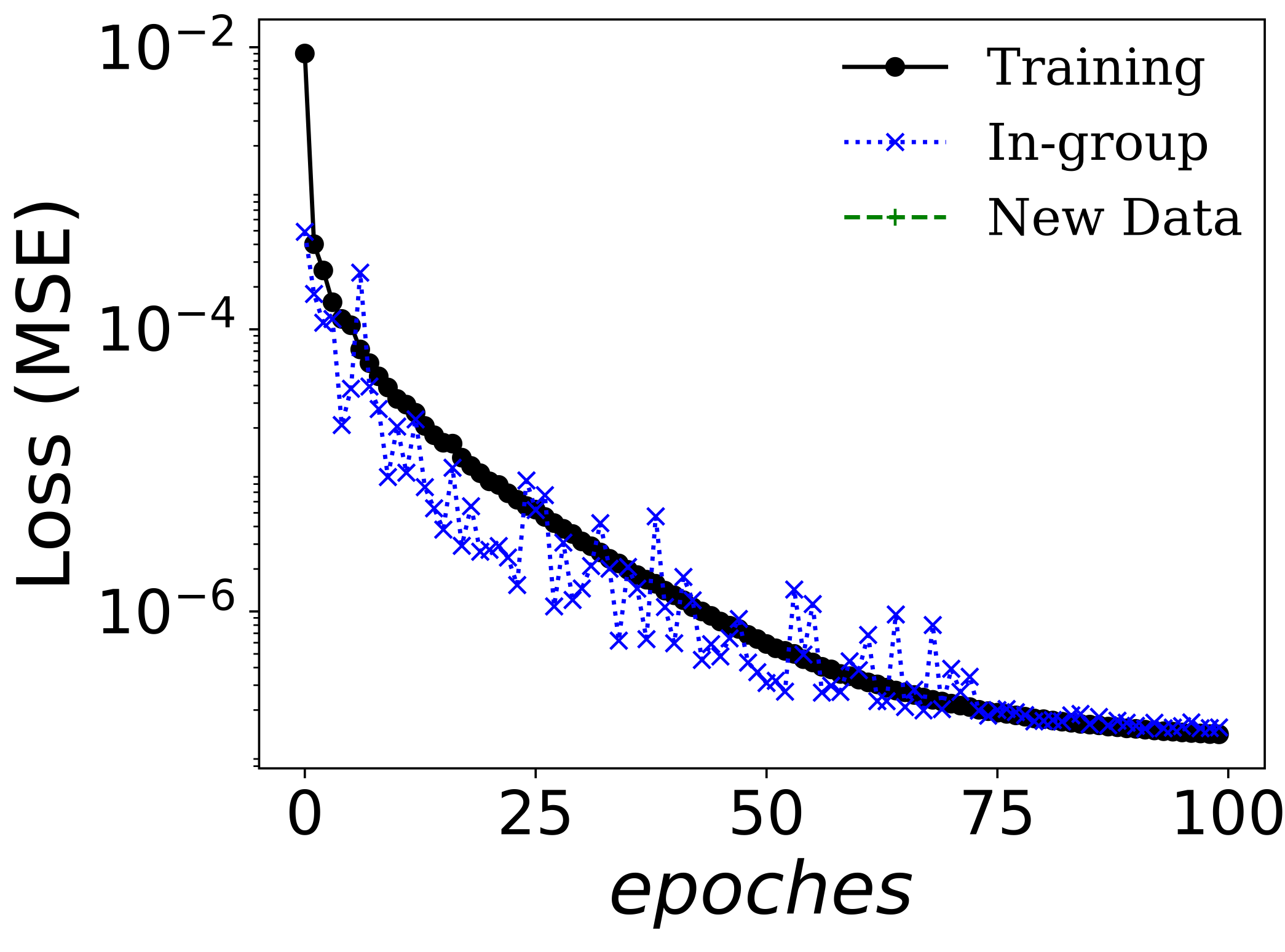}
}

\caption{\label{fig:loss_step} \acrshort{MSE} loss versus the number of iterations/epochs for the training set, in-group test set, and new data test set using different operator learning methods with the same activation function \acrshort{sRelu}. (a) \acrshort{FNO}, (b) Full-scale \acrshort{DNN}, (c) \acrshort{DNN}-\acrshort{DeepONet}, (d) \acrshort{GK}-\acrshort{RMSCNN}-\acrshort{DeepONet}, (e) Quasi-local \acrshort{DNN}. Legends for the curves shown in panels a-d are: Training Set MSE: Black solid line with circle markers; In-Group Test Set MSE: Blue Dotted Line with \text{x} markers; New data Test Set MSE: Green Dashed Line with $+$ markers.  }
\end{figure}

Figure \ref{fig:loss_step} shows the relation between the number of training iterations and the loss values (\acrshort{MSE}) on the training set, the in-group test set, and the `new data' test set. We have no `new data' test set result for the quasi-local \acrshort{DNN} method, as its training framework does not accommodate multiple test sets.  For most \acrshort{ML} methods studied in this work, the training and in-group test errors decrease concurrently and eventually converge to similar values. In contrast, the loss on the `new data' test set stops decreasing much earlier and converges to \acrshort{MSE} values that are approximately 10 to 100 times higher than those of the training and in-group test errors. This clearly indicates that all models demonstrate strong interpolation performance but significantly weaker extrapolation capabilities. As mentioned above, \acrshort{FNO} exhibits the best extrapolation performance, with the smallest gap between its interpolation and extrapolation errors.

\begin{figure}[ht!]
\centering
\vspace{-0.75\baselineskip}
%\begin{subfigure}{0.45\linewidth}
%\vspace{0\baselineskip}
%\caption{}
%\vspace{-0.4\baselineskip}
\subfloat[]{
\includegraphics[width=0.45\linewidth]{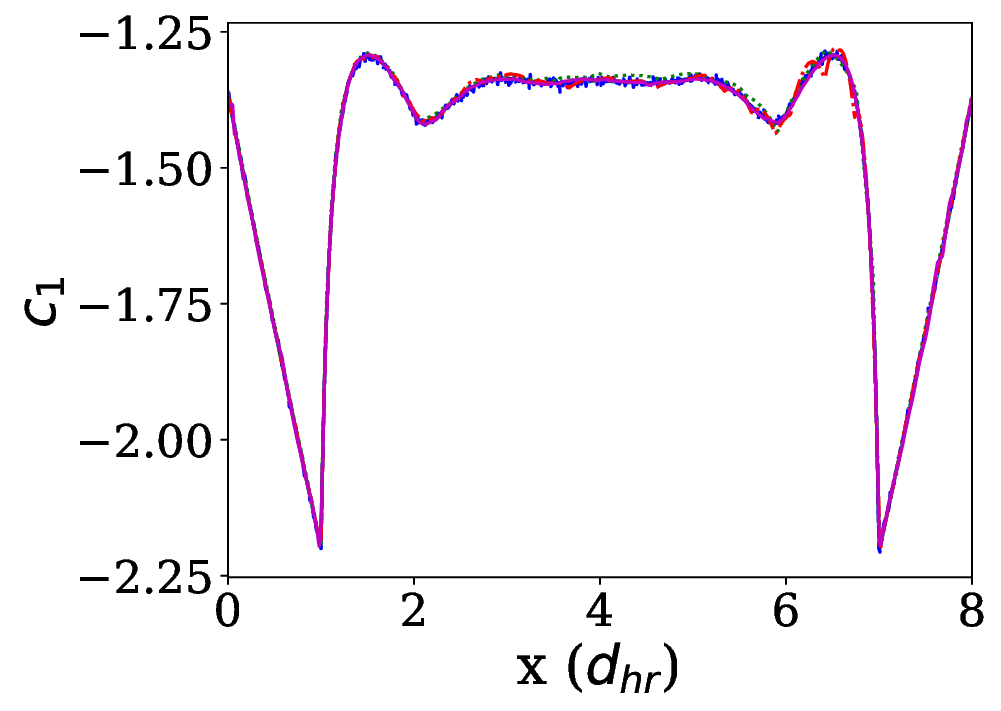}
}
%\caption{Sample 1}
%\end{subfigure}
\vspace{-0.75\baselineskip}
%\begin{subfigure}{0.45\linewidth}
%\caption{}
\subfloat[]{
\includegraphics[width=0.45\linewidth]{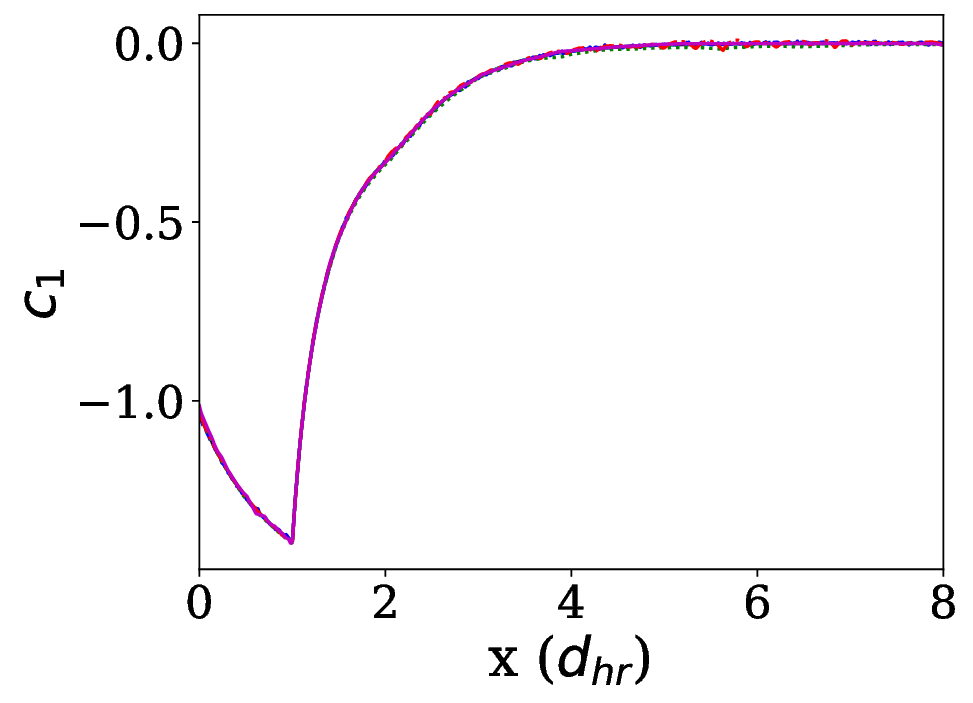}
}

%\caption{Sample 2}
%\end{subfigure}
\vspace{-0.75\baselineskip}
%\begin{subfigure}{0.45\linewidth}
%\caption{}
%\vspace{-0.4\baselineskip}
\subfloat[]{
\includegraphics[width=0.45\linewidth,trim={2pt 2pt 2pt 2pt},clip]{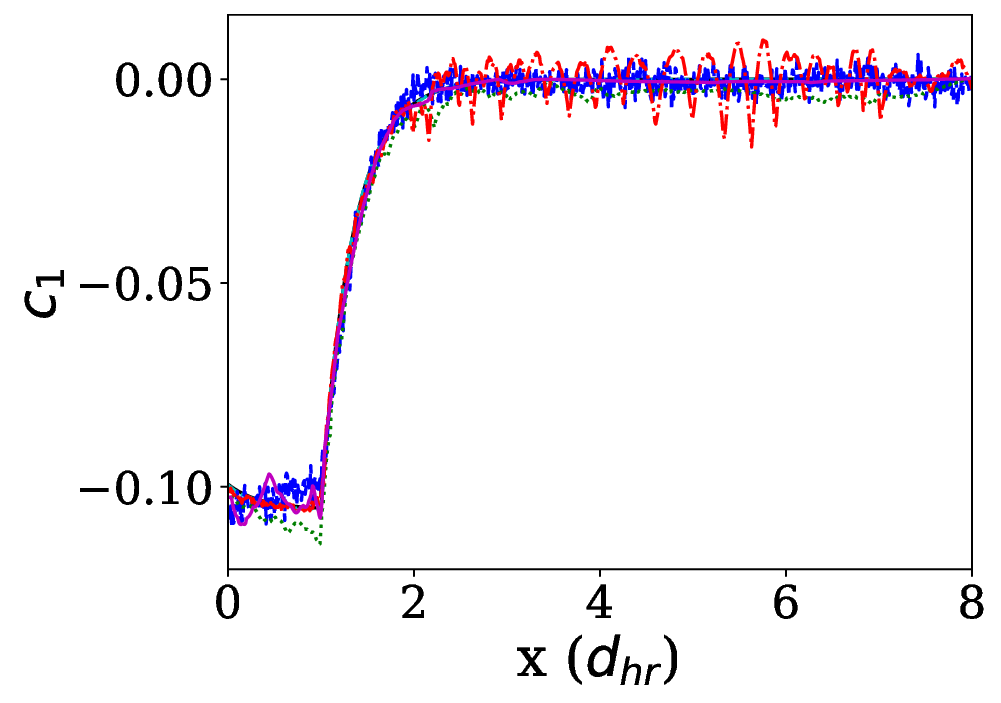}
}
%\caption{Sample 3}
%\end{subfigure}
\vspace{-0.75\baselineskip}
%\begin{subfigure}{0.45\linewidth}
%\vspace{0\baselineskip}
%\caption{}
%\vspace{-0.4\baselineskip}
\subfloat[]{
\includegraphics[width=0.45\linewidth]{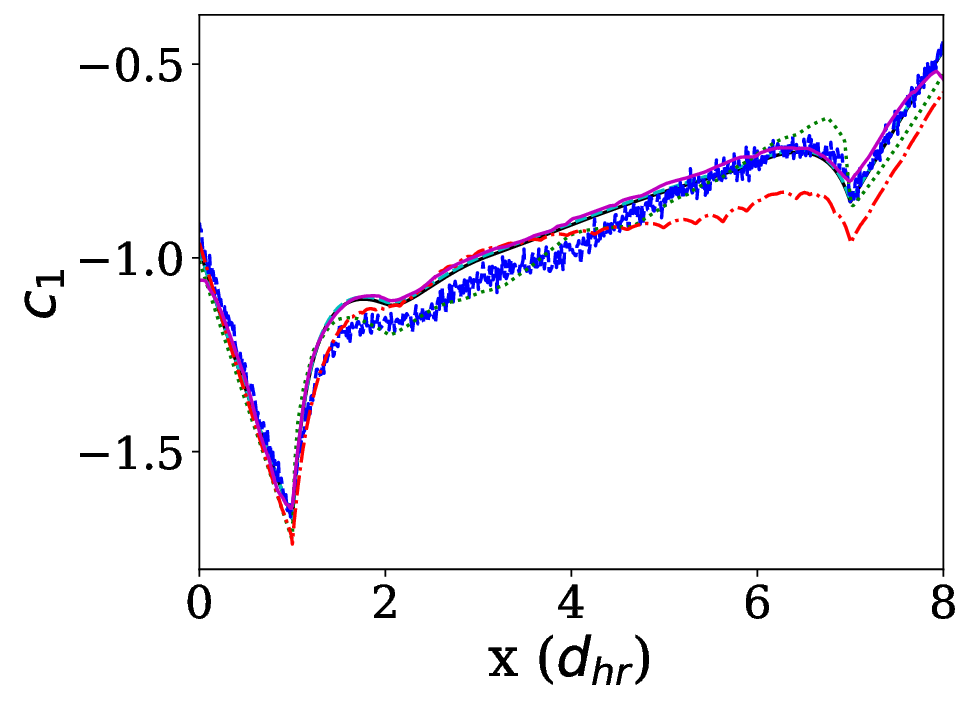}
}

%\caption{Sample 1}
%\end{subfigure}
%\begin{subfigure}{0.45\linewidth}
%\vspace{0\baselineskip}
%\caption{}
%\vspace{-0.4\baselineskip}
\vspace{-0.75\baselineskip}
\subfloat[]{
\includegraphics[width=0.45\linewidth]{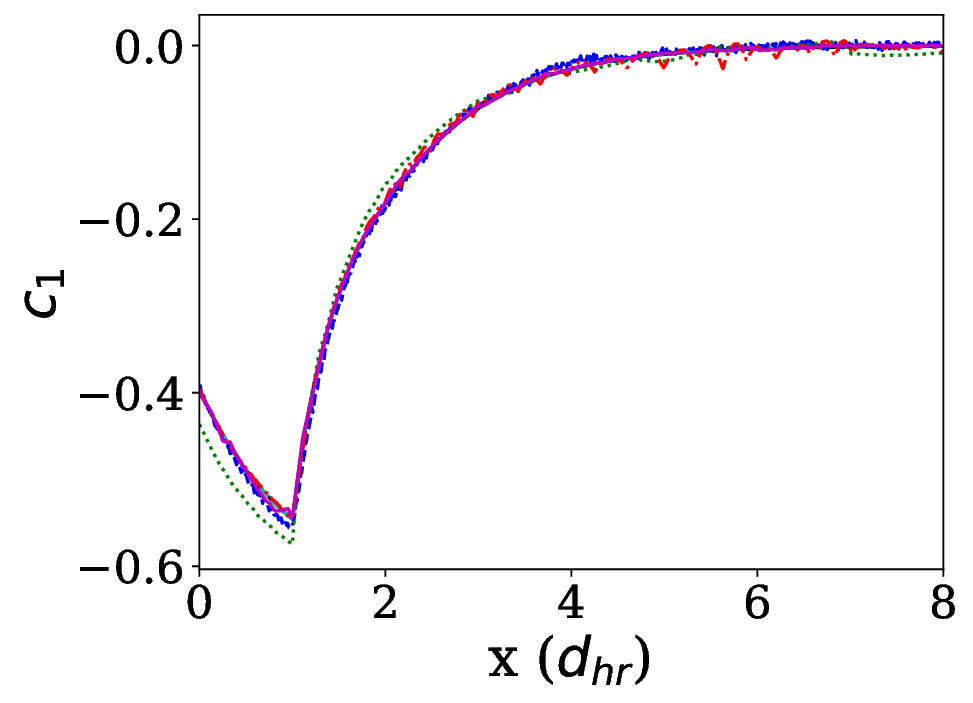}
}
%\caption{Sample 2}
%\end{subfigure}
%\begin{subfigure}{0.45\linewidth}
%\vspace{0\baselineskip}
%\caption{}
%\vspace{-0.4\baselineskip}
\subfloat[]{
\includegraphics[width=0.45\linewidth]{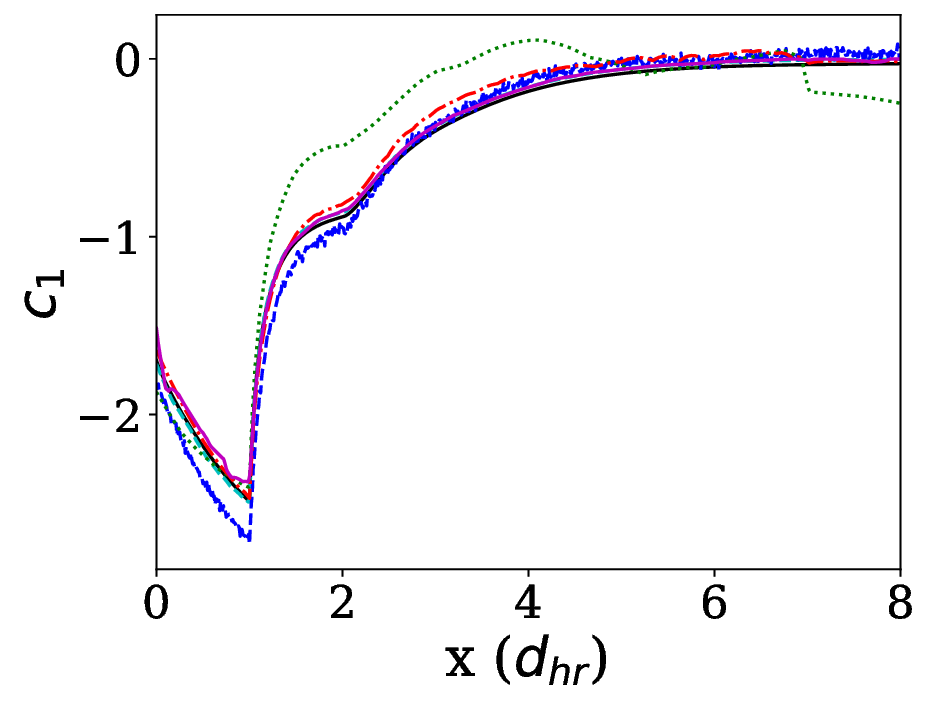}
}

%\caption{Sample 3}
%\end{subfigure}
%\begin{subfigure}{0.8\linewidth}
\vspace{-0.4\baselineskip}
\includegraphics[width=0.8\linewidth]{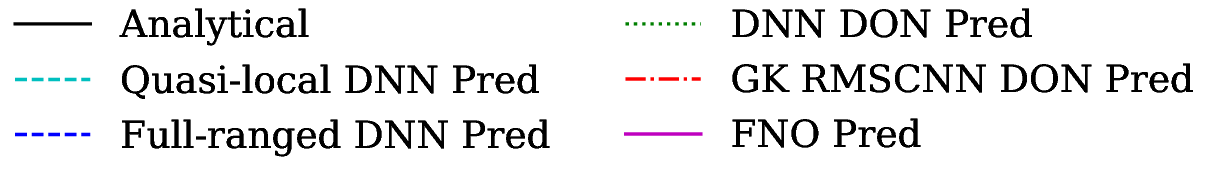}
%\caption{Sample 3}
\vspace{-0.75\baselineskip}
\caption{\label{fig:pred} A comparison of ${\corrf}(x)$ predicted by various  \acrshort{ML} methods, including full-ranged and quasi-local \acrshort{DNN}s, \acrshort{DNN}-\acrshort{DeepONet},  \acrshort{GK}-\acrshort{RMSCNN}-\acrshort{DeepONet}, \acrshort{FNO}. The preditions of \acrshort{ML} methods was labeled as "Pred", \acrshort{DeepONet} labeled as "DON" (a–c) Three examples from the in-group test set. (d–f) Three examples from the new data test set. Legend: Analytical Solution – black solid line; Quasi-local \acrshort{DNN} – cyan dashed line; Full-ranged \acrshort{DNN} – blue dashed line;\acrshort{DNN}-\acrshort{DeepONet} – green dotted line;\acrshort{GK}-\acrshort{RMSCNN}-\acrshort{DeepONet} – red dash-dotted line; \acrshort{FNO} – violet solid line}
\end{figure}

Figures \ref{fig:pred}(a–c) present representative examples from the in-group test set for the prediction of ${\corrf}$ using different \acrshort{ML} models. Overall, the models exhibit strong predictive accuracy for most samples, with high correlation to the analytical results. However, as illustrated in Figure \ref{fig:pred}(c), some inputs lead to noticeably noisy predictions or substantial deviations from the true profile.
These artifacts typically arise at short distances, where the predicted correlation function becomes erratic or oscillatory. This behavior is most noticeable in samples with low concentration from data groups II and III, particularly near sharp gradients at $x = 1$ or $x = L - 1$, where there is a steep variation of the external potential. % in near-wall regions, which poses a significant challenge for the \acrshort{ML} models. 
The large derivations in this region can likely be attributed to their low contribution to the loss. The oscillatory results of \acrshort{GK}-\acrshort{RMSCNN}-\acrshort{DeepONet} appear within a cutoff distance $x_{\text{conv,cut}}$, likely from the cut-off loss of Gaussian functions.
While many \acrshort{ML}  methods struggle to capture sharp gradients and localized features unless explicitly trained or architecturally designed for such scenarios, causing the poor performance in near-wall regions. 
%Inaccurate or insufficient representation of the external potential during training may further contribute to these discrepancies. 
As a result, the corresponding density profiles in these regions tend to show artificial fluctuations or noise, which propagate into the prediction of ${\corrf}$.

Figures \ref{fig:pred}(d–f) show the \acrshort{ML} predictions for samples from the `new data' set. These results are clearly worse than those for the in-group test set, highlighting the limited extrapolation capabilities of most \acrshort{ML} models. The discrepancies are particularly evident in regions influenced by boundary effects or strong confinement, where external potentials deviate significantly from those seen during training. 
Among all \acrshort{ML} models tested in this work, \acrshort{FNO} proves to be the most robust, demonstrating strong correlation across all samples, with only a slight decline in accuracy near $x = 1$ and $x = L - 1$, where sharp gradients are present.%in regions close to the wall
%, where the challenges noted above persist.

%\JW{A more detailed analysis of the density profiles corresponding to panels (c) and (d–f) could provide further insight into the nature of these prediction errors. Such analysis would help identify whether the observed noise originates from model limitations, input feature representations, or the underlying complexity of the external potentials in these regimes.}

\begin{figure}[ht!]
\centering
\vspace{-0.75\baselineskip}
%\begin{subfigure}{0.45\linewidth}
%\caption{}
%\vspace{-0.4\baselineskip}
\subfloat[]{
\includegraphics[width=0.45\linewidth]{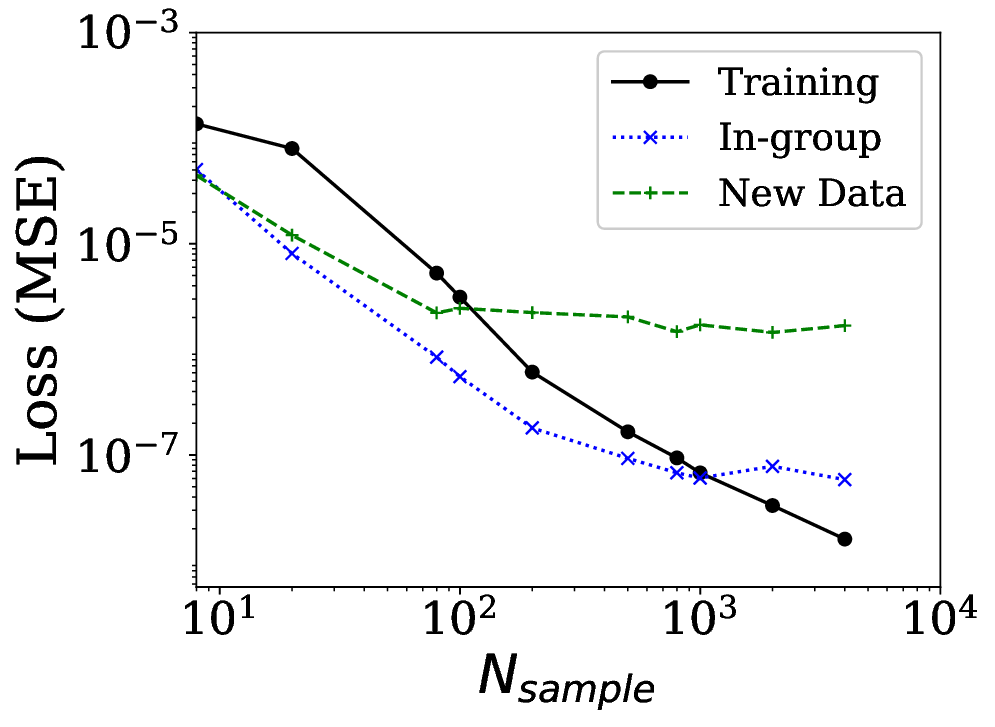} 
}
%\caption{FNO}
%\end{subfigure}
\vspace{-0.75\baselineskip}
%\begin{subfigure}{0.45\linewidth}
%\caption{}
\subfloat[]{
\includegraphics[width=0.45\linewidth,trim={5pt 5pt 5pt 5pt},clip]{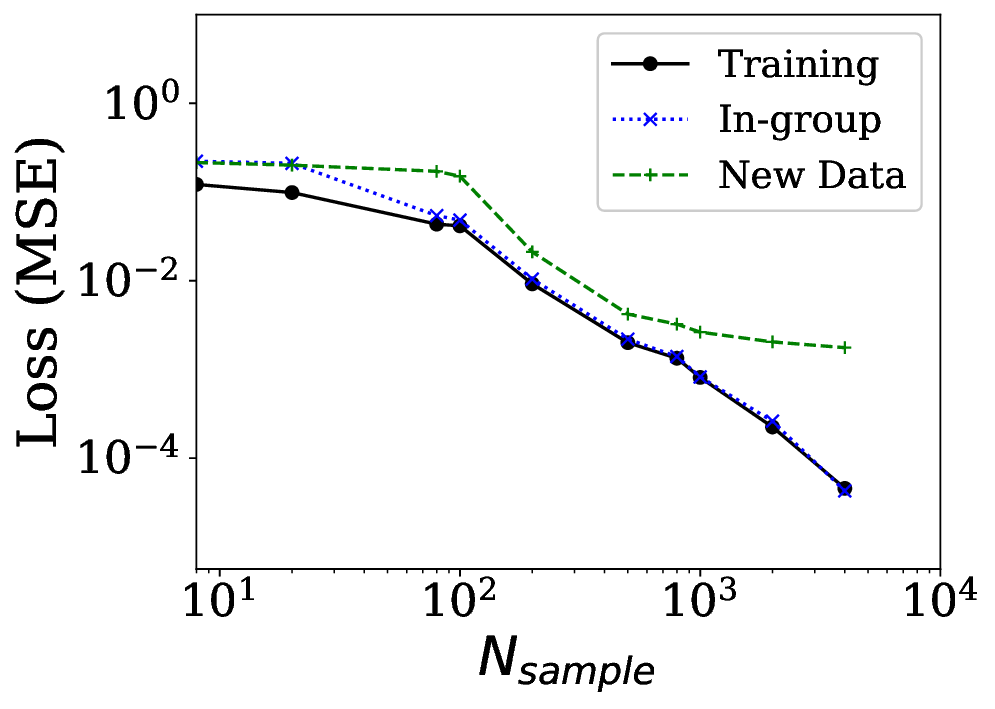} 
}

%\caption{ Vanilla DNN }
%\end{subfigure}
%\begin{subfigure}{0.45\linewidth}
%\vspace{0\baselineskip}
%\caption{}
\vspace{-0.4\baselineskip}
\subfloat[]{
\includegraphics[width=0.45\linewidth,trim={5pt 5pt 5pt 5pt},clip]{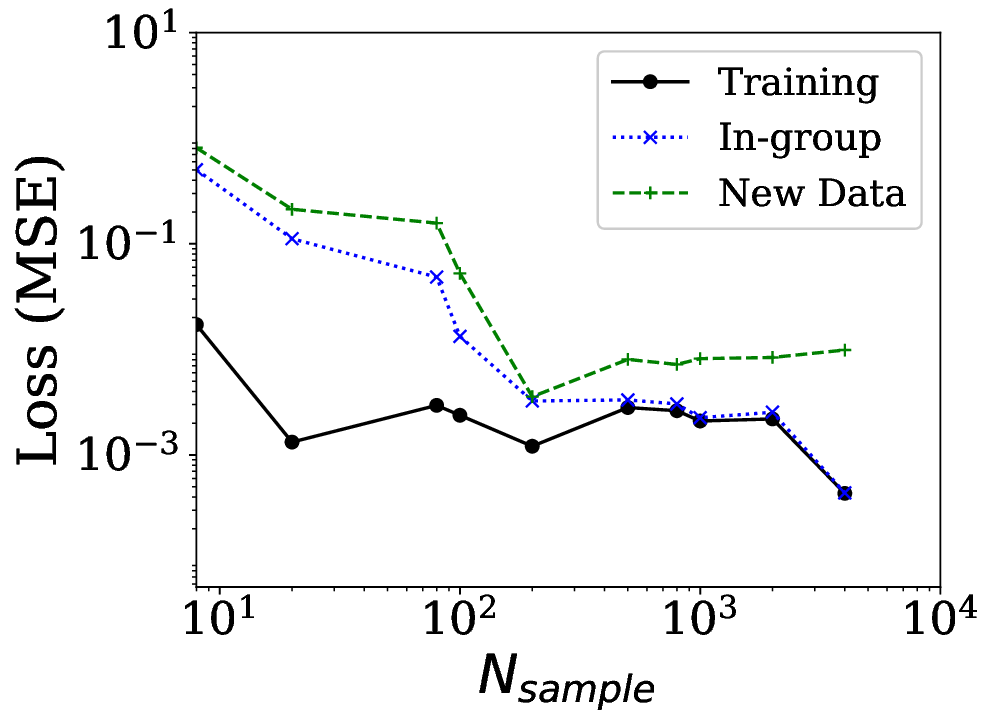} 
}
%\end{subfigure}
%\begin{subfigure}{0.45\linewidth}
%\vspace{0\baselineskip}
%\caption{}
\vspace{-0.4\baselineskip}
\subfloat[]{
\includegraphics[width=0.45\linewidth]{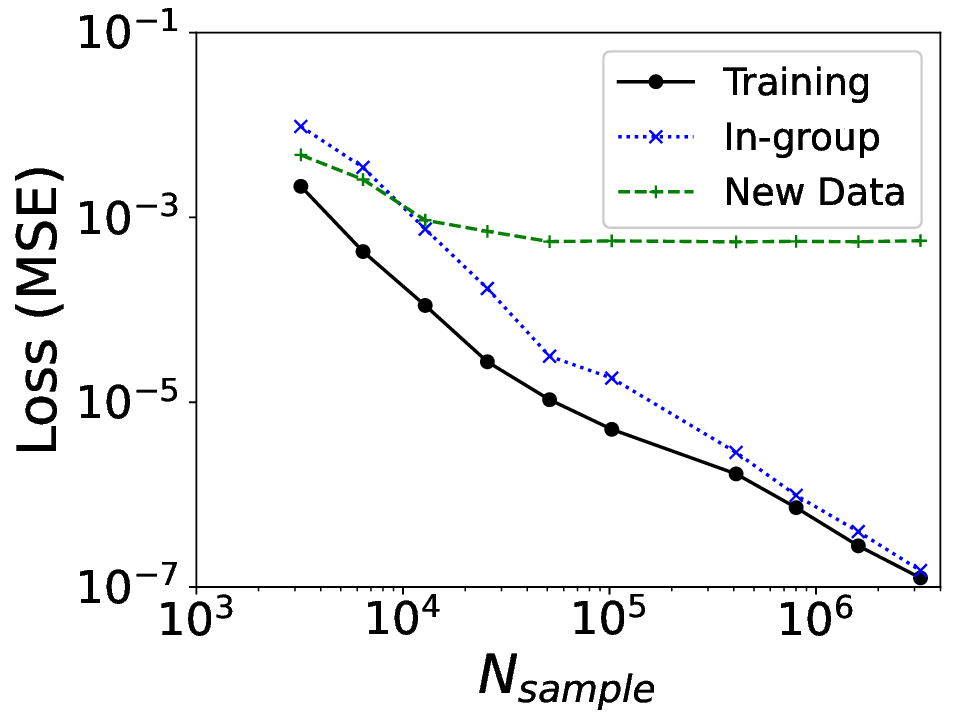}  
%\caption{ Neural-DFT }
}
%\end{subfigure}
\caption{\label{fig:loss_sample} \acrshort{MSE} loss versus training data size (\glssymbol{nsample}) on the training set, in-group test set, and new data test set, for different \acrshort{ML} methods. Training was stopped when the in-group test loss failed to decrease for 2000 steps or after reaching 100,000 steps for both \acrshort{DeepONet} and \acrshort{DNN}. For Quasi-local \acrshort{DNN}, \glssymbol{nsample} refers to the number of simulation samples. (a) \acrshort{FNO}, (b) Full-scale \acrshort{DNN}, (c) \acrshort{GK}-\acrshort{RMSCNN}-\acrshort{DeepONet}, (d) Quasi-local \acrshort{DNN}. The legend is as follows: Training Set MSE – black solid line with circle markers; In-Group Test Set MSE – blue dotted line with × markers; New Data Test Set MSE – green dashed line with $+$ markers.}
\end{figure}

Figure~\ref{fig:loss_sample} presents the relation between \acrshort{MSE} loss and the number of training samples (\glssymbol{nsample}) on a log-log scale. %, with quasi-local \acrshort{DNN} using $\nsample=N_{training,set}*\ngrid$.  
For most \acrshort{ML} models, both training loss and in-group test loss generally decrease with increasing \glssymbol{nsample}.  The \acrshort{GK}-\acrshort{RMSCNN}-\acrshort{DeepONet} exhibits noticeable fluctuations in training loss at small \glssymbol{nsample}, indicating overfitting in this regime; however, this issue is mitigated when $\nsample \ge 2000$. Across all tested \acrshort{ML} models, the loss on the `new data' test set consistently exceeds that of the in-group test set. When $\nsample \ge 2000$, training and in-group test losses become largely comparable. Both interpolation and extrapolation errors exhibit an approximately linear relation with \glssymbol{nsample} on the log-log scale for small \glssymbol{nsample}, with several models approaching a performance plateau at larger \glssymbol{nsample} values.

\begin{figure}[ht!]
\centering
\vspace{-0.75\baselineskip}
\subfloat[]{
\includegraphics[width=0.45\linewidth]{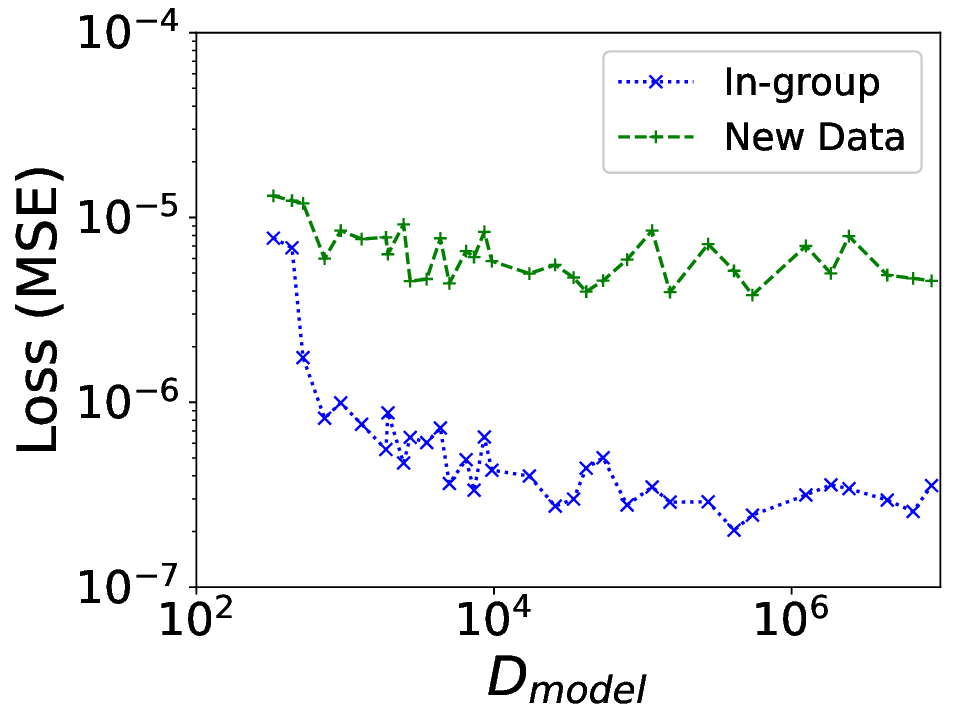}  
}
\vspace{-0.75\baselineskip}
\subfloat[]{
\includegraphics[width=0.45\linewidth]{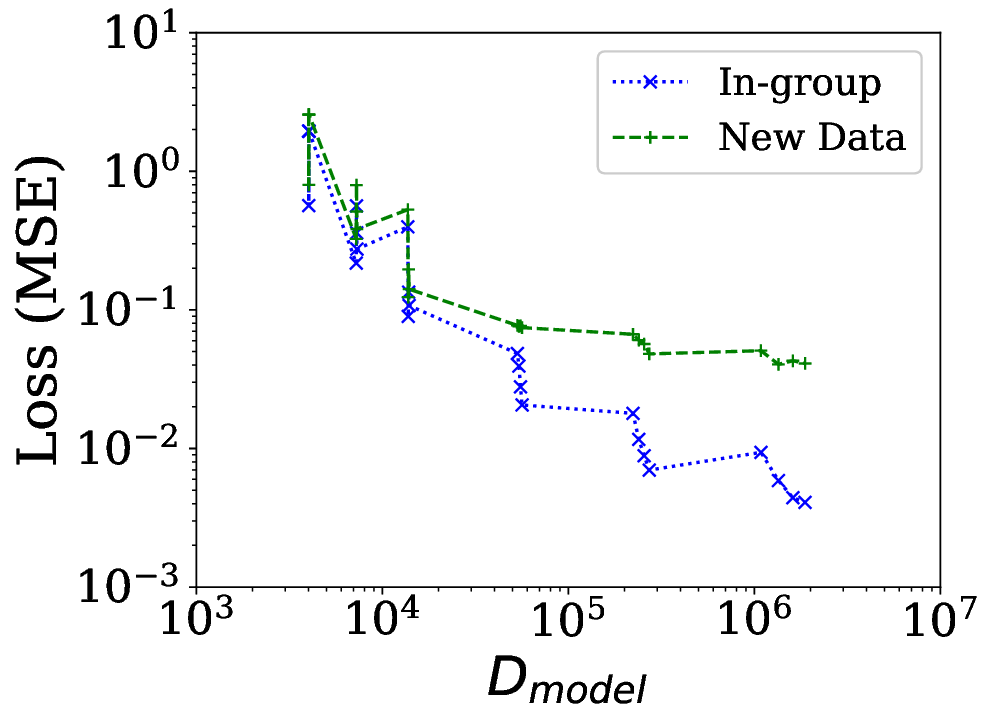} 
}

\vspace{-0.75\baselineskip}
\subfloat[]{
\includegraphics[width=0.45\linewidth]{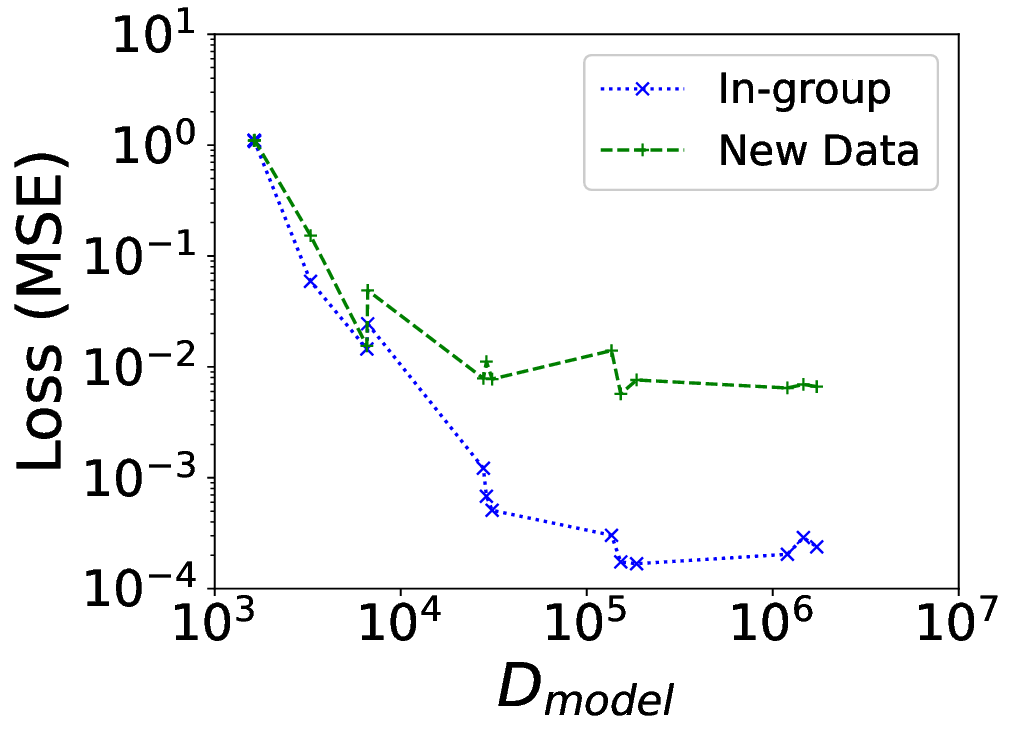} 
}
\vspace{-0.75\baselineskip}
\subfloat[]{
\includegraphics[width=0.45\linewidth]{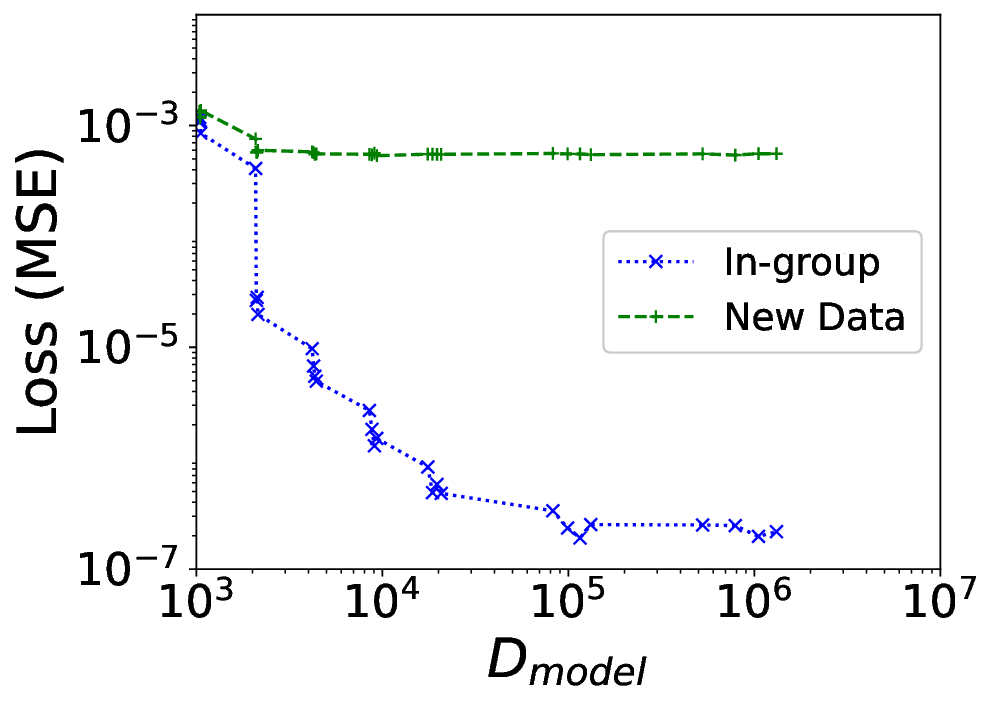}   
}

\caption{\label{fig:loss_msize} Comparison of \acrshort{MSE} loss versus model size \glssymbol{msize} for different \acrshort{ML} methods. Training was terminated when the in-group test loss did not decrease for 2000 steps or when 100,000 steps were reached, for both \acrshort{DeepONet} and \acrshort{DNN}. \acrshort{DNN}-\acrshort{DeepONet} is shown in place of \acrshort{GK}-\acrshort{RMSCNN}-\acrshort{DeepONet} due to the significantly longer training time for latter case. (a) \acrshort{FNO}, (b) Full-scale \acrshort{DNN}, (c) \acrshort{DNN}-\acrshort{DeepONet}, (d) Quasi-local \acrshort{DNN}.\\ Legends are: In-Group Test Set MSE: Blue Dotted Line with x  markers; New data Test Set MSE: Green Dashed Line with $+$  markers.}
\end{figure}

Figure~\ref{fig:loss_msize} presents the \acrshort{MSE} loss versus model size \glssymbol{msize} on a log-log scale, using various training hyperparameters as described in Section~\ref{sec:ML_method}. For each \acrshort{ML} method, the best-performing models with comparable \glssymbol{msize} values are selected for comparison. It is evident that \acrshort{FNO} consistently outperforms all other \acrshort{ML} methods in terms of accuracy across all model evaluation settings in this study. Both interpolation and extrapolation errors exhibit approximately linear trends in the log-log plot at smaller \glssymbol{msize}, and they approach a plateau as \glssymbol{msize} increases.

\subsection{Neural scaling law analysis}
To better understand the dependence of model performance on key training parameters, we analyzed the mean squared error (\acrshort{MSE}) loss using the Chinchilla scaling law. This empirical law was originally developed by researchers at OpenAI and Google\cite{kaplan2020scalinglawsneurallanguage, Bahri_2024} to describe the training of LLMs with respect to their model size and number of training tokens. It offers a unified framework for assessing the performance of various \acrshort{ML} models. While the Chinchilla scaling law has been successfully applied to tasks such as transfer learning and neural machine translation, to our knowledge, this approach has not yet been extended to \acrshort{ML} models in physical problems including \acrshort{DFT}.

The Chinchilla scaling law describes the \acrfull{NLL} of classification problems and the log-transformed regression loss, $\log(\text{MSE})$, which is proportional to \acrfull{GNLL} with constant error, as a power-law function of model size (\glssymbol{msize}) and training dataset size (\glssymbol{nsample}):
\begin{equation}\label{eqn:Chinchilla} 
\glssymbol{Loss_model} = \glssymbol{ideal_Loss_model} + \frac{A}{\nsample^\alpha} + \frac{B}{\msize^\beta} 
\end{equation} 
where $A$, $B$, $\alpha$, and $\beta$ are positive-valued scaling parameters to be inferred from data, and \glssymbol{ideal_Loss_model} represents the irreducible loss—i.e., the theoretical minimum loss achievable by an ideal model trained on an infinite dataset with infinite model capacity. This asymptotic limit serves as a fundamental measure of the generalization capability intrinsic to the class of \acrshort{ML} models under consideration, particularly in the extrapolation regime where both data and model resources are abundant.

A complementary quantity of interest in this analysis is the total training cost ($\comcost$), given by 
\begin{equation}\label{eqn:Chinchilla0} 
\comcost = C_s \times \nsample  
\end{equation} 
where $C_s$ denotes the number of floating-point operations (\acrshort{FLOPs}) required to train model parameters. For instance, it has been empirically determined that $C_s= 6\msize$ for \acrshort{LLM}s. As anticipated, the training cost scales linearly with  the size of the dataset for most \acrshort{ML} methods. For parameter tuning, the computational cost also scales linearly with the number of model parameters. When some \glssymbol{msize}-irrelevant costly steps are included in model training, we have $C_s = C_0+C_1 \msize$, with $C_0$ being the cost of those operations. 
This provides a practical constraint for model selection and optimization under limited computational budgets. 

We investigated the scaling behavior of both interpolation and extrapolation performance to evaluate and quantify the generalization capabilities of the various \acrshort{ML} methods. The results are shown in Figure~\ref{fig:loss_msize}. The fitted scaling parameters are summarized in Table~\ref{tab:scaling_law}. The high $R^2$ values confirm that the Chinchilla scaling law provides an excellent fit across all cases, with the exception of the extrapolation performance of the quasi-local \acrshort{DNN}, whose results fall entirely within a plateau region and thus do not exhibit meaningful scaling behavior. Among the methods studied, the \acrshort{FNO} achieves the lowest extrapolation \glssymbol{ideal_Loss_model}, indicating superior generalization capability in the asymptotic limit. In contrast, discrete functional learning models—such as the full-scale \acrshort{DNN} and the quasi-local \acrshort{DNN}—show markedly higher irreducible loss, reflecting significantly weaker extrapolation performance. 

%\JW{Please also discuss the results in terms of the training cost}

When training cost is taken into account, these findings offer important insights into the efficiency of each \acrshort{ML} method. The training cost, $\comcost$, which is assumed to be proportional to the training time, increases linearly with training data size and model size. For \acrshort{DNN}, \acrshort{DNN}-\acrshort{DeepONet}, \acrshort{DK}-\acrshort{CNN}-\acrshort{DeepONet}, $C_0\simeq 0$, since most of their calculation steps are \glssymbol{msize}-dependent such as forward and backward propagation. Kernel operations in \acrshort{GK} and \acrshort{FFT} operations in \acrshort{FNO} are \glssymbol{msize}-irrelevant, resulting in a measurable $C_0$ for them. 
Despite the higher computational demands of \acrshort{FNO}, its favorable scaling behavior implies that it reaches superior generalization performance more efficiently in the high-resource regime. Conversely, the plateauing behavior observed in the quasi-local \acrshort{DNN} suggests that additional computational investment yields diminishing returns, making it less cost-effective for extrapolation tasks. Overall, these results underscore the importance of considering both performance scaling and computational cost when evaluating and selecting \acrshort{ML} models for predictive tasks beyond the training domain.

\begin{table*}[ht!]
\caption{\label{tab:scaling_law} Fitted scaling law parameters from the Chinchilla analysis for \acrshort{FNO}, full-scale \acrshort{DNN}, Quasi-local \acrshort{DNN}, and \acrshort{DNN}-\acrshort{DeepONet}, evaluated on both in-group test data and a new dataset. The residual model losses are reported as $\exp (\mathcal{L}_{\infty}) = \text{MSE}_{\text{id}}$, representing the estimated \acrshort{MSE} of an ideal model (id) with infinite size and infinite training data.
} 
\centering
    \renewcommand{\arraystretch}{1.2} % Adjust row height for better readability
\begin{ruledtabular}\begin{tabular}{c c c c c c c c }

Model& Test set& $R^2$& $\exp(\mathcal{L}_{\infty})$& $\alpha$& $\beta$& A& B   \\\hline
  \multirow{2}{*}{\acrshort{FNO}}
 &In-group &0.95&$1.73\times 10^{-7}$&$0.36$&$0.26$&19.3&10.91\\
 &New Data &0.90&$2.50\times10^{-6}$&$0.38$&$0.26$&12.98&9.97\\
 \multirow{2}{*}{Full-scale \acrshort{DNN}}
 &In-group &0.85&$5.28\times 10^{-7}$&$0.54$&$0.23$&15.2&369\\
 &New Data &0.85&$4.76\times 10^{-4}$&$0.56$&$0.38$&9.6&240\\
 \multirow{2}{*}{\centering \acrshort{DNN}-\acrshort{DeepONet}}
 &In-group  &0.88 &$6.60\times 10^{-6}$ &$0.62$ &$0.21$ &21.1 &230\\
 &New Data &0.87 &$3.65\times 10^{-3}$ &$0.61$ &$0.25$ &22.2 &118\\
\multirow{2}{*}{Quasi-local \acrshort{DNN}}
 &In-group&0.95&$3.96\times 10^{-8}$&$0.53$&$0.16$& 49.3&246\\
 &New Data&0.62&$1.61\times 10^{-3}$&$0.19$&$0.42$&363&76\\
\end{tabular}\end{ruledtabular}
\end{table*}

\subsection{Excess free energy calculations}
From the operator mapping between the density profile and the direct correlation function, \(\glssymbol{corrfoperator}: \rho(x) \mapsto \corrf\), we can readily calculate the excess free energy \({\Fex}\) corresponding to any given density profile \(\rho(x)\) by numerical integration:
\begin{equation} \label{eqn:cdft:Fex}
\beta \Fex[\rho] = -\int_0^L dx \, \rho(x) \int_0^1 d\lambda \, \corrf(x, [\lambda\rho(x)]).
\end{equation}
The resulting values predicted by different \acrshort{ML} models are compared against the exact excess free energy for the \acrshort{HR} system:\cite{RN158} 
\begin{equation} \label{eqn:F_ex}
\beta {\Fex}[\rho]=-\int_0^L dx\rho(x) \ln\Big[ 1-\int_{x-\HRdia}^{x}dx'\rho(x')\Big]
\end{equation}

\begin{figure}[ht!]
\centering
\vspace{-0.75\baselineskip}
\subfloat[]{
\includegraphics[width=0.45\linewidth]{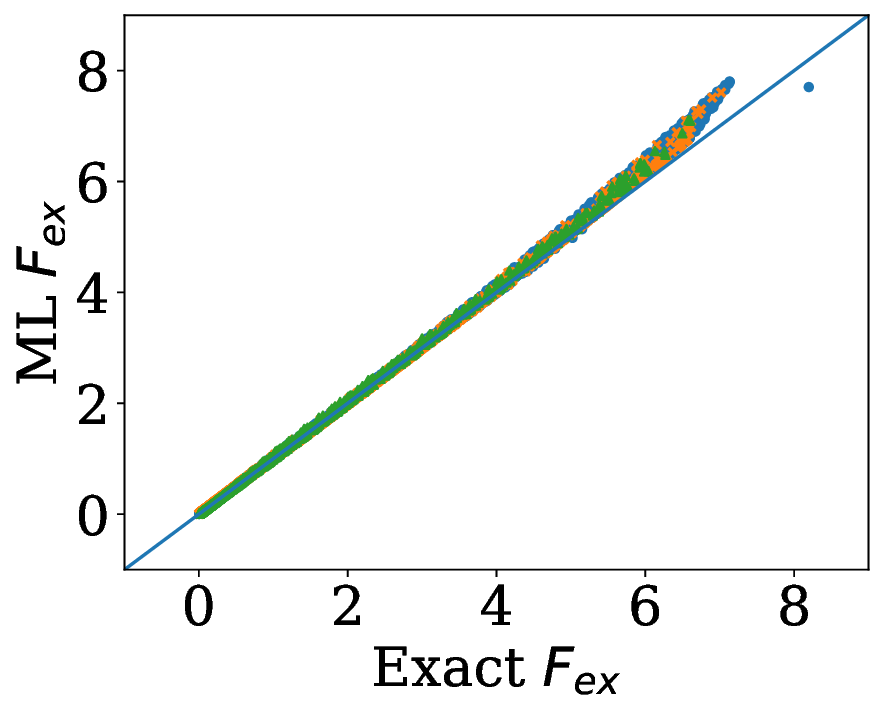}  
%\caption{Fourier Neural Operator(FNO)}
}
\vspace{-0.75\baselineskip}
\subfloat[]{
\includegraphics[width=0.45\linewidth]{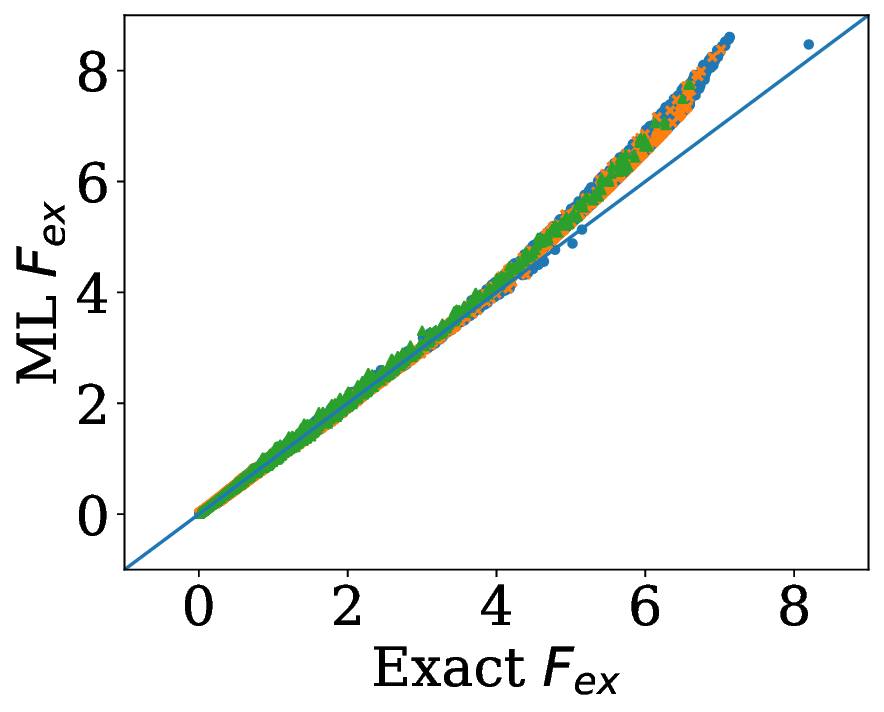} 
%\caption{ Vanilla DNN }
}

\vspace{-0.75\baselineskip}
\subfloat[]{
\includegraphics[width=0.45\linewidth,trim={10pt 0pt 3.2em 10pt},clip]{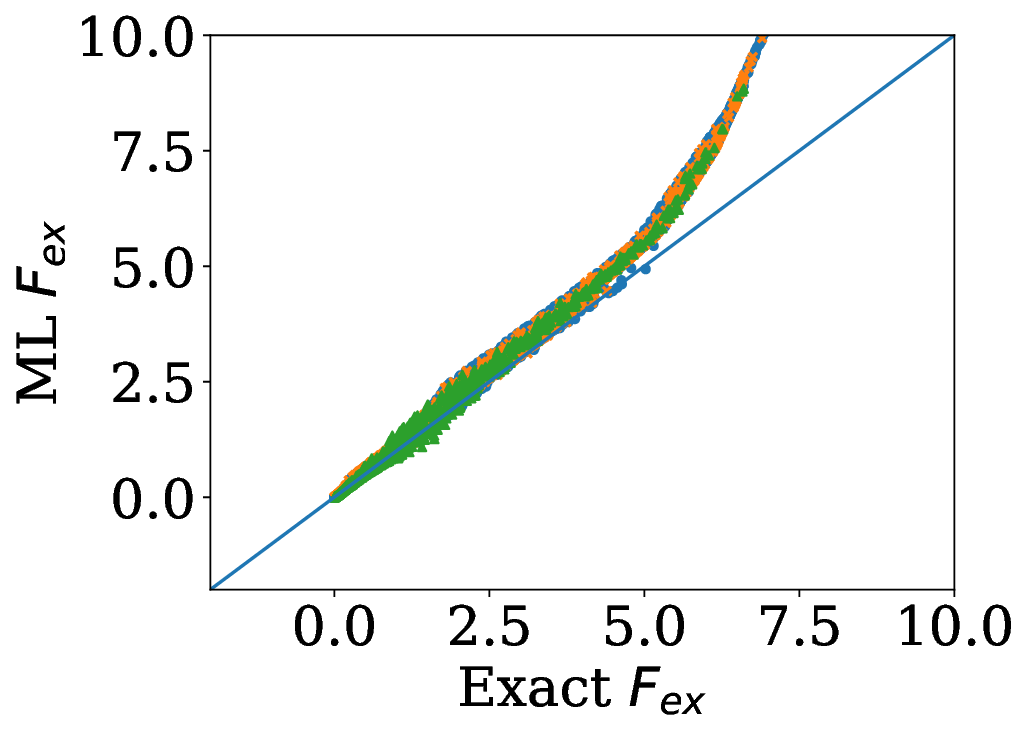} 
}
%\caption{G-MS-CNN DeepONet}
\vspace{-0.75\baselineskip}
\subfloat[]{
\includegraphics[width=0.45\linewidth]{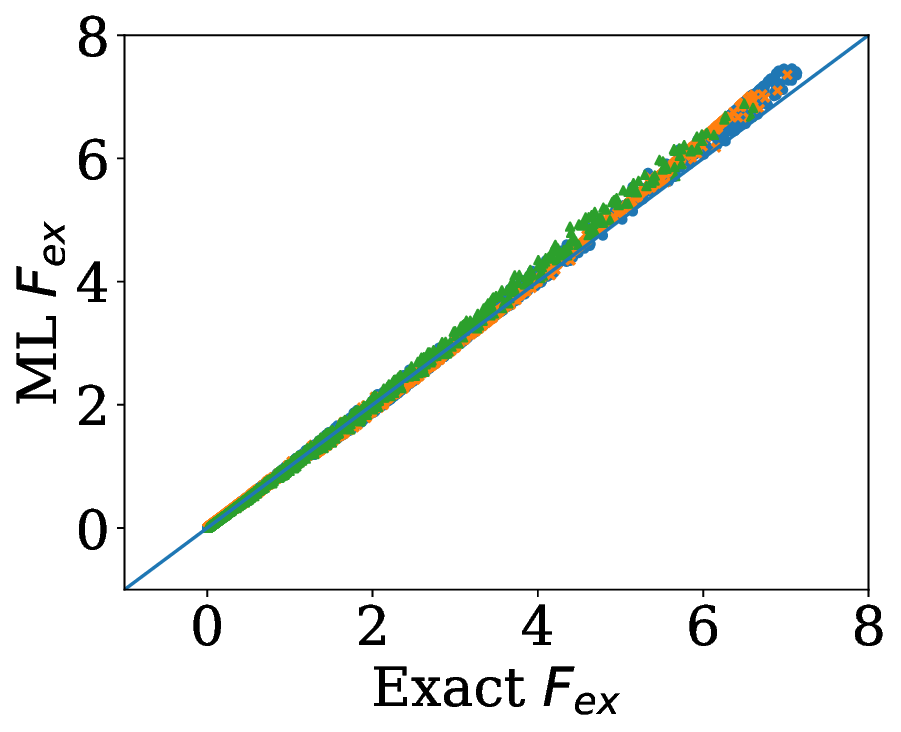} 
%\caption{Neural DFT}
}

\vspace{-0.75\baselineskip}
\subfloat[]{
\includegraphics[width=0.45\linewidth]{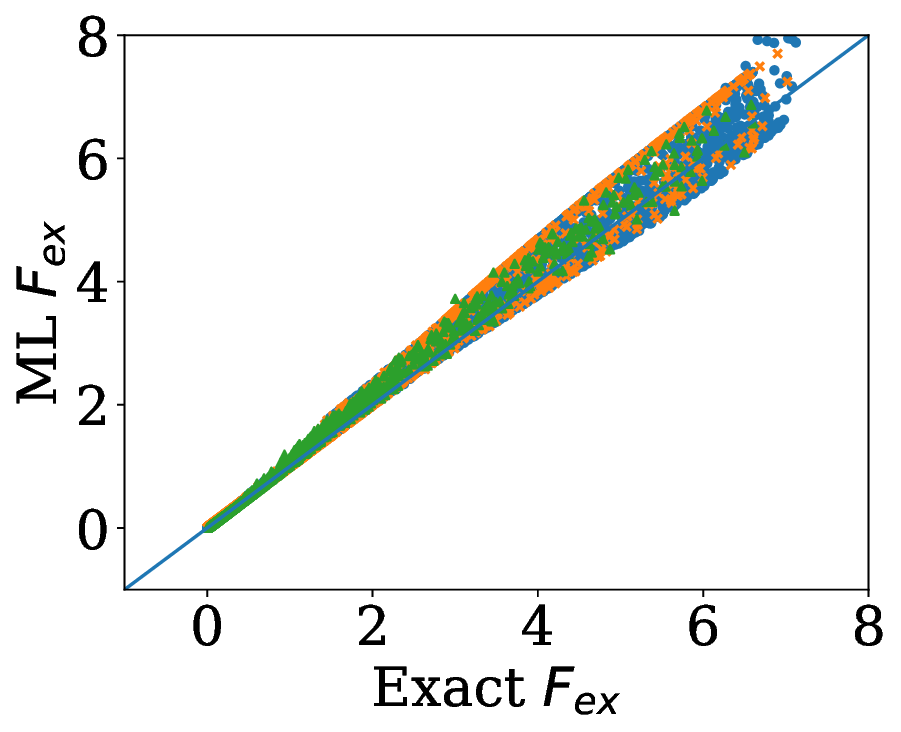} 
%\caption{G-MS-CNN DeepONet}
}
\subfloat{
\includegraphics[width=0.3\linewidth]{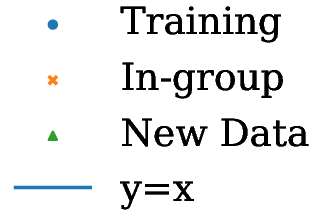} 
}

\caption{\label{fig:Fex} Comparison of the excess Helmholtz energy \glssymbol{Fex} predicted by different \acrshort{ML} methods with the exact results. (a) \acrshort{FNO} (b) Full-scale \acrshort{DNN}(c)  \acrshort{DNN}-\acrshort{DeepONet} (d) \acrshort{GK}-\acrshort{RMSCNN}-\acrshort{DeepONet} (e) Quasi-local \acrshort{DNN}.\\ Legends are: Training Set: Blue; In-Group Test Set MSE: Orange; New data Test Set MSE:Green. The Diagonal line: Blue line.%\JW{use "Exact $F_{ex}$" and "ML $F_{ex}$" for the titles of x-axis and y-axis.}
}
\end{figure}

Figure~\ref{fig:Fex} compares the \glssymbol{Fex} values computed using various \acrshort{ML}-learned \glssymbol{corrfoperator} via Eqn.~\ref{eqn:cdft:Fex}, against the analytical solutions obtained from Eqn.~\ref{eqn:F_ex}, for identical density profiles from the training, in-group test, and new data test sets. We observe that all \acrshort{ML} methods tend to overestimate \glssymbol{Fex} in the high-density regime, with no appreciable differences across the data sets. This systematic deviation at large densities may be attributed to the scarcity of training data. Notably, the accuracy of \acrshort{ML} predictions of \glssymbol{Fex} remains consistent across training, in-group test, and new test samples, suggesting that the learned \glssymbol{corrfoperator} generalizes well to unseen density profiles, particularly in terms of integrated functional values.

\begin{table}[ht!]
\centering
\caption{\label{tab:Fex_compare_ONs} Comparison of $R^2$ score and ${\Fex}$ calculation time for different \acrshort{ML} models. 
}
\renewcommand{\arraystretch}{1.2}
\begin{ruledtabular}\begin{tabular}{m{5em} m{4em}<{\centering} m{4em}<{\centering} m{4em}<{\centering} m{5em}<{\centering} m{4em}<{\centering}}

 & Full-scale \acrshort{DNN} &Quasi-local \acrshort{DNN} &\acrshort{DNN}-\acrshort{DeepONet} &\acrshort{GK}-\acrshort{RMSCNN}-\acrshort{DeepONet}&\acrshort{FNO} \\
\hline
Training & 0.970 &0.958& 0.910                    & 0.9956              &    0.9992              \\
In-group & 0.964         &0.960& 0.894        & 0.9977             &    0.9991                    \\
New Data &   0.972       &0.968&    0.932      & 0.9941              &    0.9990              \\
Time (s)& 0.2& 31& 0.6& 2.1&0.5\\

\end{tabular}\end{ruledtabular}
\end{table}
The squared $R$ scores ($R^2$) of \glssymbol{Fex} predictions across different data sets, along with the computation times for various \acrshort{ML} models, are summarized in Table~\ref{tab:Fex_compare_ONs}. Higher $R^2$ values indicate better predictive performance. Among all methods, \acrshort{FNO} achieves the highest accuracy in computing \glssymbol{Fex}, followed by \acrshort{GK}-\acrshort{RMSCNN}-\acrshort{DeepONet}. Overall, the accuracy of \glssymbol{Fex} predictions appears to be correlated with both interpolation and extrapolation capabilities of the \acrshort{ML} models, and no substantial differences are observed across different sources of $\rho(x)$. This consistency may be attributed to the integral nature of \glssymbol{Fex}, which depends on accurate $\corrf$ predictions even in the low-density regime—i.e., values of $\corrf(x,[a\rho(x)])$ for $a \in [0,1]$—that often lie outside the training data distribution.

\subsection{Choice of activation function}\label{sec:compare_actis}
Activation functions play a critical role in the training and performance of operator learning models, as they directly influence the model’s expressiveness, stability, and convergence behavior. For both \acrshort{DeepONet} and \acrshort{FNO}, the choice of activation function determines their ability to capture complex, nonlinear input–output mappings across function spaces. Unlike traditional pointwise regression tasks, operator learning requires handling high-dimensional and often nonlocal dependencies, placing additional demands on the smoothness and differentiability properties of the activation function. Moreover, certain activation functions can significantly affect the conditioning of gradients during training, impacting both the speed and robustness of optimization. As such, a careful evaluation of activation functions is essential for balancing model accuracy, training efficiency, and generalization performance, particularly when extending models to out-of-distribution or extrapolation regimes.

\begin{table*}[ht!]
\caption{\label{tab:compare_actis} Variation of \acrshort{MSE} loss and training time with the choice of activation function. The metrics $\corrf$-\text{MSE}$T$, $\corrf$-\text{MSE}${IG}$, and $\corrf$-\text{MSE}$_{ND}$ denote the \acrshort{MSE} loss evaluated on the training set, in-group test set, and out-of-distribution (New Data) test set, respectively.}
\centering
\renewcommand{\arraystretch}{1.2}
\begin{ruledtabular}\begin{tabular}{m{5em}<{\centering} l c c c c c }

    Model &  & sRelu & Relu  & Selu & Logistic   \\ 
    \hline
    \multirow{4}{*}{Full-scale DNN}
    & $\corrf-\text{MSE}_T$ & $5.8 \times {10}^{-4} $& $ 6.7 \times {10}^{-4} $ &  $ 2.2 \times {10}^{-3} $& $ 2.6 \times {10}^{-3} $  \\
    &$\corrf-\text{MSE}_{IG}$&$ 5.8 \times {10}^{-4} $ &$ 6.5 \times {10}^{-3} $   &   $ 2.1 \times {10}^{-3} $  & $ 2.8 \times {10}^{-3} $\\
    &$\corrf-\text{MSE}_{ND}$& $ 2.2 \times {10}^{-2} $ &$ 1.9 \times {10}^{-2} $   &   $ 6.3 \times {10}^{-3} $  & $ 7.3 \times {10}^{-3} $\\
    & time(s)  &1923& 1854  & 1923 & 2402 \\ 
    %\hline
    \multirow{4}{5em}{\centering DNN-DeepONet}
    &$\corrf-\text{MSE}_T$& $ 8.7 \times {10}^{-5} $ & $ 5.5 \times {10}^{-4} $  &$ 5.4 \times {10}^{-3} $ & $ 3.8 \times {10}^{-4} $  \\
    &$\corrf-\text{MSE}_{IG}$& $ 2.2 \times {10}^{-4} $ &$ 5.6 \times {10}^{-4} $  &   $ 5.36 \times {10}^{-3} $  & $ 3.4 \times {10}^{-4} $ \\
    &$\corrf-\text{MSE}_{ND}$& $ 8.6 \times {10}^{-3} $ &$ 5.7 \times {10}^{-3} $ & $ 5.4 \times {10}^{-3} $   & $ 4.5 \times {10}^{-3} $  \\
    & time(s)  &2285 &1577  & 2514   &  2825\\ 
    %\hline
    \multirow{4}{5em}{\centering GK-RMSCNN-DeepONet}
    &$\corrf-\text{MSE}_T$& $ 3.4 \times {10}^{-5} $ &$ 1.1 \times {10}^{-4} $  &  $ 7.8 \times {10}^{-4} $ & $ 1.9 \times {10}^{-3} $ \\
    &$\corrf-\text{MSE}_{IG}$&$ 4.4 \times {10}^{-5} $ &$ 1.15 \times {10}^{-4} $    &  $ 7.3 \times {10}^{-4} $  &$ 1.6 \times {10}^{-3} $\\
    &$\corrf-\text{MSE}_{ND}$&$ 9.5 \times {10}^{-4} $ & $ 3.5 \times {10}^{-3} $    &    $ 4.9 \times {10}^{-3} $  & $ 6.8 \times {10}^{-3} $\\
    & time(s)  & 11145 &9543    & 14827   & 14749  \\ 
    %\hline
    \multirow{4}{*}{FNO}
    &$\corrf-\text{MSE}_T$&$ 5.3 \times {10}^{-7} $& $ 1.5 \times {10}^{-7} $   &$ 2.6 \times {10}^{-7} $ & $ 1.2 \times {10}^{-6} $ \\
    &$\corrf-\text{MSE}_{IG}$& $ 5.1 \times {10}^{-7} $ & $ 2.3 \times {10}^{-7} $    & $ 8.3 \times {10}^{-7} $ & $ 3.6 \times {10}^{-6} $ \\
    &$\corrf-\text{MSE}_{ND}$&$ 7.8 \times {10}^{-6} $& $ 6.1 \times {10}^{-5} $    & $ 5.0 \times {10}^{-5} $  & $ 1.9 \times {10}^{-5} $ \\
    & time(s)  &6430 & 5631   & 6933   &  7643 \\ 

\end{tabular}\end{ruledtabular}

\end{table*}

Four types of activation functions were tested in this study: (1) \acrfull{sRelu}, (2) \acrfull{Relu}, (3) \acrfull{Selu}, and (4) the logistic function. The mathematical definitions of these functions are provided in Appendix~\ref{appendix:activations}. Table~\ref{tab:compare_actis} summarizes the performance of different \acrshort{ML} models in terms of \acrshort{MSE} loss and training time. The quasi-local \acrshort{DNN} was excluded from this comparison due to its prohibitively long training time. As expected, the results are sensitive to the choice of activation function. In general, \acrshort{sRelu} yields superior extrapolation performance across most models. Notable exceptions include the full-scale \acrshort{DNN} with \acrshort{Selu}, and the \acrshort{DNN}-\acrshort{DeepONet} with the logistic activation function, both of which outperform their \acrshort{sRelu}-based counterparts. However, these two configurations still underperform relative to \acrshort{GK}-\acrshort{CNN}-\acrshort{DeepONet} and \acrshort{FNO}, which remain the top-performing models in terms of generalization to unseen data.

\subsection{cDFT solvers} \label{sec:density}
As mentioned above, neural operators can also be trained to predict \(\rho(x)\) from external potential \(V_{\text{ext}}(x)\) or background potential \(V_{\text{loc}}(x)\).  In this case, the primary objective is to calculate the density profile directly from the external potential, thus circumventing the energy-based computations typically required by numerical solvers. In other words,  the density operator \(\glssymbol{rhooperator}: V_{\text{loc}} \mapsto \rho\) acts as a surrogate for solving the \acrshort{EL} equation or direct minimization of the grand potential \glssymbol{Omega}. %This \acrshort{ML}-based strategy can be implemented using the excess free energy functional or the one-body direct correlation function, derived either analytically or through functional learning. 

Picard’s iteration remains the state-of-the-art method for most current applications of \acrshort{cDFT}. This method is thus used as the baseline in this study. In the operator learning framework for \(\glssymbol{rhooperator}: V_{\text{loc}} \mapsto \rho\) mapping, we consider two strategies: (1) solving the \acrshort{EL} equation using Picard’s iteration, and (2) minimizing the grand potential \glssymbol{Omega} using the \acrfull{L-BFGS} algorithm. Both approaches are driven by the learned functional $\glssymbol{corrfoperator} : \rho \mapsto \corrf$. This approach is referred to as operator-\acrshort{DFT}, with the variant implemented via \acrshort{FNO} denoted as \acrshort{FNO}-\acrshort{DFT}, which achieves the best overall performance. \acrshort{ML}-\acrshort{DFT} via quasi-local \acrshort{DNN} is referred to as “Neural-DFT,” following the implementation by Sammüller et al.\cite{Sammüller_2024}. In contrast, for direct learning of the operator $\glssymbol{rhooperator} : V_{\text{loc}} \mapsto \rho$, the \acrshort{FNO}-based model again yields the most accurate results and is presented as the representative example of direct density-based operator learning. %Results for other models trained to learn $\Goperator_{\rho}$ are provided in the \Cref{appendix:additional} .

In a previous study,\cite{RN139} we demonstrated that the \acrfull{ALEC} framework provides an effective surrogate method for solving the \acrshort{EL} equation. Here we use \acrshort{ALEC} with \acrfull{GPR} for learning the operator \glssymbol{rhooperator} as a point of comparison. As a sampling-based active learning strategy, \acrshort{ALEC} requires a sample generator to provide training data of $V_{loc}\mapsto \rho$. \acrshort{ML}-based \acrshort{DFT} methods, such as  Neural-DFT using Quasi-local \acrshort{DNN} and operator-\acrshort{DFT}s, are suitable candidates for this purpose. In our evaluation, we present an example of \acrshort{ALEC}-\acrshort{GPR} using \acrshort{FNO}-\acrshort{DFT} as the sample generator.

\begin{table}[ht!]
\caption{Comparison of different numerical methods for computing the density profiles of 1D-\acrshort{HR} fluids from $V_{loc}(x)$.}
\label{tab:RMSE_rho}
\centering
\renewcommand{\arraystretch}{1.2}
\begin{ruledtabular}\begin{tabular}{>{\centering\arraybackslash}m{10em} c c c }
 Method&  $\rho-\text{MSE}_{IG}$ &  $\rho-\text{MSE}_{ND}$ & time(s) \\ \hline
 Analytical Solution& baseline  & baseline  & 6 \\
 Numerical Solver (NS)& $ 2.1 \times {10}^{-9} $  & $ 3.2 \times {10}^{-9} $  & $ 6.3 \times {10}^{4} $\\
 FNO Density Operator& $ 7.6 \times {10}^{-5} $ &$ 9.5 \times {10}^{-4} $ & 12\\
 GK-RMSCNN-DeepONet Density Operator& $ 9.3 \times {10}^{-4} $ &$ 8.4 \times {10}^{-3} $ & 14\\
 ALEC-GPR-NS&   $ 2.4 \times {10}^{-6} $  & $ 1.0 \times {10}^{-5} $   &$ 1.2 \times {10}^{2} $\\
 ALEC-GPR-FNO-DFT&   $ 2.5 \times {10}^{-6} $  & $ 1.0 \times {10}^{-5} $ &$ 5.3 \times {10}^{2} $\\
 FNO-DFT& $ 2.4 \times {10}^{-5} $  &  $ 4.8 \times {10}^{-5} $ & $ 8.8 \times {10}^{4} $\\
 GK-RMSCNN-DeepONet-DFT& $ 2.5 \times {10}^{-4} $  &  $ 2.9 \times {10}^{-4} $ & $ 1.1 \times {10}^{5} $\\
 Neural-DFT\footnotemark[1] & / & / & $ 2.1 \times {10}^{6} $ \\ 

\end{tabular}\end{ruledtabular}
 \footnotetext[1]{The computation time for Neural-DFT is estimated from 100 samples. No prediction error was listed because of the long computation time.}
\end{table}

Table~\ref{tab:RMSE_rho} presents the \acrshort{MSE} and computation time for density predictions under various combinations of chemical potential ($\mu$) and external potential ($V_{\text{ext}}(x)$), evaluated using both the in-group and `new data' test sets. The metrics $\rho$-$\text{MSE}_{IG}$ and $\rho$-$\text{MSE}_{ND}$ represent the mean squared error on the in-group and `new data' test sets, respectively. The methods labeled as \acrshort{NS}, Neural-\acrshort{DFT}, and \acrshort{FNO}-\acrshort{DFT} solve the \acrshort{EL} equation using either Picard iteration or the \acrshort{L-BFGS} algorithm. In contrast, the \acrshort{ALEC}-\acrshort{GPR} and the direct density operator approach based on \acrshort{FNO} predict the density profile directly, bypassing the variational procedure. The \acrshort{ALEC}-\acrshort{GPR} results are shown with different sample generators, as indicated in the method name (e.g., \acrshort{ALEC}-\acrshort{GPR}-\acrshort{NS} and \acrshort{ALEC}-\acrshort{GPR}-\acrshort{FNO}-\acrshort{DFT}).

The results reveal that the neural operators \glssymbol{rhooperator} exhibit significant extrapolation errors on the `new data' test set, even when using the best-performing \acrshort{ML} model. Specifically, the lowest \acrshort{MSE} on the `new data' set—$9.5 \times 10^{-4}$—achieved by the \acrshort{FNO} density operator, still reflects inadequate predictive accuracy. This performance gap suggests that neural operators face challenges in capturing the complex mapping from $V_{\text{ext}}$ to $\rho$, especially when generalizing to out-of-distribution data. %A likely explanation is that the long-range interactions encoded in \glssymbol{rhooperator} vary significantly with the spatial structure of the potential, limiting the model’s extrapolation capability. 
The external potential $V_{ext}$ influences the density $\rho$ via an explicit exponential relationship, inducing a global coupling between these two functions. This global effect encoded in \glssymbol{rhooperator} is likely the reason that limits the model’s extrapolation capability.

The computational cost for predicting the density profiles with operator-\acrshort{DFT}s is comparable to that of \acrshort{NS}. This is somewhat expected because both require solving the \acrshort{EL} equation on the entire space. The \acrshort{MSE} of \acrshort{FNO}-\acrshort{DFT} is slightly larger for the `new data' test set compared to the in-group test set, the error level shows acceptable prediction ability for extrapolation. 

The exact prediction error of the neural-\acrshort{DFT} is omitted from Table~\ref{tab:RMSE_rho} due to its significantly longer runtime in generating $\rho(x)$. Although neural-\acrshort{DFT} also solves the self-consistent \acrshort{EL} equation like the \acrshort{NS} and operator-\acrshort{DFT} approaches, its use of discrete, quasi-local predictions for ${\corrf}$ leads to substantially higher computational cost. Unlike other methods that compute ${\corrf}$ as a vectorized operation, neural-\acrshort{DFT} evaluates ${\corrf}$ point-by-point, resulting in a much slower runtime. This inefficiency similarly affects its performance in computing \glssymbol{Fex}, making it less practical for large-scale applications.

Among all methods tested, \acrshort{ALEC}-\acrshort{GPR} demonstrates the best overall performance, achieving low prediction error, strong extrapolation capability, and efficient computation time. For density profile prediction, it remains the most effective surrogate for solving the \acrshort{EL} equation. Notably, the \acrshort{FNO}-\acrshort{DFT} model performs well as a data generator for \acrshort{ALEC}, enabling accurate and efficient active learning.

\begin{figure}[ht!]
\centering
\vspace{-0\baselineskip}
\subfloat[]{
\includegraphics[width=0.45\linewidth]{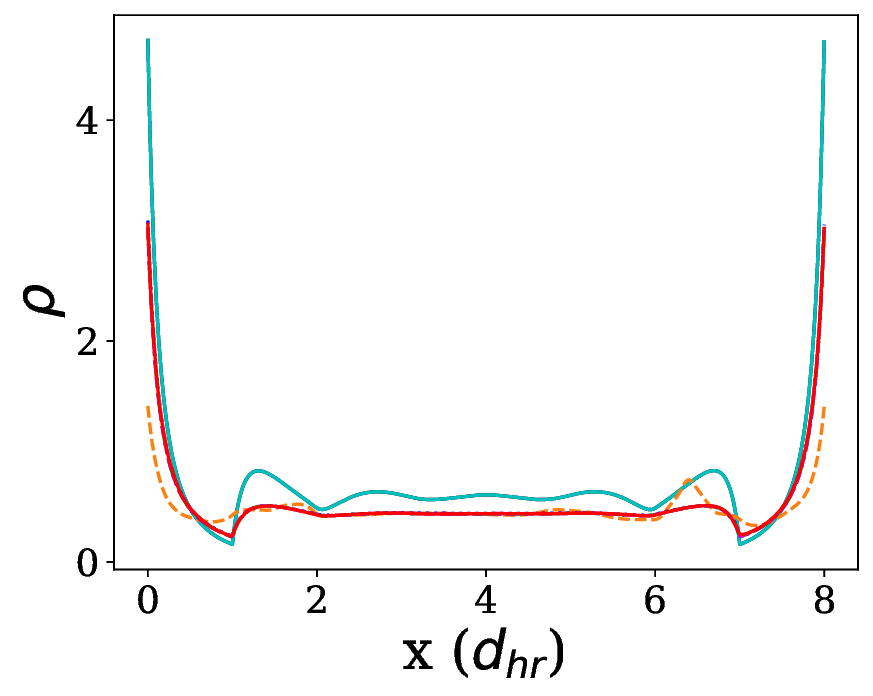} 
}
%\caption{Sample from In-group test set}
\vspace{-0\baselineskip}
\subfloat[]{
\includegraphics[width=0.45\linewidth]{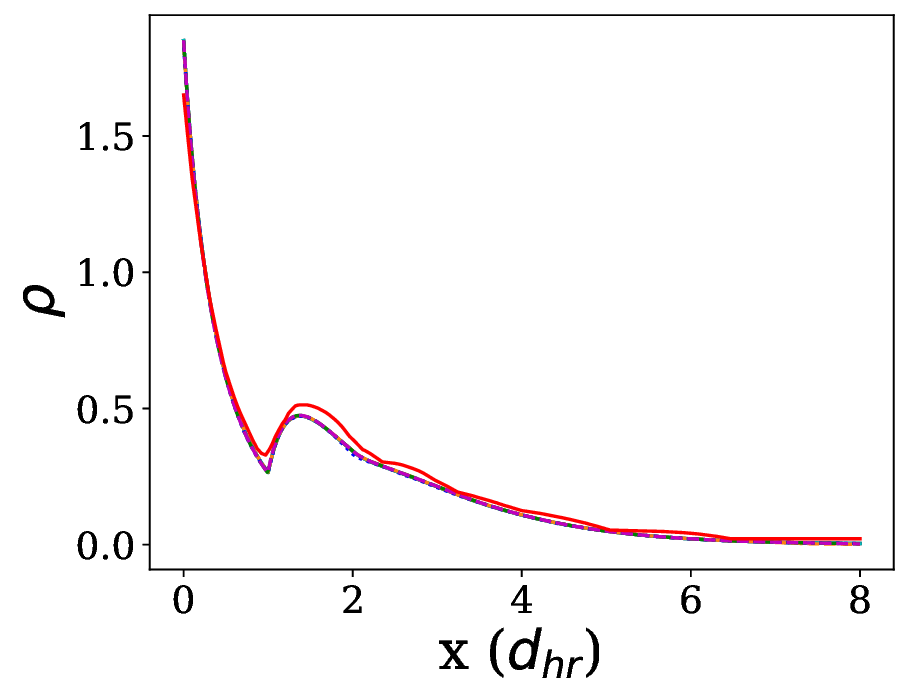}  
}
%\caption{sample from new dataset}

\includegraphics[width=0.85\linewidth]{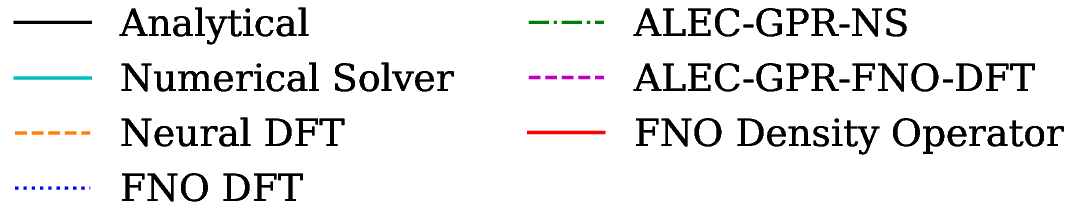} 
\vspace{-0.75\baselineskip}
\caption{\label{fig:density} Some representative examples for the predictions of density profiles under different $V_{loc}$ using various methods. (a) Prediction for a sample from the in-group test set; (b) Prediction for a sample from the new data test set. The legend is as follows: Analytical Solution– black solid line; Numerical Solver (NS)– cyan solid line; Neural-DFT– orange dashed line; \acrshort{FNO}-DFT– blue dotted line; \acrshort{ALEC}-\acrshort{GPR} with numerical solver generator– green dash-dotted line; \acrshort{ALEC}-\acrshort{GPR} with \acrshort{FNO}-DFT generator– violet dashed line;  \acrshort{FNO} Direct Density Operator– red solid line.}
\end{figure}

Figure~\ref{fig:density} presents the predicted density profiles for representative samples from data groups III and V across different \acrshort{ML} methods. In Figure~\ref{fig:density}(a), the quasi-local \acrshort{DFT} exhibits significant errors, particularly near the confining walls. In Figure~\ref{fig:density}(b), the \acrshort{FNO}-based direct density operator shows relatively larger errors across the entire spatial domain compared to other approaches. For the remaining methods, the prediction errors are not readily noticeable in the plotted profiles, indicating generally better agreement with the reference results.

%\acrfull{NS}, Neural-DFT, and FNO-DFT are obtained by solving the \acrshort{EL} equation—either explicitly (\acrshort{NS}) or using a learned mapping $\rho \mapsto \corrf$—via Picard iteration and/or minimization of the grand potential $\Omega$ using the \acrshort{L-BFGS} method. \acrfull{ALEC}-\acrfull{GPR} and the neural-operator-based direct density operator (using \acrshort{FNO}) directly predict the density profile without iterative solution of the \acrshort{EL} equation. \acrshort{ALEC}-\acrshort{GPR} results are shown for two different sample generators (\acrshort{NS} and \acrshort{FNO}-\acrshort{DFT}).

\subsection{ Performances of \glsentrylong{FNO} on additional examples}

The \acrshort{FNO} model demonstrates superior performance in learning both the direct correlation operator \glssymbol{corrfoperator} and the density operator \glssymbol{rhooperator}. However, its extrapolation capability for \glssymbol{rhooperator} is limited, making it unsuitable as a direct surrogate for the numerical solver (\acrshort{NS}) in density profile calculations. For this task, sampling-based active learning strategies such as \acrshort{ALEC} are preferred due to their robustness and adaptability. In this context, \acrshort{FNO}-\glssymbol{corrfoperator} serves effectively as a sample generator, offering a computationally efficient alternative to the time-intensive \acrshort{NS}.

In order to evaluate the general performance of this method, \acrshort{FNO} models with optimized hyperparameters were tested across a range of system sizes, using datasets \glssymbol{Dset} with $L = 4, 5, 6, 7, 8, 10, 12$ and grid resolution of $\Delta x=0.01$. \glssymbol{msize} is constant for \acrshort{FNO} regardless of simulation box size, since its trainable parameters are either channel-wise or in Fourier space.  To show its performance on more complex systems with realistic intermolecular interactions, we applied the model to a \acrshort{LJ} fluid with the data generated from the traditional \acrshort{cDFT} calculations using the weighted density approximation\cite{TANG2001149,Soares2023}, with the same box dimensions and external potential \glssymbol{Vext}. The LJ fluid simulations employed an atomic diameter $\sigma_{LJ}=1$ and temperature $T=0.6T_c$, allowing for the possibility of phase transitions. Results indicate that \acrshort{FNO} effectively captures the behavior of systems with intermolecular interactions. A summary of performance metrics is presented in Table~\ref{tab:FNO_performances}. 
\begin{figure}
\centering
\vspace{-0\baselineskip}
\subfloat[]{
\includegraphics[width=0.45\linewidth]{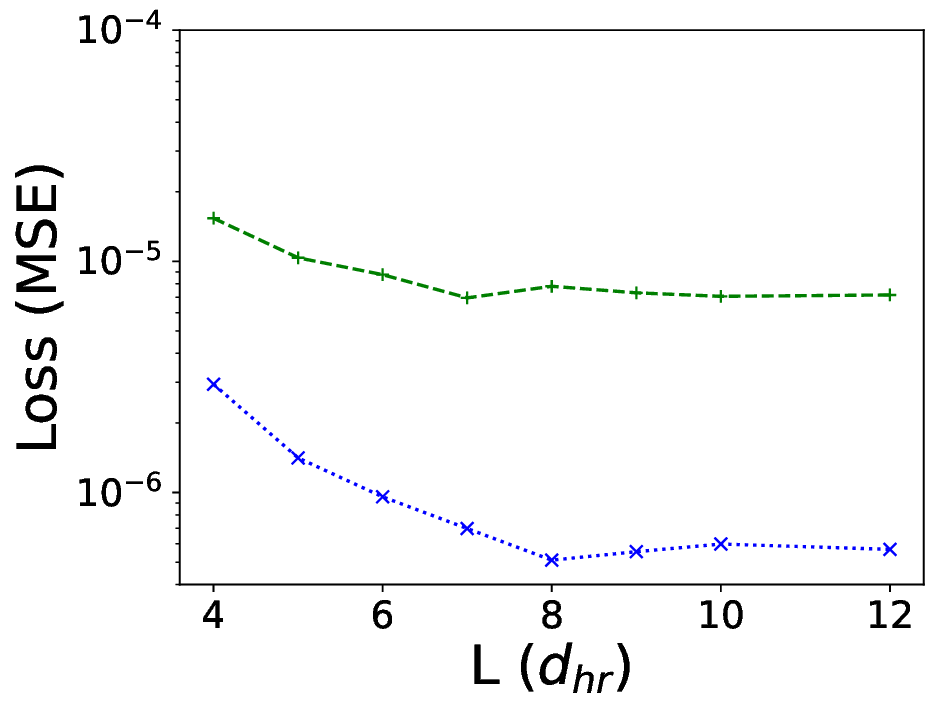} 
}
%\caption{Sample from In-group test set}
\vspace{-0\baselineskip}
\subfloat[]{
\includegraphics[width=0.45\linewidth]{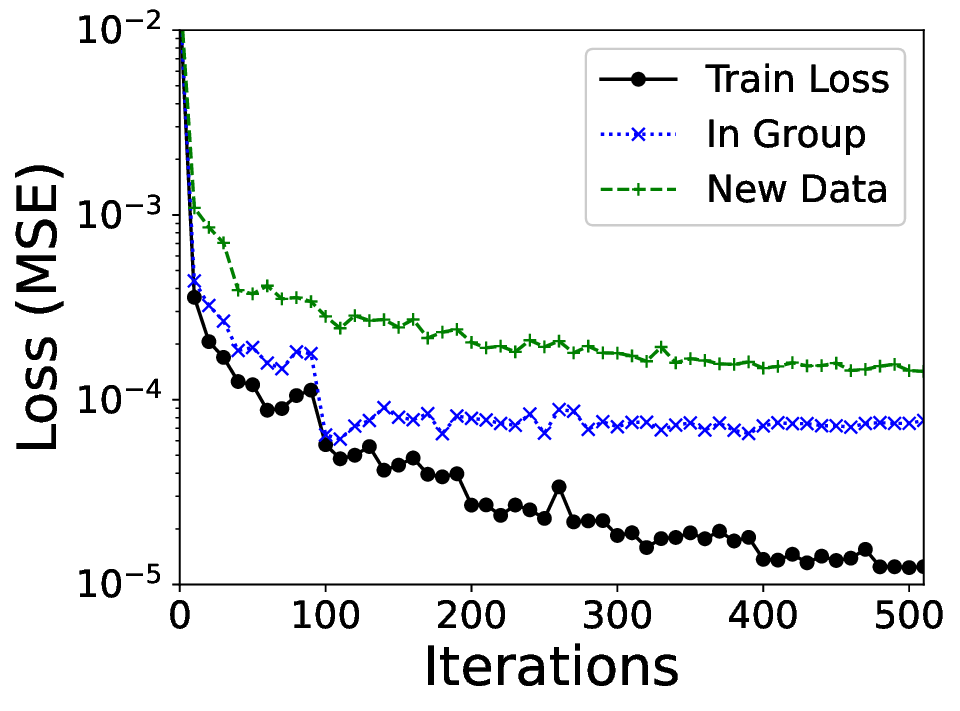}  
}
%\caption{sample from new dataset}
\caption{Performance of \acrshort{FNO} in learning the direct correlation operator \glssymbol{corrfoperator}. (a) \acrshort{MSE} loss versus box sizes \glssymbol{L} in learning \glssymbol{corrfoperator} in 1D \acrshort{HR} fluid. (b) \acrshort{MSE} loss versus training iteration for \acrshort{LJ} fluids confined between hard walls at system size $\glssymbol{L} = 8$. \\
Legends — 
Training Set MSE: Black solid line with circle markers
In‑Group Test Set MSE: blue dotted line with “×” markers; 
New‑Data Test Set MSE: green dashed line with “+” markers.}
\label{fig:boxsize_comparisons}
\end{figure}

Figure \ref{fig:boxsize_comparisons}(a) shows that the direct correlation operator \glssymbol{corrfoperator} achieves consistently low losses for both interpolation and extrapolation across a range of system sizes, demonstrating robust predictive capability. The \acrshort{MSE} losses are slightly higher for smaller box sizes, likely due to the increased relative influence of sharp gradients in the near-wall region.  It is worth noting that, for the FNO model, the number of trainable parameters remains constant regardless of box size. Figure \ref{fig:boxsize_comparisons}(b) presents the \acrshort{MSE} loss as a function of training iterations for \acrshort{LJ} fluids confined between hard walls. The FNO performance in this case closely matches that observed for the one-dimensional \acrshort{HR} fluid, as shown in Figure \ref{fig:loss_step}. The convergence behavior reflects consistent generalization across both familiar and unseen configurations. 

\begin{figure}
\centering
\vspace{-0\baselineskip}
\subfloat[]{
\includegraphics[width=0.45\linewidth]{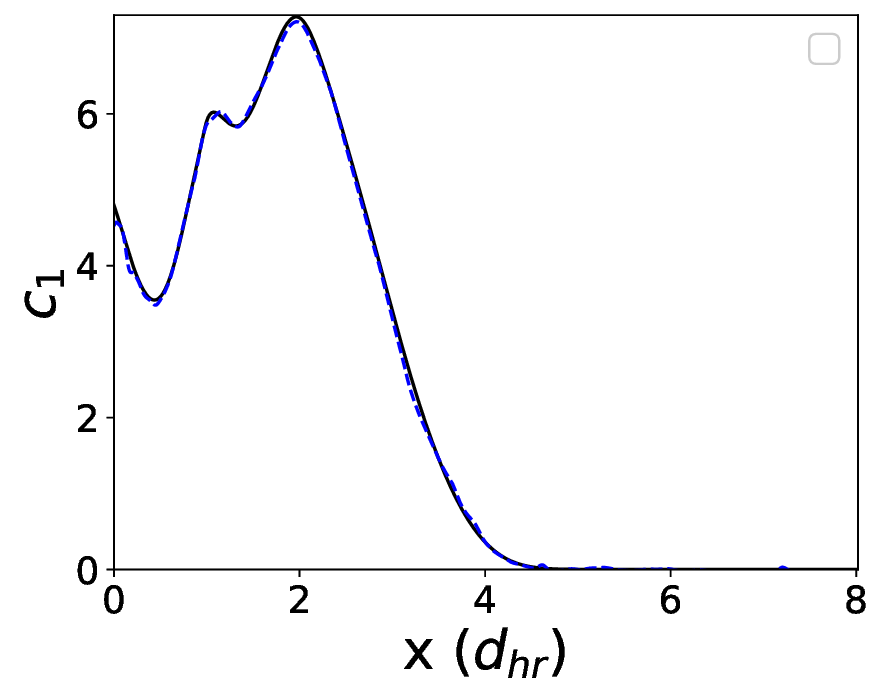} 
}
%\caption{Sample from In-group test set}
\vspace{-0\baselineskip}
\subfloat[]{
\includegraphics[width=0.45\linewidth]{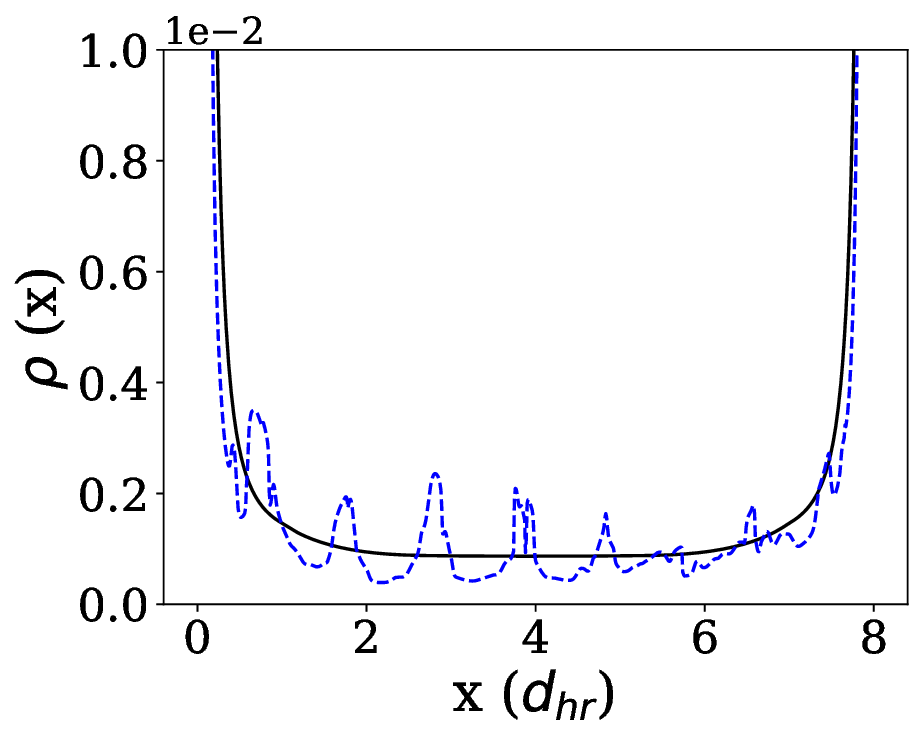}  
}
\vspace{-0\baselineskip}
\subfloat[]{
\includegraphics[width=0.45\linewidth]{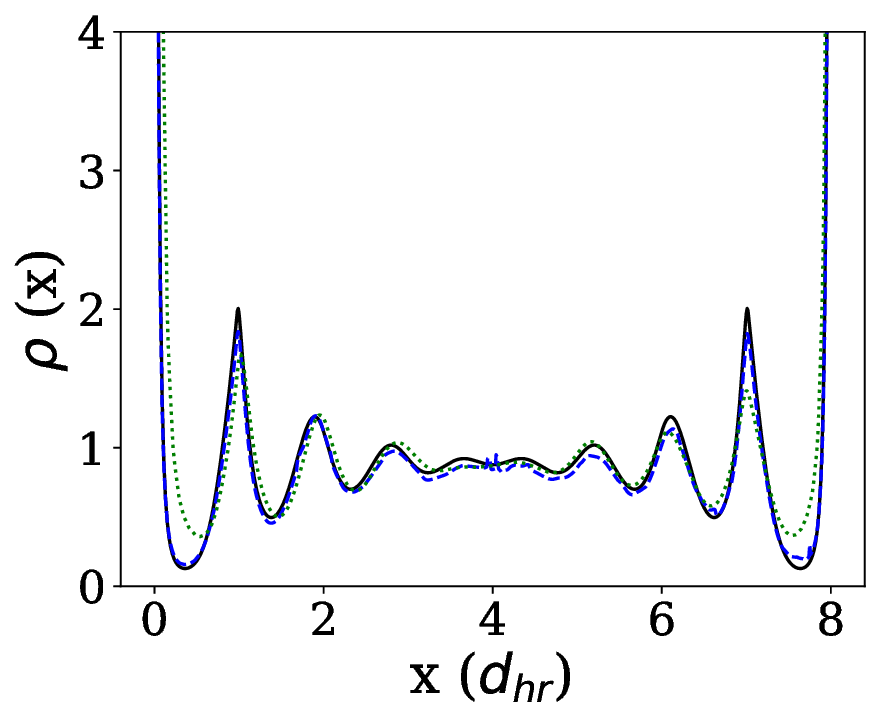}  
}
\vspace{-0\baselineskip}
\subfloat[]{
\includegraphics[width=0.45\linewidth]{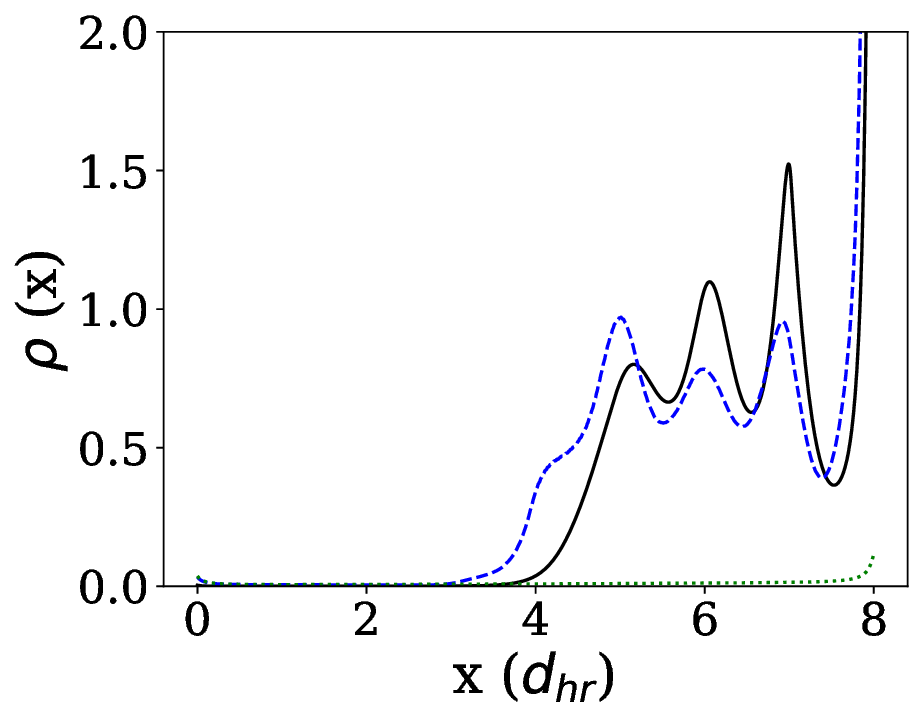}  
}

\includegraphics[width=0.85\linewidth]{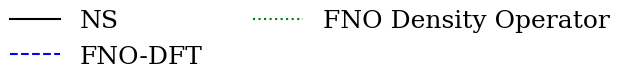} 
\vspace{-0.75\baselineskip}
%\caption{sample from new dataset}
\caption{Performance of \acrshort{FNO} in learning the direct correlation operator \glssymbol{corrfoperator} for one-dimensional \acrshort{LJ} fluids confined between hard walls at system size $\glssymbol{L} = 8$. (a) Representative prediction of the direct correlation function \glssymbol{corrf}, demonstrating the model’s accuracy in capturing spatial features under confinement. (b,c) Density profiles predicted by \acrshort{FNO}-\acrshort{cDFT} compared with \acrshort{NS} results for phase transition scenarios at $T = 0.6T_c$ and bulk density $\rho_b = 0.88$ with attractive walls (\Cref{appendix:V_ext}, Group~II, $\epsilon = 0.5$), illustrating distinct thermodynamic states: (b) gas phase (the \acrshort{FNO} direct density operator yields no solution; and (c) liquid phase. 
(d) Gas–liquid coexistence profile at $\rho_b = 0.88$ under a combined external potential of attractive walls and a linear term (\Cref{appendix:V_ext}, Group~III) with $\epsilon = 0.25$ and $V_{\mathrm{ext}}(x) = 0.8 - 0.15x$.\\ 
Legends : Numerical Solver: Black Solid line;  \acrshort{FNO}-DFT: Blue dashed line;\acrshort{FNO} Direct Density Operator– Green dotted line.}
\label{fig:1DLJ}
\end{figure}
%1D LJ fluid 

%Table concludes the performance of FNO leaning \glssymbol{rhooperator} for different
\begin{table}[]
\centering
\caption{Summary of \acrshort{FNO} performance across different datasets. Unless otherwise noted, all models are trained to learn the direct correlation operator \glssymbol{corrfoperator}; exceptions are explicitly indicated for models targeting the density operator \glssymbol{rhooperator}. Here, \text{HR} denotes one-dimensional hard-rod fluid, and \text{LJ} refers to one-dimensional Lennard-Jones fluid with $\sigma_{LJ} = 1$. }
\label{tab:FNO_performances}
\centering
\renewcommand{\arraystretch}{1.2}
\begin{ruledtabular}
\begin{tabular}{c c c c}
Data  & $\text{MSE}_{IG}$ & $\text{MSE}_{ND}$ & time (s)\\\hline
\text{HR}-$L=4$&$3.5\times 10^{-6}$ & $1.5\times 10^{-5}$&6221\\
\text{HR}-$L=8$& $5.1\times 10^{-7}$& $7.8\times 10^{-6}$ &6430\\
\text{HR}-$L=12$& $5.6\times 10^{-7}$ & $7.2\times 10^{-6}$ &6512\\
\text{LJ}-$L=8$& $3.8\times 10^{-5}$ & $7.1\times 10^{-5}$ &6321\\
\glssymbol{rhooperator}-\text{HR}-$L=8$& $7.6\times 10^{-5}$ & $9.5\times 10^{-4}$&6522\\
\glssymbol{rhooperator}-\text{LJ}-$L=8$& $ 1.05\times 10^{-5}$ & $7.32\times 10^{-4}$&6321\\
\end{tabular}
\end{ruledtabular}
\end{table}

%phase?
For the 1D-\acrshort{LJ} fluid with $\sigma_{LJ} = 1$, the extrapolation loss in learning the direct correlation operator \glssymbol{corrfoperator} is $7.1 \times 10^{-5}$. While this is higher than the corresponding performance on \acrshort{HR} fluids, it remains within a reasonable range. However, the poor extrapolation behavior observed in learning the density operator \glssymbol{rhooperator} persists for \acrshort{LJ} fluids, limiting its reliability in using as a surrogate for solving the \acrshort{EL} equation. %As shown in Figure \ref{fig:1DLJ}(a), the extrapolation loss exhibits significant degradation, underscoring the challenge of generalizing beyond the training distribution. 
A representative prediction of the direct correlation function \glssymbol{corrf} is presented in Figure \ref{fig:1DLJ}(b), demonstrating the model’s ability to capture key structural features despite extrapolation limitations.

Phase transition phenomena are a central focus in molecular simulations and hold particular relevance for operator learning within \acrshort{cDFT}. To examine such behavior, we employ a 1D-\acrshort{LJ} fluid as a representative system, since phase transitions are absent in 1D-\acrshort{HR} fluids due to their purely repulsive interactions. In this study, we examine a bulk density of $\rho_b=0.88$ under confinement by van der Waals–like attractive walls (Appendix B, Group II) at a system size of $L = 8$. This setup facilitates the investigation of interfacial structures and thermodynamic signatures characteristic of phase transitions in low-dimensional systems. As shown in Figure \ref{fig:1DLJ}(d), when subjected to an external potential of the type defined in Appendix B, Group V, which comprises the combined effects of attractive walls and a linear external potential, the direct correlation operator \glssymbol{corrfoperator} successfully captures liquid–gas phase separation, further supporting its predictive capability. In contrast, \acrshort{FNO}-\glssymbol{rhooperator}, which directly learns the mapping from external potential to density, fails to resolve these distinct phases, producing only the liquid-phase solution in Figures \ref{fig:1DLJ}(b,c) and the gas-phase solution in Figure \ref{fig:1DLJ}(d).

\subsection{Transferability of operator mapping}\label{sec:Operator_ops}
Operator learning methods aim to predict an infinite-dimensional output function $v$ from a given input function $u$, offering flexibility that is often unattainable with discrete functional learning methods. In comparison to functional learning methods, a key advantage of operator learning lies in its compatibility with newly positioned sensors for output measurements, thereby enabling model migration across different sensing configurations.

For instance, in the case of \acrshort{DeepONet}, the output function can be evaluated at arbitrary sensor locations, provided the input function is sampled at consistent positions. This flexibility is further enhanced in smooth-kernel \acrshort{CNN}-based \acrshort{DeepONet} architectures, such as \acrshort{GK}-\acrshort{CNN}-\acrshort{DeepONet} and \acrshort{GK}-\acrshort{RMSCNN}-\acrshort{DeepONet}, which can tolerate changes in the input sensor positions as long as the sampling resolution satisfies $\Delta x \leq x_{conv,cut} = 0.1$. In contrast, \acrshort{FNO} requires uniform sampling for both input and output functions, which limits its ability to focus on regions of physical significance. Nevertheless, \acrshort{FNO} supports model transfer across different sensor resolutions within the same spatial domain.

This capacity for flexible sampling is particularly relevant to \acrshort{cDFT} problems, where spatial significance often varies across the domain. Regions near phase interfaces or solid boundaries exhibit rapid variations in density $\rho$ and other physical properties, necessitating higher sampling frequencies to capture local behavior accurately. Conversely, bulk-like regions with nearly uniform external potentials tend to exhibit lower variability and can be adequately described with coarser sampling.

\begin{figure}[ht!]
\centering
\vspace{-0.75\baselineskip}
\subfloat[]{
\includegraphics[width=0.45\linewidth]{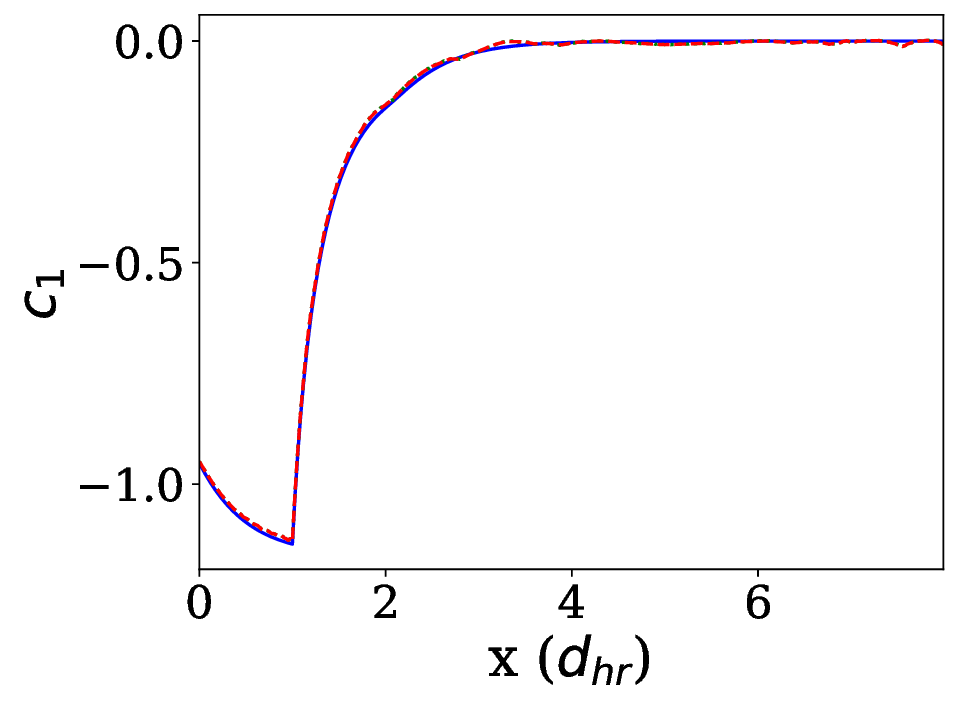} 
%\caption{G-MSCNN DeepONet Significance-relevant prediction, example from data group 2}
}
\vspace{-0.75\baselineskip}
\subfloat[]{
\includegraphics[width=0.45\linewidth]{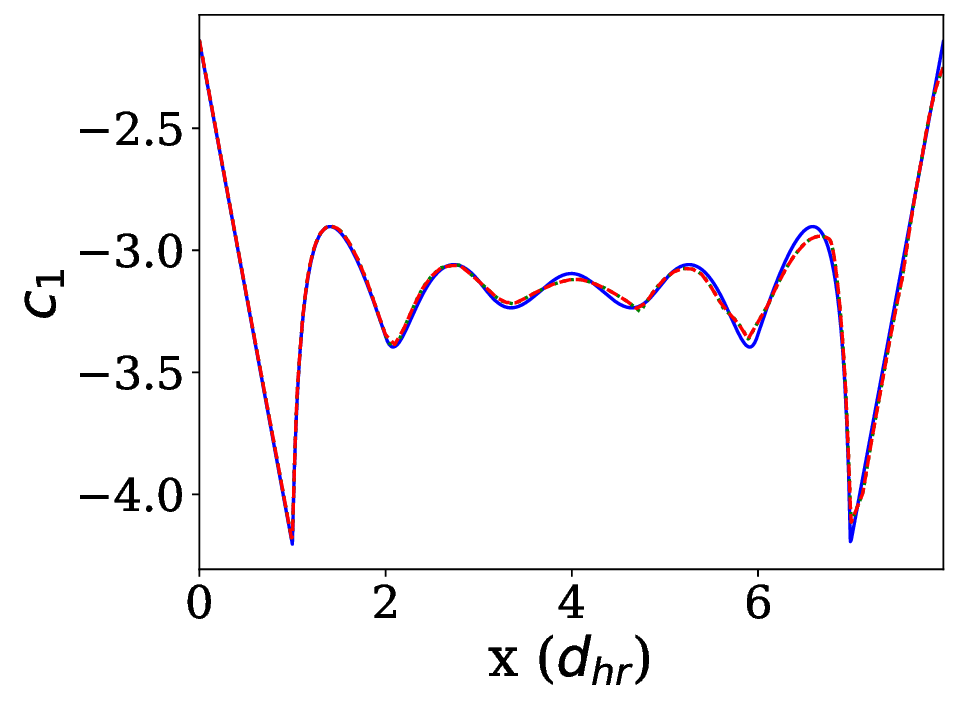} 
}
%\caption{G-MSCNN DeepONet Significance-relevant prediction, example from data group 3}

\vspace{-0.75\baselineskip}
\subfloat[]{
\includegraphics[width=0.45\linewidth]{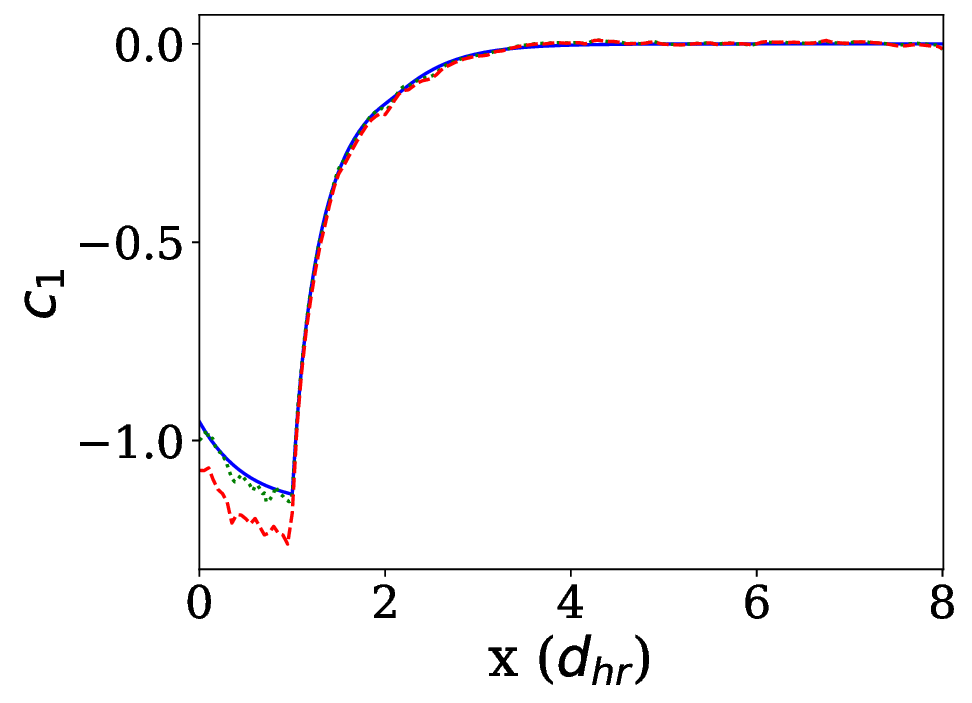}  
%\caption{FNO Migration of model from  example from data group 2}
}
\subfloat[]{
\includegraphics[width=0.45\linewidth]{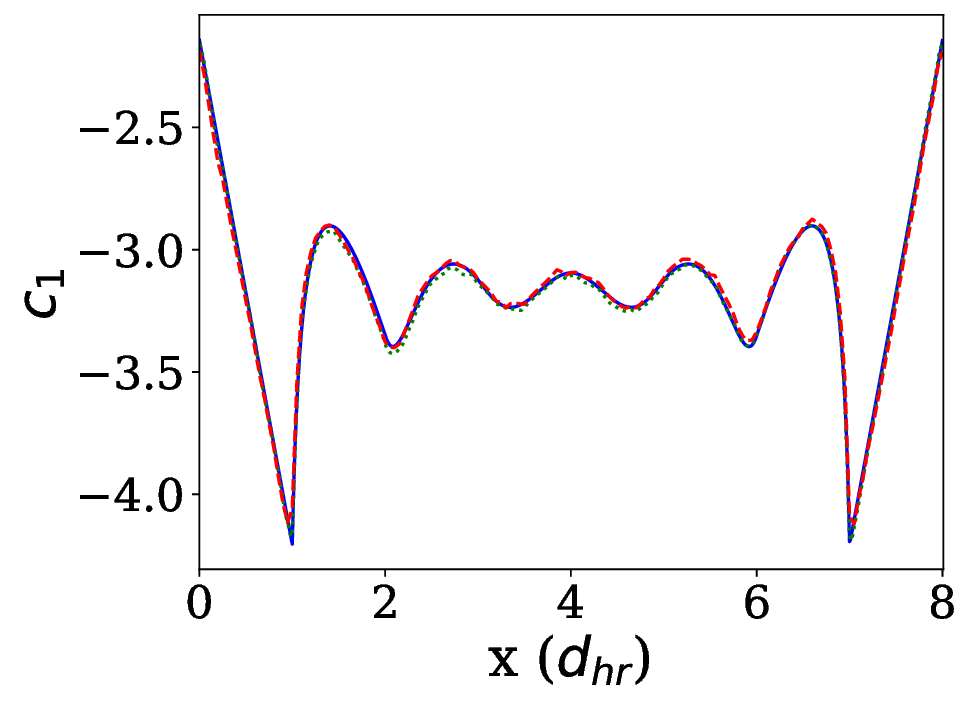} 
%\caption{FNO Migration of model from  example from data group 3}
}

\includegraphics[width=0.8\linewidth]{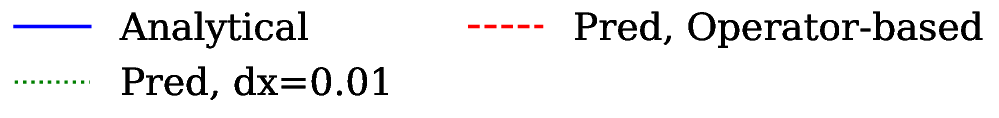} 
\vspace{-0.75\baselineskip}
\caption{\label{fig:out_as_functional} Function prediction by operator-\acrshort{DFT} models.
(a, b) Significance-relevant  predictions by \acrshort{GK}-\acrshort{RMSCNN}-\acrshort{DeepONet}.
(c, d) Resolution migration of \acrshort{FNO} models from high-resolution data ($dz = 0.01$) to low-resolution data ($dz = 0.05$).
(a, c) Sample from data group III; (b, d) Sample from data group V.
"Pred" denotes the model prediction. 
"Operator-based" denotes significance-relevant resolution for (ab), $\Delta x = 0.01$ for $x \in (0, 1.5) \cup (L - 1.5, L)$, $\Delta x = 0.1$ for $x \in (1.5,L - 1.5)$. For (cd) it denotes migrated resolution $\Delta x = 0.05$. The legend is as follows: Analytical Solution – blue solid line; Model Prediction with default grid resolution $\Delta x = 0.01$ – green dotted line; Operator-based – red dashed line.}  
\end{figure}

To demonstrate the transferability of \acrshort{DeepONet} operators, several samples were selected from the test sets of data groups III and V. Figure \ref{fig:out_as_functional} presents the significance-relevant prediction utilizing the \acrshort{GK}-\acrshort{RMSCNN}-\acrshort{DeepONet} framework. The same samples are also employed for \acrshort{FNO} resolution migration. The regions near the walls, \(x \in (0, 1.5) \cup (L - 1.5, L)\), are designated as high-significance regions, with a grid resolution of \(\Delta x = 0.01\). In the low-significance region, \(x \in (1.5, L - 1.5)\), a coarser grid resolution of \(\Delta x = 0.1\) is used. The results are compared to predictions made on uniformly discretized grids with \(\Delta x = 0.01\) and to the analytical solution for validation. Predictions from the \acrshort{FNO} model, using high-sensor-resolution input data with \(\Delta x = 0.01\) and low-sensor-resolution input data with \(\Delta x = 0.05\), are also compared with the analytical solutions. The results in Figure \ref{fig:out_as_functional} demonstrate that both significance-relevant predictions by \acrshort{GK}-\acrshort{RMSCNN}-\acrshort{DeepONet} and resolution migration using \acrshort{FNO} perform well for \(\corrf\) predictions.

\begin{figure}[ht!]
\centering
\subfloat[]{
\includegraphics[width=0.45\linewidth]{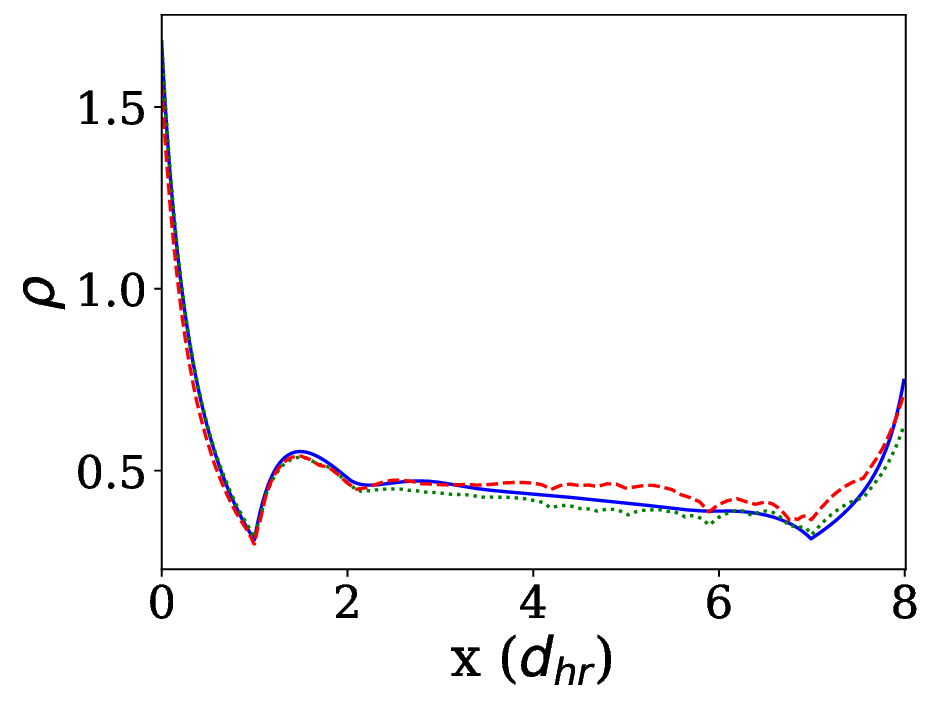}  
%\caption{G-MSCNN-DeepONet-cDFT with significance-relevant  uneven grid }
}
\subfloat[]{
\includegraphics[width=0.45\linewidth]{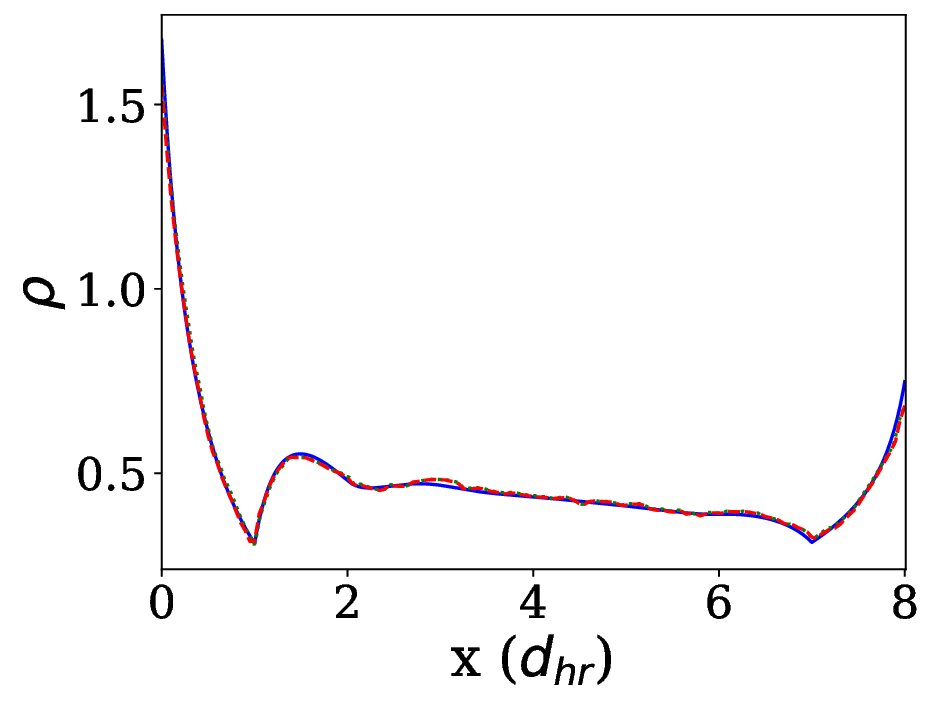} 
%\caption{FNO-cDFT with resolution migration from $dz=0.01$ to $dz=0.05$ }
}

\includegraphics[width=0.8\linewidth]{Figure11_12leg.eps} 
\vspace{-0.75\baselineskip}
\caption{\label{fig:application} Illustrative examples for the density profiles  obtained from operator-\acrshort{DFT} with operator-based functions, compared against analytical results and direct operator-\acrshort{DFT} calculations. (a) Significance-relevant prediction by \acrshort{GK}-\acrshort{RMSCNN}-\acrshort{DeepONet}. (b) Resolution migration of \acrshort{FNO} models across different sensor grids. Here, "Pred" denotes the \acrshort{ML} prediction.
"Operator-based" denotes significance-relevant resolution for (a), $\Delta x = 0.01$ for $x \in (0, 1.5) \cup (L - 1.5, L)$, $\Delta x = 0.1$ for $x \in (1.5,L - 1.5)$. For (b) it denotes migrated resolution $\Delta x = 0.05$. The legend is as follows: Analytical Solution – blue solid line; Model Prediction with default grid resolution $\Delta x = 0.01$ – green dotted line; Operator-based – red dashed line.}
\end{figure}

The capability of neural operators to enable model migration across varying input sensor grids implies that the underlying \acrshort{cDFT} framework and its associated calculations are also consistently applicable and executable across these diverse grid structures. To illustrate this, Figure \ref{fig:application}~(a) presents the significance-relevant prediction of \(\rho(x)\) using the \acrshort{GK}-\acrshort{RMSCNN}-\acrshort{DeepONet} framework on an uneven grid—identical to that used in Figure \ref{fig:out_as_functional}~(a,b). Meanwhile, Figure \ref{fig:application}~(b) shows the \(\rho(x)\) prediction from the migration of a trained \acrshort{FNO} model, where the input sensor resolution transitions from \(\Delta x = 0.01\) to \(\Delta x = 0.05\). These results are compared against same-grid \(\rho(x)\) predictions obtained from the same trained model using the same \(V_{\mathrm{loc}}\), as well as with the analytical solution. The results demonstrate that resolution migration is also effective for density prediction.

\section{Conclusion}

In this study, we investigated the application of various neural-operator architectures—including multiple variants of \acrshort{DeepONet} and \acrshort{FNO}—for learning the mapping from the density profile \(\rho(x)\) to the direct correlation function \({\corrf}(x)\) within the \acrshort{cDFT} framework. The results are compared against baseline models based on discrete functional mappings using \acrshort{DNN}, trained with either full-range data or quasi-local data (Neural-DFT). We analyzed and discussed in detail the influence of the selected \acrshort{ML} architectures and activation functions on the model training and performance.

For numerical efficiency, we used the training and testing data from analytical solutions of the one-dimensional \acrshort{HR} fluid subjected to various external potentials. To evaluate both interpolation and extrapolation capabilities of the \acrshort{ML} models, we constructed five distinct groups of external potentials, with the training samples randomly selected from four distinctive data groups (I-IV). The in-group test set is generated within the same data distribution as the training set. An additional test set, referred to as the `new data' test set, is drawn from data group V to assess the extrapolation performance of the models. In addition to evaluating the mean squared error (\acrshort{MSE}) of the predicted output function \({\corrf}\), we further assess model performance based on the calculated excess free energy, \(\Fex\), given a specified \(\rho(x)\), and the predicted \(\rho(x)\) under a new background potential \(V_{\mathrm{loc}}(x)\).

The results show that the \acrshort{sRelu} activation function outperforms all other activation functions across the different \acrshort{ML} models. \acrshort{FNO} is the best neural operator model in every metric of model evaluation, including on \acrshort{MSE} loss for In-group and New Data test loss, as well as the prediction of \glssymbol{Fex}.  Among the DeepONet variants tested in this work, the model using a Gaussian Kernel Multi-scale Residual Convolutional Neural Network as its branch net (\acrshort{GK}-\acrshort{RMSCNN}-\acrshort{DeepONet}) delivers the best performance.
%For variances of \acrshort{DeepONet}s,  \acrshort{DeepONet} with Gaussian Kernel Multi-scale Residual Convolution Neural Network as its rranch net (\acrshort{GK}-\acrshort{RMSCNN}-\acrshort{DeepONet}) performs the best among various \acrshort{DeepONet}s tested in this work. 
Neural scaling law analysis is used to estimate the generalization ability of ideal models with infinite trainable parameters \glssymbol{msize} and infinite training samples \glssymbol{nsample}. All neural operators closely follow the Chinchilla scaling law, with \acrshort{FNO} exhibiting the strongest ideal generalization performance.
%Neural scaling law analysis is used for the estimation of generalization ability for ideal models with infinite trainable parameters \glssymbol{msize} and infinite training samples \glssymbol{nsample}. All neural operators fit well to Chinchilla scaling, with \acrshort{FNO} also being the best one in ideal generalization ability.

We also demonstrate that neural operators can be used to compute density profiles by solving the self-consistent \acrshort{EL} equation. However, they struggle to directly learn the mapping $V_{\text{ext}} \mapsto \rho$, exhibiting poor extrapolation performance. Among the surrogate methods tested, \acrshort{ALEC}-\acrshort{GPR} proves to be the most effective alternative to solving the \acrshort{EL} equation. Because active learning requires an efficient sample generation strategy, operator-DFT offers a promising surrogate for \acrshort{NS} in this context.
%We also show that neural operators can be used to solve the density profiles by solving the self-consistent \acrshort{EL} equation. We find that neural operators cannot learn the direct mapping of $V_{ext}\mapsto\rho$ well, showing poor extrapolation performances. \acrshort{ALEC}-\acrshort{GPR} is the best surrogate method of the numerical solver (NS) in solving \acrshort{EL} equation. The active learning method requires a sample generation method, operator-DFT can be an effective surrogate of \acrshort{NS} there. 
In contrast to standard neural networks, a key advantage of neural operators is their ability to perform operator-based computations. These models learn infinite-dimensional function mappings, enabling evaluation at arbitrary positions—a capability inherently supported by \acrshort{DeepONet} architectures. Furthermore, scalable convolution layers that use smooth kernels, such as those in Gaussian Kernel CNNs (\acrshort{GK}-\acrshort{CNN}), allow flexibility in the input sensor grid of the first convolution layer. This feature enables models like \acrshort{GK}-\acrshort{RMSCNN}-DeepONet (and \acrshort{GK}-\acrshort{CNN}-DeepONet) to make predictions on uneven grids while maintaining relevance to the underlying functional structure. Unlike \acrshort{DeepONet}, the \acrshort{FNO} architecture restricts both input and output functions to the same domain with uniform discretization. As a result, it does not support evaluation at arbitrary positions but excels in transferring across sensor resolutions—making it suitable for tasks involving model migration. These operator-based capabilities also extend to other \acrshort{cDFT} computations, such as predicting $\rho(x)$ under new external potentials $V_{\text{loc}}(x)$.

%One major benefit of using neural operators is the ability to perform operator-based operations. The output learned is an infinite-dimensional function, and its value can be measured at any possible position, which is allowed in DeepONet methods. What's more, scalable convolution layers with a smooth function as its kernel allow the changing of the input sensor grid in the 1st convolution layer, while \acrshort{GK}-\acrshort{CNN} are examples of them. This allows the \acrshort{GK}-\acrshort{RMSCNN}-DeepONet (and \acrshort{GK}-\acrshort{CNN}-DeepONet) to perform significance-relevant prediction with uneven grid. \acrshort{FNO} limits the input and output functions defined on the same domain, and using the same even discretization, so measuring the output function at different positions is not possible. It allows and performs well for model migration between different sensor resolutions. These operator-based operations can also be applied during other \acrshort{cDFT} calculations such as $\rho(x)$ predictions with new $V_{loc}$.

Although in this work we evaluated the performance of neural operators for \acrshort{cDFT} using data from analytical solutions of the 1D \acrshort{HR} fluid, these \acrshort{ML} methods can be readily extended to more complex systems lacking satisfactory analytical density functional approximations, such as multi-component systems and those with many-body interactions. \cite{doi:10.1021/acs.jctc.5c00484} A central limitation in the broader application of cDFT lies in the lack of accurate analytical expressions for the direct correlation functional. ML-based approximations of \glssymbol{corrf} offer a promising avenue to overcome this challenge, thereby enhancing the applicability of \acrshort{cDFT} across a wider class of physical systems. Besides, the operator learning models are inherently designed for multiple-input, multiple-output mappings, facilitating straightforward application to multi-component systems.\cite{RN173,RN170}.

Finally, the full-scale mapping employed in this study restricts the spatial domain of the trained model to $\Dset$, limiting the applicability of the operator learning methods. In \acrshort{ML}-enhanced \acrshort{cDFT} calculations, we may approximate the dependence of ${\corrf}(x,\rho)$ on $\rho(x)$ via the quasi-local mapping: 
\begin{equation}
    {\rho^{(k)}(x')}_{x-\rcutoff}^{x+\rcutoff} \mapsto {\corrf}^{(k)}(x)
\end{equation}
where $\rcutoff$ is a cutoff distance. In that case, only $\rho(x')$ within the position window centered at $x$ with length $2*\rcutoff$ impacts the $\corrf$ value at position $x$.  While neural-DFT models the local value of $\corrf$ with a segment of $\rho$ as the input,\cite{RN152,RN141} the quasi-local operator learning treats $\corrf$ as a function rather than as discrete data points, thereby preserving the operator relationship.\cite{doi:10.1021/acs.jctc.5c00484}  Besides, \acrshort{FNO} and its theoretical foundation, the graph neural operator (GNO), can effectively utilize training data at varying resolutions. By combining low-resolution data with physically informed neural operators (PINO), it becomes possible to accurately predict solutions at higher resolutions for solving partial differential equations.\cite{RN181,RN182} A major limitation in \acrshort{ML}-based approach is the high cost of generating high-quality, high-resolution simulation data. PINO models trained on low-resolution data but guided by high-resolution physical constraints offer a promising approach to efficiently develop accurate \acrshort{ML} correlation functions, particularly for complex molecular systems involving many-body interactions.

\begin{acknowledgments}
This research is made possible through the financial support from the NSF-DFG Lead Agency Activity in Chemistry and Transport in Confined Spaces under Grant No. NSF 2234013.
\end{acknowledgments}

\section*{Data Availability Statement}
The data and codes that support the findings of this study will be openly available at https://github.com/prttt26/Operator-DFT-Full-Range, with the 1D \acrshort{HR} data available at https://github.com/prttt26/1D-Hard-Rod-Analytical.

\printnoidxglossary[type=acronym, title={List of Abbreviations}]

\printnoidxglossary[sort=use,type=symbol, title=symbols,nonumberlist]

\bibliography{jcp_cDFT_on}   

%merlin.mbs aipnum4-1.bst 2010-07-25 4.21a (PWD, AO, DPC) hacked
%Control: key (0)
%Control: author (8) initials jnrlst
%Control: editor formatted (1) identically to author
%Control: production of article title (-1) disabled
%Control: page (0) single
%Control: year (1) truncated
%Control: production of eprint (0) enabled
\begin{thebibliography}{57}%
\makeatletter
\providecommand \@ifxundefined [1]{%
 \@ifx{#1\undefined}
}%
\providecommand \@ifnum [1]{%
 \ifnum #1\expandafter \@firstoftwo
 \else \expandafter \@secondoftwo
 \fi
}%
\providecommand \@ifx [1]{%
 \ifx #1\expandafter \@firstoftwo
 \else \expandafter \@secondoftwo
 \fi
}%
\providecommand \natexlab [1]{#1}%
\providecommand \enquote  [1]{``#1''}%
\providecommand \bibnamefont  [1]{#1}%
\providecommand \bibfnamefont [1]{#1}%
\providecommand \citenamefont [1]{#1}%
\providecommand \href@noop [0]{\@secondoftwo}%
\providecommand \href [0]{\begingroup \@sanitize@url \@href}%
\providecommand \@href[1]{\@@startlink{#1}\@@href}%
\providecommand \@@href[1]{\endgroup#1\@@endlink}%
\providecommand \@sanitize@url [0]{\catcode `\\12\catcode `\$12\catcode `\&12\catcode `\#12\catcode `\^12\catcode `\_12\catcode `\%12\relax}%
\providecommand \@@startlink[1]{}%
\providecommand \@@endlink[0]{}%
\providecommand \url  [0]{\begingroup\@sanitize@url \@url }%
\providecommand \@url [1]{\endgroup\@href {#1}{\urlprefix }}%
\providecommand \urlprefix  [0]{URL }%
\providecommand \Eprint [0]{\href }%
\providecommand \doibase [0]{http://dx.doi.org/}%
\providecommand \selectlanguage [0]{\@gobble}%
\providecommand \bibinfo  [0]{\@secondoftwo}%
\providecommand \bibfield  [0]{\@secondoftwo}%
\providecommand \translation [1]{[#1]}%
\providecommand \BibitemOpen [0]{}%
\providecommand \bibitemStop [0]{}%
\providecommand \bibitemNoStop [0]{.\EOS\space}%
\providecommand \EOS [0]{\spacefactor3000\relax}%
\providecommand \BibitemShut  [1]{\csname bibitem#1\endcsname}%
\let\auto@bib@innerbib\@empty
%</preamble>
\bibitem [{\citenamefont {Evans}(1979)}]{Evans79}%
  \BibitemOpen
  \bibfield  {author} {\bibinfo {author} {\bibfnamefont {R.}~\bibnamefont {Evans}},\ }\href {\doibase 10.1080/00018737900101365} {\bibfield  {journal} {\bibinfo  {journal} {Advances in Physics}\ }\textbf {\bibinfo {volume} {28}},\ \bibinfo {pages} {143} (\bibinfo {year} {1979})},\ \Eprint {http://arxiv.org/abs/https://doi.org/10.1080/00018737900101365} {https://doi.org/10.1080/00018737900101365} \BibitemShut {NoStop}%
\bibitem [{\citenamefont {Jeanmairet}\ \emph {et~al.}(2014)\citenamefont {Jeanmairet}, \citenamefont {Levy}, \citenamefont {Levesque},\ and\ \citenamefont {Borgis}}]{doi:10.1021/ed500049m}%
  \BibitemOpen
  \bibfield  {author} {\bibinfo {author} {\bibfnamefont {G.}~\bibnamefont {Jeanmairet}}, \bibinfo {author} {\bibfnamefont {N.}~\bibnamefont {Levy}}, \bibinfo {author} {\bibfnamefont {M.}~\bibnamefont {Levesque}}, \ and\ \bibinfo {author} {\bibfnamefont {D.}~\bibnamefont {Borgis}},\ }\href {\doibase 10.1021/ed500049m} {\bibfield  {journal} {\bibinfo  {journal} {Journal of Chemical Education}\ }\textbf {\bibinfo {volume} {91}},\ \bibinfo {pages} {2112} (\bibinfo {year} {2014})},\ \Eprint {http://arxiv.org/abs/https://doi.org/10.1021/ed500049m} {https://doi.org/10.1021/ed500049m} \BibitemShut {NoStop}%
\bibitem [{\citenamefont {Wu}(2017)}]{RN8386}%
  \BibitemOpen
  \bibfield  {author} {\bibinfo {author} {\bibfnamefont {J.}~\bibnamefont {Wu}},\ }\href {https://link.springer.com/book/10.1007/978-981-10-2502-0} {\emph {\bibinfo {title} {Variational Methods in Molecular Modeling}}}\ (\bibinfo  {publisher} {Springer Berlin Heidelberg},\ \bibinfo {address} {New York, NY},\ \bibinfo {year} {2017})\ p.\ \bibinfo {pages} {324}\BibitemShut {NoStop}%
\bibitem [{\citenamefont {Wu}(2006)}]{wu2006density}%
  \BibitemOpen
  \bibfield  {author} {\bibinfo {author} {\bibfnamefont {J.}~\bibnamefont {Wu}},\ }\href {\doibase https://doi.org/10.1002/aic.10713} {\bibfield  {journal} {\bibinfo  {journal} {AIChE Journal}\ }\textbf {\bibinfo {volume} {52}},\ \bibinfo {pages} {1169} (\bibinfo {year} {2006})},\ \Eprint {http://arxiv.org/abs/https://aiche.onlinelibrary.wiley.com/doi/pdf/10.1002/aic.10713} {https://aiche.onlinelibrary.wiley.com/doi/pdf/10.1002/aic.10713} \BibitemShut {NoStop}%
\bibitem [{\citenamefont {Wu}\ and\ \citenamefont {Li}(2007)}]{wu2007}%
  \BibitemOpen
  \bibfield  {author} {\bibinfo {author} {\bibfnamefont {J.}~\bibnamefont {Wu}}\ and\ \bibinfo {author} {\bibfnamefont {Z.}~\bibnamefont {Li}},\ }\href {\doibase 10.1146/annurev.physchem.58.032806.104650} {\bibfield  {journal} {\bibinfo  {journal} {Annual Review of Physical Chemistry}\ }\textbf {\bibinfo {volume} {58}},\ \bibinfo {pages} {85} (\bibinfo {year} {2007})},\ \Eprint {http://arxiv.org/abs/https://doi.org/10.1146/annurev.physchem.58.032806.104650} {https://doi.org/10.1146/annurev.physchem.58.032806.104650} \BibitemShut {NoStop}%
\bibitem [{\citenamefont {de~Morais~Sermoud}\ \emph {et~al.}(2024)\citenamefont {de~Morais~Sermoud}, \citenamefont {{de Freitas Gonçalves}}, \citenamefont {{Barreto Jr.}}, \citenamefont {Franco}, \citenamefont {Tavares},\ and\ \citenamefont {Castier}}]{SERMOUD2024114177}%
  \BibitemOpen
  \bibfield  {author} {\bibinfo {author} {\bibfnamefont {V.}~\bibnamefont {de~Morais~Sermoud}}, \bibinfo {author} {\bibfnamefont {A.}~\bibnamefont {{de Freitas Gonçalves}}}, \bibinfo {author} {\bibfnamefont {A.~G.}\ \bibnamefont {{Barreto Jr.}}}, \bibinfo {author} {\bibfnamefont {L.~F.~M.}\ \bibnamefont {Franco}}, \bibinfo {author} {\bibfnamefont {F.~W.}\ \bibnamefont {Tavares}}, \ and\ \bibinfo {author} {\bibfnamefont {M.}~\bibnamefont {Castier}},\ }\href {\doibase https://doi.org/10.1016/j.fluid.2024.114177} {\bibfield  {journal} {\bibinfo  {journal} {Fluid Phase Equilibria}\ }\textbf {\bibinfo {volume} {586}},\ \bibinfo {pages} {114177} (\bibinfo {year} {2024})}\BibitemShut {NoStop}%
\bibitem [{\citenamefont {Sammüller}, \citenamefont {Hermann},\ and\ \citenamefont {Schmidt}(2024)}]{Sammüller_2024}%
  \BibitemOpen
  \bibfield  {author} {\bibinfo {author} {\bibfnamefont {F.}~\bibnamefont {Sammüller}}, \bibinfo {author} {\bibfnamefont {S.}~\bibnamefont {Hermann}}, \ and\ \bibinfo {author} {\bibfnamefont {M.}~\bibnamefont {Schmidt}},\ }\href {\doibase 10.1088/1361-648X/ad326f} {\bibfield  {journal} {\bibinfo  {journal} {Journal of Physics: Condensed Matter}\ }\textbf {\bibinfo {volume} {36}},\ \bibinfo {pages} {243002} (\bibinfo {year} {2024})}\BibitemShut {NoStop}%
\bibitem [{\citenamefont {Simon}\ and\ \citenamefont {Oettel}(2024)}]{simon2024}%
  \BibitemOpen
  \bibfield  {author} {\bibinfo {author} {\bibfnamefont {A.}~\bibnamefont {Simon}}\ and\ \bibinfo {author} {\bibfnamefont {M.}~\bibnamefont {Oettel}},\ }\href {https://arxiv.org/abs/2406.07345} {\enquote {\bibinfo {title} {Machine learning approaches to classical density functional theory},}\ } (\bibinfo {year} {2024}),\ \Eprint {http://arxiv.org/abs/2406.07345} {arXiv:2406.07345 [cond-mat.stat-mech]} \BibitemShut {NoStop}%
\bibitem [{\citenamefont {Wu}\ and\ \citenamefont {Gu}(2023)}]{doi:10.1021/acs.jpclett.3c02804}%
  \BibitemOpen
  \bibfield  {author} {\bibinfo {author} {\bibfnamefont {J.}~\bibnamefont {Wu}}\ and\ \bibinfo {author} {\bibfnamefont {M.}~\bibnamefont {Gu}},\ }\href {\doibase 10.1021/acs.jpclett.3c02804} {\bibfield  {journal} {\bibinfo  {journal} {The Journal of Physical Chemistry Letters}\ }\textbf {\bibinfo {volume} {14}},\ \bibinfo {pages} {10545} (\bibinfo {year} {2023})},\ \Eprint {http://arxiv.org/abs/https://doi.org/10.1021/acs.jpclett.3c02804} {https://doi.org/10.1021/acs.jpclett.3c02804} \BibitemShut {NoStop}%
\bibitem [{\citenamefont {Lin}\ and\ \citenamefont {Oettel}(2019)}]{10.21468/SciPostPhys.6.2.025}%
  \BibitemOpen
  \bibfield  {author} {\bibinfo {author} {\bibfnamefont {S.-C.}\ \bibnamefont {Lin}}\ and\ \bibinfo {author} {\bibfnamefont {M.}~\bibnamefont {Oettel}},\ }\href {\doibase 10.21468/SciPostPhys.6.2.025} {\bibfield  {journal} {\bibinfo  {journal} {SciPost Phys.}\ }\textbf {\bibinfo {volume} {6}},\ \bibinfo {pages} {025} (\bibinfo {year} {2019})}\BibitemShut {NoStop}%
\bibitem [{\citenamefont {Lin}, \citenamefont {Martius},\ and\ \citenamefont {Oettel}(2020)}]{10.1063/1.5135919}%
  \BibitemOpen
  \bibfield  {author} {\bibinfo {author} {\bibfnamefont {S.-C.}\ \bibnamefont {Lin}}, \bibinfo {author} {\bibfnamefont {G.}~\bibnamefont {Martius}}, \ and\ \bibinfo {author} {\bibfnamefont {M.}~\bibnamefont {Oettel}},\ }\href {\doibase 10.1063/1.5135919} {\bibfield  {journal} {\bibinfo  {journal} {The Journal of Chemical Physics}\ }\textbf {\bibinfo {volume} {152}},\ \bibinfo {pages} {021102} (\bibinfo {year} {2020})},\ \Eprint {http://arxiv.org/abs/https://pubs.aip.org/aip/jcp/article-pdf/doi/10.1063/1.5135919/15568654/021102\_1\_online.pdf} {https://pubs.aip.org/aip/jcp/article-pdf/doi/10.1063/1.5135919/15568654/021102\_1\_online.pdf} \BibitemShut {NoStop}%
\bibitem [{\citenamefont {{Dillon}}\ \emph {et~al.}(2017)\citenamefont {{Dillon}}, \citenamefont {{Langmore}}, \citenamefont {{Tran}}, \citenamefont {{Brevdo}}, \citenamefont {{Vasudevan}}, \citenamefont {{Moore}}, \citenamefont {{Patton}}, \citenamefont {{Alemi}}, \citenamefont {{Hoffman}},\ and\ \citenamefont {{Saurous}}}]{2017arXiv171110604D}%
  \BibitemOpen
  \bibfield  {author} {\bibinfo {author} {\bibfnamefont {J.~V.}\ \bibnamefont {{Dillon}}}, \bibinfo {author} {\bibfnamefont {I.}~\bibnamefont {{Langmore}}}, \bibinfo {author} {\bibfnamefont {D.}~\bibnamefont {{Tran}}}, \bibinfo {author} {\bibfnamefont {E.}~\bibnamefont {{Brevdo}}}, \bibinfo {author} {\bibfnamefont {S.}~\bibnamefont {{Vasudevan}}}, \bibinfo {author} {\bibfnamefont {D.}~\bibnamefont {{Moore}}}, \bibinfo {author} {\bibfnamefont {B.}~\bibnamefont {{Patton}}}, \bibinfo {author} {\bibfnamefont {A.}~\bibnamefont {{Alemi}}}, \bibinfo {author} {\bibfnamefont {M.}~\bibnamefont {{Hoffman}}}, \ and\ \bibinfo {author} {\bibfnamefont {R.~A.}\ \bibnamefont {{Saurous}}},\ }\href {\doibase 10.48550/arXiv.1711.10604} {\bibfield  {journal} {\bibinfo  {journal} {arXiv e-prints}\ ,\ \bibinfo {eid} {arXiv:1711.10604}} (\bibinfo {year} {2017})},\ \Eprint {http://arxiv.org/abs/1711.10604} {arXiv:1711.10604 [cs.LG]} \BibitemShut {NoStop}%
\bibitem [{\citenamefont {Kelley}\ \emph {et~al.}(2024)\citenamefont {Kelley}, \citenamefont {Quinton}, \citenamefont {Fazel}, \citenamefont {Karimitari}, \citenamefont {Sutton},\ and\ \citenamefont {Sundararaman}}]{10.1063/5.0223792}%
  \BibitemOpen
  \bibfield  {author} {\bibinfo {author} {\bibfnamefont {M.~M.}\ \bibnamefont {Kelley}}, \bibinfo {author} {\bibfnamefont {J.}~\bibnamefont {Quinton}}, \bibinfo {author} {\bibfnamefont {K.}~\bibnamefont {Fazel}}, \bibinfo {author} {\bibfnamefont {N.}~\bibnamefont {Karimitari}}, \bibinfo {author} {\bibfnamefont {C.}~\bibnamefont {Sutton}}, \ and\ \bibinfo {author} {\bibfnamefont {R.}~\bibnamefont {Sundararaman}},\ }\href {\doibase 10.1063/5.0223792} {\bibfield  {journal} {\bibinfo  {journal} {The Journal of Chemical Physics}\ }\textbf {\bibinfo {volume} {161}},\ \bibinfo {pages} {144101} (\bibinfo {year} {2024})},\ \Eprint {http://arxiv.org/abs/https://pubs.aip.org/aip/jcp/article-pdf/doi/10.1063/5.0223792/20198541/144101\_1\_5.0223792.pdf} {https://pubs.aip.org/aip/jcp/article-pdf/doi/10.1063/5.0223792/20198541/144101\_1\_5.0223792.pdf} \BibitemShut {NoStop}%
\bibitem [{\citenamefont {Malpica-Morales}\ \emph {et~al.}(2023)\citenamefont {Malpica-Morales}, \citenamefont {Yatsyshin}, \citenamefont {Durán-Olivencia},\ and\ \citenamefont {Kalliadasis}}]{10.1063/5.0146920}%
  \BibitemOpen
  \bibfield  {author} {\bibinfo {author} {\bibfnamefont {A.}~\bibnamefont {Malpica-Morales}}, \bibinfo {author} {\bibfnamefont {P.}~\bibnamefont {Yatsyshin}}, \bibinfo {author} {\bibfnamefont {M.~A.}\ \bibnamefont {Durán-Olivencia}}, \ and\ \bibinfo {author} {\bibfnamefont {S.}~\bibnamefont {Kalliadasis}},\ }\href {\doibase 10.1063/5.0146920} {\bibfield  {journal} {\bibinfo  {journal} {The Journal of Chemical Physics}\ }\textbf {\bibinfo {volume} {159}},\ \bibinfo {pages} {104109} (\bibinfo {year} {2023})},\ \Eprint {http://arxiv.org/abs/https://pubs.aip.org/aip/jcp/article-pdf/doi/10.1063/5.0146920/18122563/104109\_1\_5.0146920.pdf} {https://pubs.aip.org/aip/jcp/article-pdf/doi/10.1063/5.0146920/18122563/104109\_1\_5.0146920.pdf} \BibitemShut {NoStop}%
\bibitem [{\citenamefont {Yatsyshin}, \citenamefont {Kalliadasis},\ and\ \citenamefont {Duncan}(2022)}]{10.1063/5.0071629}%
  \BibitemOpen
  \bibfield  {author} {\bibinfo {author} {\bibfnamefont {P.}~\bibnamefont {Yatsyshin}}, \bibinfo {author} {\bibfnamefont {S.}~\bibnamefont {Kalliadasis}}, \ and\ \bibinfo {author} {\bibfnamefont {A.~B.}\ \bibnamefont {Duncan}},\ }\href {\doibase 10.1063/5.0071629} {\bibfield  {journal} {\bibinfo  {journal} {The Journal of Chemical Physics}\ }\textbf {\bibinfo {volume} {156}},\ \bibinfo {pages} {074105} (\bibinfo {year} {2022})},\ \Eprint {http://arxiv.org/abs/https://pubs.aip.org/aip/jcp/article-pdf/doi/10.1063/5.0071629/16536031/074105\_1\_online.pdf} {https://pubs.aip.org/aip/jcp/article-pdf/doi/10.1063/5.0071629/16536031/074105\_1\_online.pdf} \BibitemShut {NoStop}%
\bibitem [{\citenamefont {Cats}\ \emph {et~al.}(2021)\citenamefont {Cats}, \citenamefont {Kuipers}, \citenamefont {de~Wind}, \citenamefont {van Damme}, \citenamefont {Coli}, \citenamefont {Dijkstra},\ and\ \citenamefont {van Roij}}]{10.1063/5.0042558}%
  \BibitemOpen
  \bibfield  {author} {\bibinfo {author} {\bibfnamefont {P.}~\bibnamefont {Cats}}, \bibinfo {author} {\bibfnamefont {S.}~\bibnamefont {Kuipers}}, \bibinfo {author} {\bibfnamefont {S.}~\bibnamefont {de~Wind}}, \bibinfo {author} {\bibfnamefont {R.}~\bibnamefont {van Damme}}, \bibinfo {author} {\bibfnamefont {G.~M.}\ \bibnamefont {Coli}}, \bibinfo {author} {\bibfnamefont {M.}~\bibnamefont {Dijkstra}}, \ and\ \bibinfo {author} {\bibfnamefont {R.}~\bibnamefont {van Roij}},\ }\href {\doibase 10.1063/5.0042558} {\bibfield  {journal} {\bibinfo  {journal} {APL Materials}\ }\textbf {\bibinfo {volume} {9}},\ \bibinfo {pages} {031109} (\bibinfo {year} {2021})},\ \Eprint {http://arxiv.org/abs/https://pubs.aip.org/aip/apm/article-pdf/doi/10.1063/5.0042558/14564986/031109\_1\_online.pdf} {https://pubs.aip.org/aip/apm/article-pdf/doi/10.1063/5.0042558/14564986/031109\_1\_online.pdf} \BibitemShut {NoStop}%
\bibitem [{\citenamefont {Dijkman}\ \emph {et~al.}(2025)\citenamefont {Dijkman}, \citenamefont {Dijkstra}, \citenamefont {van Roij}, \citenamefont {Welling}, \citenamefont {van~de Meent},\ and\ \citenamefont {Ensing}}]{PhysRevLett.134.056103}%
  \BibitemOpen
  \bibfield  {author} {\bibinfo {author} {\bibfnamefont {J.}~\bibnamefont {Dijkman}}, \bibinfo {author} {\bibfnamefont {M.}~\bibnamefont {Dijkstra}}, \bibinfo {author} {\bibfnamefont {R.}~\bibnamefont {van Roij}}, \bibinfo {author} {\bibfnamefont {M.}~\bibnamefont {Welling}}, \bibinfo {author} {\bibfnamefont {J.-W.}\ \bibnamefont {van~de Meent}}, \ and\ \bibinfo {author} {\bibfnamefont {B.}~\bibnamefont {Ensing}},\ }\href {\doibase 10.1103/PhysRevLett.134.056103} {\bibfield  {journal} {\bibinfo  {journal} {Phys. Rev. Lett.}\ }\textbf {\bibinfo {volume} {134}},\ \bibinfo {pages} {056103} (\bibinfo {year} {2025})}\BibitemShut {NoStop}%
\bibitem [{\citenamefont {Samm\"uller}\ and\ \citenamefont {Schmidt}(2024)}]{PhysRevE.110.L032601}%
  \BibitemOpen
  \bibfield  {author} {\bibinfo {author} {\bibfnamefont {F.}~\bibnamefont {Samm\"uller}}\ and\ \bibinfo {author} {\bibfnamefont {M.}~\bibnamefont {Schmidt}},\ }\href {\doibase 10.1103/PhysRevE.110.L032601} {\bibfield  {journal} {\bibinfo  {journal} {Phys. Rev. E}\ }\textbf {\bibinfo {volume} {110}},\ \bibinfo {pages} {L032601} (\bibinfo {year} {2024})}\BibitemShut {NoStop}%
\bibitem [{\citenamefont {Sammüller}\ \emph {et~al.}(2023)\citenamefont {Sammüller}, \citenamefont {Hermann}, \citenamefont {de~Las~Heras},\ and\ \citenamefont {Schmidt}}]{RN141}%
  \BibitemOpen
  \bibfield  {author} {\bibinfo {author} {\bibfnamefont {F.}~\bibnamefont {Sammüller}}, \bibinfo {author} {\bibfnamefont {S.}~\bibnamefont {Hermann}}, \bibinfo {author} {\bibfnamefont {D.}~\bibnamefont {de~Las~Heras}}, \ and\ \bibinfo {author} {\bibfnamefont {M.}~\bibnamefont {Schmidt}},\ }\href {\doibase 10.1073/pnas.2312484120} {\bibfield  {journal} {\bibinfo  {journal} {Proc Natl Acad Sci U S A}\ }\textbf {\bibinfo {volume} {120}},\ \bibinfo {pages} {e2312484120} (\bibinfo {year} {2023})}\BibitemShut {NoStop}%
\bibitem [{\citenamefont {Samm\"uller}, \citenamefont {Schmidt},\ and\ \citenamefont {Evans}(2025)}]{PhysRevX.15.011013}%
  \BibitemOpen
  \bibfield  {author} {\bibinfo {author} {\bibfnamefont {F.}~\bibnamefont {Samm\"uller}}, \bibinfo {author} {\bibfnamefont {M.}~\bibnamefont {Schmidt}}, \ and\ \bibinfo {author} {\bibfnamefont {R.}~\bibnamefont {Evans}},\ }\href {\doibase 10.1103/PhysRevX.15.011013} {\bibfield  {journal} {\bibinfo  {journal} {Phys. Rev. X}\ }\textbf {\bibinfo {volume} {15}},\ \bibinfo {pages} {011013} (\bibinfo {year} {2025})}\BibitemShut {NoStop}%
\bibitem [{\citenamefont {Bui}\ and\ \citenamefont {Cox}(2025)}]{PhysRevLett.134.148001}%
  \BibitemOpen
  \bibfield  {author} {\bibinfo {author} {\bibfnamefont {A.~T.}\ \bibnamefont {Bui}}\ and\ \bibinfo {author} {\bibfnamefont {S.~J.}\ \bibnamefont {Cox}},\ }\href {\doibase 10.1103/PhysRevLett.134.148001} {\bibfield  {journal} {\bibinfo  {journal} {Phys. Rev. Lett.}\ }\textbf {\bibinfo {volume} {134}},\ \bibinfo {pages} {148001} (\bibinfo {year} {2025})}\BibitemShut {NoStop}%
\bibitem [{\citenamefont {Yang}\ \emph {et~al.}(0)\citenamefont {Yang}, \citenamefont {Pan}, \citenamefont {Sun},\ and\ \citenamefont {Wu}}]{doi:10.1021/acs.jctc.5c00484}%
  \BibitemOpen
  \bibfield  {author} {\bibinfo {author} {\bibfnamefont {J.}~\bibnamefont {Yang}}, \bibinfo {author} {\bibfnamefont {R.}~\bibnamefont {Pan}}, \bibinfo {author} {\bibfnamefont {J.}~\bibnamefont {Sun}}, \ and\ \bibinfo {author} {\bibfnamefont {J.}~\bibnamefont {Wu}},\ }\href {\doibase 10.1021/acs.jctc.5c00484} {\bibfield  {journal} {\bibinfo  {journal} {Journal of Chemical Theory and Computation}\ }\textbf {\bibinfo {volume} {0}},\ \bibinfo {pages} {null} (\bibinfo {year} {0})},\ \Eprint {http://arxiv.org/abs/https://doi.org/10.1021/acs.jctc.5c00484} {https://doi.org/10.1021/acs.jctc.5c00484} \BibitemShut {NoStop}%
\bibitem [{\citenamefont {Santos-Silva}\ \emph {et~al.}(2014)\citenamefont {Santos-Silva}, \citenamefont {Teixeira}, \citenamefont {Anquetil-Deck},\ and\ \citenamefont {Cleaver}}]{PhysRevE.89.053316}%
  \BibitemOpen
  \bibfield  {author} {\bibinfo {author} {\bibfnamefont {T.}~\bibnamefont {Santos-Silva}}, \bibinfo {author} {\bibfnamefont {P.~I.~C.}\ \bibnamefont {Teixeira}}, \bibinfo {author} {\bibfnamefont {C.}~\bibnamefont {Anquetil-Deck}}, \ and\ \bibinfo {author} {\bibfnamefont {D.~J.}\ \bibnamefont {Cleaver}},\ }\href {\doibase 10.1103/PhysRevE.89.053316} {\bibfield  {journal} {\bibinfo  {journal} {Phys. Rev. E}\ }\textbf {\bibinfo {volume} {89}},\ \bibinfo {pages} {053316} (\bibinfo {year} {2014})}\BibitemShut {NoStop}%
\bibitem [{\citenamefont {Fang}, \citenamefont {Gu},\ and\ \citenamefont {Wu}(2022)}]{RN139}%
  \BibitemOpen
  \bibfield  {author} {\bibinfo {author} {\bibfnamefont {X.}~\bibnamefont {Fang}}, \bibinfo {author} {\bibfnamefont {M.}~\bibnamefont {Gu}}, \ and\ \bibinfo {author} {\bibfnamefont {J.}~\bibnamefont {Wu}},\ }\href {\doibase 10.1063/5.0121805} {\bibfield  {journal} {\bibinfo  {journal} {J Chem Phys}\ }\textbf {\bibinfo {volume} {157}},\ \bibinfo {pages} {214109} (\bibinfo {year} {2022})}\BibitemShut {NoStop}%
\bibitem [{\citenamefont {Li}\ \emph {et~al.}(2020{\natexlab{a}})\citenamefont {Li}, \citenamefont {Kovachki}, \citenamefont {Azizzadenesheli}, \citenamefont {Liu}, \citenamefont {Bhattacharya}, \citenamefont {Stuart},\ and\ \citenamefont {Anandkumar}}]{li2020fourier}%
  \BibitemOpen
  \bibfield  {author} {\bibinfo {author} {\bibfnamefont {Z.}~\bibnamefont {Li}}, \bibinfo {author} {\bibfnamefont {N.}~\bibnamefont {Kovachki}}, \bibinfo {author} {\bibfnamefont {K.}~\bibnamefont {Azizzadenesheli}}, \bibinfo {author} {\bibfnamefont {B.}~\bibnamefont {Liu}}, \bibinfo {author} {\bibfnamefont {K.}~\bibnamefont {Bhattacharya}}, \bibinfo {author} {\bibfnamefont {A.}~\bibnamefont {Stuart}}, \ and\ \bibinfo {author} {\bibfnamefont {A.}~\bibnamefont {Anandkumar}},\ }\href@noop {} {\bibfield  {journal} {\bibinfo  {journal} {arXiv preprint arXiv:2010.08895}\ } (\bibinfo {year} {2020}{\natexlab{a}})}\BibitemShut {NoStop}%
\bibitem [{\citenamefont {Wang}\ \emph {et~al.}(2023)\citenamefont {Wang}, \citenamefont {Fu}, \citenamefont {Du}, \citenamefont {Gao}, \citenamefont {Huang}, \citenamefont {Liu}, \citenamefont {Chandak}, \citenamefont {Liu}, \citenamefont {Van~Katwyk}, \citenamefont {Deac} \emph {et~al.}}]{wang2023scientific}%
  \BibitemOpen
  \bibfield  {author} {\bibinfo {author} {\bibfnamefont {H.}~\bibnamefont {Wang}}, \bibinfo {author} {\bibfnamefont {T.}~\bibnamefont {Fu}}, \bibinfo {author} {\bibfnamefont {Y.}~\bibnamefont {Du}}, \bibinfo {author} {\bibfnamefont {W.}~\bibnamefont {Gao}}, \bibinfo {author} {\bibfnamefont {K.}~\bibnamefont {Huang}}, \bibinfo {author} {\bibfnamefont {Z.}~\bibnamefont {Liu}}, \bibinfo {author} {\bibfnamefont {P.}~\bibnamefont {Chandak}}, \bibinfo {author} {\bibfnamefont {S.}~\bibnamefont {Liu}}, \bibinfo {author} {\bibfnamefont {P.}~\bibnamefont {Van~Katwyk}}, \bibinfo {author} {\bibfnamefont {A.}~\bibnamefont {Deac}},  \emph {et~al.},\ }\href@noop {} {\bibfield  {journal} {\bibinfo  {journal} {Nature}\ }\textbf {\bibinfo {volume} {620}},\ \bibinfo {pages} {47} (\bibinfo {year} {2023})}\BibitemShut {NoStop}%
\bibitem [{\citenamefont {Kovachki}\ \emph {et~al.}(2023)\citenamefont {Kovachki}, \citenamefont {Li}, \citenamefont {Liu}, \citenamefont {Azizzadenesheli}, \citenamefont {Bhattacharya}, \citenamefont {Stuart},\ and\ \citenamefont {Anandkumar}}]{JMLR2023}%
  \BibitemOpen
  \bibfield  {author} {\bibinfo {author} {\bibfnamefont {N.}~\bibnamefont {Kovachki}}, \bibinfo {author} {\bibfnamefont {Z.}~\bibnamefont {Li}}, \bibinfo {author} {\bibfnamefont {B.}~\bibnamefont {Liu}}, \bibinfo {author} {\bibfnamefont {K.}~\bibnamefont {Azizzadenesheli}}, \bibinfo {author} {\bibfnamefont {K.}~\bibnamefont {Bhattacharya}}, \bibinfo {author} {\bibfnamefont {A.}~\bibnamefont {Stuart}}, \ and\ \bibinfo {author} {\bibfnamefont {A.}~\bibnamefont {Anandkumar}},\ }\href {http://jmlr.org/papers/v24/21-1524.html} {\bibfield  {journal} {\bibinfo  {journal} {Journal of Machine Learning Research}\ }\textbf {\bibinfo {volume} {24}},\ \bibinfo {pages} {1} (\bibinfo {year} {2023})}\BibitemShut {NoStop}%
\bibitem [{\citenamefont {Vanderlick}, \citenamefont {Scriven},\ and\ \citenamefont {Davis}(1986)}]{RN183}%
  \BibitemOpen
  \bibfield  {author} {\bibinfo {author} {\bibfnamefont {T.~K.}\ \bibnamefont {Vanderlick}}, \bibinfo {author} {\bibfnamefont {L.~E.}\ \bibnamefont {Scriven}}, \ and\ \bibinfo {author} {\bibfnamefont {H.~T.}\ \bibnamefont {Davis}},\ }\href {\doibase 10.1103/physreva.34.5130} {\bibfield  {journal} {\bibinfo  {journal} {Phys Rev A Gen Phys}\ }\textbf {\bibinfo {volume} {34}},\ \bibinfo {pages} {5130} (\bibinfo {year} {1986})}\BibitemShut {NoStop}%
\bibitem [{\citenamefont {Muckley}\ \emph {et~al.}(2023)\citenamefont {Muckley}, \citenamefont {Saal}, \citenamefont {Meredig}, \citenamefont {Roper},\ and\ \citenamefont {Martin}}]{RN144}%
  \BibitemOpen
  \bibfield  {author} {\bibinfo {author} {\bibfnamefont {E.~S.}\ \bibnamefont {Muckley}}, \bibinfo {author} {\bibfnamefont {J.~E.}\ \bibnamefont {Saal}}, \bibinfo {author} {\bibfnamefont {B.}~\bibnamefont {Meredig}}, \bibinfo {author} {\bibfnamefont {C.~S.}\ \bibnamefont {Roper}}, \ and\ \bibinfo {author} {\bibfnamefont {J.~H.}\ \bibnamefont {Martin}},\ }\href {\doibase 10.1039/d3dd00082f} {\bibfield  {journal} {\bibinfo  {journal} {Digital Discovery}\ }\textbf {\bibinfo {volume} {2}},\ \bibinfo {pages} {1425} (\bibinfo {year} {2023})}\BibitemShut {NoStop}%
\bibitem [{\citenamefont {Kaplan}\ \emph {et~al.}(2020)\citenamefont {Kaplan}, \citenamefont {McCandlish}, \citenamefont {Henighan}, \citenamefont {Brown}, \citenamefont {Chess}, \citenamefont {Child}, \citenamefont {Gray}, \citenamefont {Radford}, \citenamefont {Wu},\ and\ \citenamefont {Amodei}}]{kaplan2020scalinglawsneurallanguage}%
  \BibitemOpen
  \bibfield  {author} {\bibinfo {author} {\bibfnamefont {J.}~\bibnamefont {Kaplan}}, \bibinfo {author} {\bibfnamefont {S.}~\bibnamefont {McCandlish}}, \bibinfo {author} {\bibfnamefont {T.}~\bibnamefont {Henighan}}, \bibinfo {author} {\bibfnamefont {T.~B.}\ \bibnamefont {Brown}}, \bibinfo {author} {\bibfnamefont {B.}~\bibnamefont {Chess}}, \bibinfo {author} {\bibfnamefont {R.}~\bibnamefont {Child}}, \bibinfo {author} {\bibfnamefont {S.}~\bibnamefont {Gray}}, \bibinfo {author} {\bibfnamefont {A.}~\bibnamefont {Radford}}, \bibinfo {author} {\bibfnamefont {J.}~\bibnamefont {Wu}}, \ and\ \bibinfo {author} {\bibfnamefont {D.}~\bibnamefont {Amodei}},\ }\href {https://arxiv.org/abs/2001.08361} {\enquote {\bibinfo {title} {Scaling laws for neural language models},}\ } (\bibinfo {year} {2020}),\ \Eprint {http://arxiv.org/abs/2001.08361} {arXiv:2001.08361 [cs.LG]} \BibitemShut {NoStop}%
\bibitem [{\citenamefont {Hohenberg}\ and\ \citenamefont {Kohn}(1964)}]{RN40}%
  \BibitemOpen
  \bibfield  {author} {\bibinfo {author} {\bibfnamefont {P.}~\bibnamefont {Hohenberg}}\ and\ \bibinfo {author} {\bibfnamefont {W.}~\bibnamefont {Kohn}},\ }\href@noop {} {\bibfield  {journal} {\bibinfo  {journal} {Phys. Rev. B}\ }\textbf {\bibinfo {volume} {136}},\ \bibinfo {pages} {B864} (\bibinfo {year} {1964})}\BibitemShut {NoStop}%
\bibitem [{\citenamefont {Mermin}(1965)}]{RN71}%
  \BibitemOpen
  \bibfield  {author} {\bibinfo {author} {\bibfnamefont {N.~D.}\ \bibnamefont {Mermin}},\ }\href@noop {} {\bibfield  {journal} {\bibinfo  {journal} {Phys. Rev.}\ }\textbf {\bibinfo {volume} {137}},\ \bibinfo {pages} {A1441} (\bibinfo {year} {1965})}\BibitemShut {NoStop}%
\bibitem [{\citenamefont {Percus}(1988)}]{RN158}%
  \BibitemOpen
  \bibfield  {author} {\bibinfo {author} {\bibfnamefont {J.~K.}\ \bibnamefont {Percus}},\ }\href {\doibase 10.1007/bf01011639} {\bibfield  {journal} {\bibinfo  {journal} {Journal of Statistical Physics}\ }\textbf {\bibinfo {volume} {52}},\ \bibinfo {pages} {1157} (\bibinfo {year} {1988})}\BibitemShut {NoStop}%
\bibitem [{\citenamefont {Lu}\ \emph {et~al.}(2022{\natexlab{a}})\citenamefont {Lu}, \citenamefont {Meng}, \citenamefont {Cai}, \citenamefont {Mao}, \citenamefont {Goswami}, \citenamefont {Zhang},\ and\ \citenamefont {Karniadakis}}]{LU2022114778}%
  \BibitemOpen
  \bibfield  {author} {\bibinfo {author} {\bibfnamefont {L.}~\bibnamefont {Lu}}, \bibinfo {author} {\bibfnamefont {X.}~\bibnamefont {Meng}}, \bibinfo {author} {\bibfnamefont {S.}~\bibnamefont {Cai}}, \bibinfo {author} {\bibfnamefont {Z.}~\bibnamefont {Mao}}, \bibinfo {author} {\bibfnamefont {S.}~\bibnamefont {Goswami}}, \bibinfo {author} {\bibfnamefont {Z.}~\bibnamefont {Zhang}}, \ and\ \bibinfo {author} {\bibfnamefont {G.~E.}\ \bibnamefont {Karniadakis}},\ }\href {\doibase https://doi.org/10.1016/j.cma.2022.114778} {\bibfield  {journal} {\bibinfo  {journal} {Computer Methods in Applied Mechanics and Engineering}\ }\textbf {\bibinfo {volume} {393}},\ \bibinfo {pages} {114778} (\bibinfo {year} {2022}{\natexlab{a}})}\BibitemShut {NoStop}%
\bibitem [{\citenamefont {Li}\ \emph {et~al.}(2021{\natexlab{a}})\citenamefont {Li}, \citenamefont {Kovachki}, \citenamefont {Azizzadenesheli}, \citenamefont {Liu}, \citenamefont {Bhattacharya}, \citenamefont {Stuart},\ and\ \citenamefont {Anandkumar}}]{li2021}%
  \BibitemOpen
  \bibfield  {author} {\bibinfo {author} {\bibfnamefont {Z.}~\bibnamefont {Li}}, \bibinfo {author} {\bibfnamefont {N.}~\bibnamefont {Kovachki}}, \bibinfo {author} {\bibfnamefont {K.}~\bibnamefont {Azizzadenesheli}}, \bibinfo {author} {\bibfnamefont {B.}~\bibnamefont {Liu}}, \bibinfo {author} {\bibfnamefont {K.}~\bibnamefont {Bhattacharya}}, \bibinfo {author} {\bibfnamefont {A.}~\bibnamefont {Stuart}}, \ and\ \bibinfo {author} {\bibfnamefont {A.}~\bibnamefont {Anandkumar}},\ }\href {https://arxiv.org/abs/2010.08895} {\enquote {\bibinfo {title} {Fourier neural operator for parametric partial differential equations},}\ } (\bibinfo {year} {2021}{\natexlab{a}}),\ \Eprint {http://arxiv.org/abs/2010.08895} {arXiv:2010.08895 [cs.LG]} \BibitemShut {NoStop}%
\bibitem [{\citenamefont {Diab}\ and\ \citenamefont {Al~Kobaisi}(2024)}]{U-DeepONet}%
  \BibitemOpen
  \bibfield  {author} {\bibinfo {author} {\bibfnamefont {W.}~\bibnamefont {Diab}}\ and\ \bibinfo {author} {\bibfnamefont {M.}~\bibnamefont {Al~Kobaisi}},\ }\href {\doibase 10.1038/s41598-024-72393-0} {\bibfield  {journal} {\bibinfo  {journal} {Sci Rep}\ }\textbf {\bibinfo {volume} {14}},\ \bibinfo {pages} {21298} (\bibinfo {year} {2024})}\BibitemShut {NoStop}%
\bibitem [{\citenamefont {Kontolati}\ \emph {et~al.}(2023)\citenamefont {Kontolati}, \citenamefont {Goswami}, \citenamefont {Karniadakis},\ and\ \citenamefont {Shields}}]{RN171}%
  \BibitemOpen
  \bibfield  {author} {\bibinfo {author} {\bibfnamefont {K.}~\bibnamefont {Kontolati}}, \bibinfo {author} {\bibfnamefont {S.}~\bibnamefont {Goswami}}, \bibinfo {author} {\bibfnamefont {G.~E.}\ \bibnamefont {Karniadakis}}, \ and\ \bibinfo {author} {\bibfnamefont {M.~D.}\ \bibnamefont {Shields}},\ }\href {\doibase 10.48550/arXiv.2304.07599} {\enquote {\bibinfo {title} {Learning in latent spaces improves the predictive accuracy of deep neural operators},}\ } (\bibinfo {year} {2023})\BibitemShut {NoStop}%
\bibitem [{\citenamefont {Lin}\ \emph {et~al.}(2021)\citenamefont {Lin}, \citenamefont {Li}, \citenamefont {Lu}, \citenamefont {Cai}, \citenamefont {Maxey},\ and\ \citenamefont {Karniadakis}}]{RN148}%
  \BibitemOpen
  \bibfield  {author} {\bibinfo {author} {\bibfnamefont {C.}~\bibnamefont {Lin}}, \bibinfo {author} {\bibfnamefont {Z.}~\bibnamefont {Li}}, \bibinfo {author} {\bibfnamefont {L.}~\bibnamefont {Lu}}, \bibinfo {author} {\bibfnamefont {S.}~\bibnamefont {Cai}}, \bibinfo {author} {\bibfnamefont {M.}~\bibnamefont {Maxey}}, \ and\ \bibinfo {author} {\bibfnamefont {G.~E.}\ \bibnamefont {Karniadakis}},\ }\href {\doibase 10.1063/5.0041203} {\bibfield  {journal} {\bibinfo  {journal} {J Chem Phys}\ }\textbf {\bibinfo {volume} {154}},\ \bibinfo {pages} {104118} (\bibinfo {year} {2021})}\BibitemShut {NoStop}%
\bibitem [{\citenamefont {Lu}\ \emph {et~al.}(2022{\natexlab{b}})\citenamefont {Lu}, \citenamefont {Meng}, \citenamefont {Cai}, \citenamefont {Mao}, \citenamefont {Goswami}, \citenamefont {Zhang},\ and\ \citenamefont {Karniadakis}}]{RN143}%
  \BibitemOpen
  \bibfield  {author} {\bibinfo {author} {\bibfnamefont {L.}~\bibnamefont {Lu}}, \bibinfo {author} {\bibfnamefont {X.}~\bibnamefont {Meng}}, \bibinfo {author} {\bibfnamefont {S.}~\bibnamefont {Cai}}, \bibinfo {author} {\bibfnamefont {Z.}~\bibnamefont {Mao}}, \bibinfo {author} {\bibfnamefont {S.}~\bibnamefont {Goswami}}, \bibinfo {author} {\bibfnamefont {Z.}~\bibnamefont {Zhang}}, \ and\ \bibinfo {author} {\bibfnamefont {G.~E.}\ \bibnamefont {Karniadakis}},\ }\href {\doibase 10.1016/j.cma.2022.114778} {\bibfield  {journal} {\bibinfo  {journal} {Computer Methods in Applied Mechanics and Engineering}\ }\textbf {\bibinfo {volume} {393}} (\bibinfo {year} {2022}{\natexlab{b}}),\ 10.1016/j.cma.2022.114778}\BibitemShut {NoStop}%
\bibitem [{\citenamefont {Diakogiannis}\ \emph {et~al.}(2020)\citenamefont {Diakogiannis}, \citenamefont {Waldner}, \citenamefont {Caccetta},\ and\ \citenamefont {Wu}}]{ResUNet}%
  \BibitemOpen
  \bibfield  {author} {\bibinfo {author} {\bibfnamefont {F.~I.}\ \bibnamefont {Diakogiannis}}, \bibinfo {author} {\bibfnamefont {F.}~\bibnamefont {Waldner}}, \bibinfo {author} {\bibfnamefont {P.}~\bibnamefont {Caccetta}}, \ and\ \bibinfo {author} {\bibfnamefont {C.}~\bibnamefont {Wu}},\ }\href {\doibase 10.1016/j.isprsjprs.2020.01.013} {\bibfield  {journal} {\bibinfo  {journal} {ISPRS Journal of Photogrammetry and Remote Sensing}\ }\textbf {\bibinfo {volume} {162}},\ \bibinfo {pages} {94} (\bibinfo {year} {2020})}\BibitemShut {NoStop}%
\bibitem [{\citenamefont {Gao}\ \emph {et~al.}(2021)\citenamefont {Gao}, \citenamefont {Cheng}, \citenamefont {Zhao}, \citenamefont {Zhang}, \citenamefont {Yang},\ and\ \citenamefont {Torr}}]{Res2Net_2021}%
  \BibitemOpen
  \bibfield  {author} {\bibinfo {author} {\bibfnamefont {S.-H.}\ \bibnamefont {Gao}}, \bibinfo {author} {\bibfnamefont {M.-M.}\ \bibnamefont {Cheng}}, \bibinfo {author} {\bibfnamefont {K.}~\bibnamefont {Zhao}}, \bibinfo {author} {\bibfnamefont {X.-Y.}\ \bibnamefont {Zhang}}, \bibinfo {author} {\bibfnamefont {M.-H.}\ \bibnamefont {Yang}}, \ and\ \bibinfo {author} {\bibfnamefont {P.}~\bibnamefont {Torr}},\ }\href {\doibase 10.1109/tpami.2019.2938758} {\bibfield  {journal} {\bibinfo  {journal} {IEEE Transactions on Pattern Analysis and Machine Intelligence}\ }\textbf {\bibinfo {volume} {43}},\ \bibinfo {pages} {652–662} (\bibinfo {year} {2021})}\BibitemShut {NoStop}%
\bibitem [{\citenamefont {Sosnovik}, \citenamefont {Moskalev},\ and\ \citenamefont {Smeulders}(2021)}]{sosnovik2021disco}%
  \BibitemOpen
  \bibfield  {author} {\bibinfo {author} {\bibfnamefont {I.}~\bibnamefont {Sosnovik}}, \bibinfo {author} {\bibfnamefont {A.}~\bibnamefont {Moskalev}}, \ and\ \bibinfo {author} {\bibfnamefont {A.}~\bibnamefont {Smeulders}},\ }\href@noop {} {\bibfield  {journal} {\bibinfo  {journal} {arXiv preprint arXiv:2106.02733}\ } (\bibinfo {year} {2021})}\BibitemShut {NoStop}%
\bibitem [{\citenamefont {Bahri}\ \emph {et~al.}(2024)\citenamefont {Bahri}, \citenamefont {Dyer}, \citenamefont {Kaplan}, \citenamefont {Lee},\ and\ \citenamefont {Sharma}}]{Bahri_2024}%
  \BibitemOpen
  \bibfield  {author} {\bibinfo {author} {\bibfnamefont {Y.}~\bibnamefont {Bahri}}, \bibinfo {author} {\bibfnamefont {E.}~\bibnamefont {Dyer}}, \bibinfo {author} {\bibfnamefont {J.}~\bibnamefont {Kaplan}}, \bibinfo {author} {\bibfnamefont {J.}~\bibnamefont {Lee}}, \ and\ \bibinfo {author} {\bibfnamefont {U.}~\bibnamefont {Sharma}},\ }\href {\doibase 10.1073/pnas.2311878121} {\bibfield  {journal} {\bibinfo  {journal} {Proceedings of the National Academy of Sciences}\ }\textbf {\bibinfo {volume} {121}} (\bibinfo {year} {2024}),\ 10.1073/pnas.2311878121}\BibitemShut {NoStop}%
\bibitem [{\citenamefont {Wang}, \citenamefont {Xu},\ and\ \citenamefont {Zhu}(2022)}]{CNN_compact_acti}%
  \BibitemOpen
  \bibfield  {author} {\bibinfo {author} {\bibfnamefont {J.~D.}\ \bibnamefont {Wang}}, \bibinfo {author} {\bibfnamefont {J.~C.}\ \bibnamefont {Xu}}, \ and\ \bibinfo {author} {\bibfnamefont {J.~Q.}\ \bibnamefont {Zhu}},\ }\href {\doibase 10.1007/978-3-031-08754-7_40} {\bibfield  {journal} {\bibinfo  {journal} {Computational Science, Iccs 2022, Pt Ii}\ ,\ \bibinfo {pages} {319}} (\bibinfo {year} {2022})}\BibitemShut {NoStop}%
\bibitem [{\citenamefont {Lu}\ \emph {et~al.}(2021)\citenamefont {Lu}, \citenamefont {Meng}, \citenamefont {Mao},\ and\ \citenamefont {Karniadakis}}]{lu2021deepxde}%
  \BibitemOpen
  \bibfield  {author} {\bibinfo {author} {\bibfnamefont {L.}~\bibnamefont {Lu}}, \bibinfo {author} {\bibfnamefont {X.}~\bibnamefont {Meng}}, \bibinfo {author} {\bibfnamefont {Z.}~\bibnamefont {Mao}}, \ and\ \bibinfo {author} {\bibfnamefont {G.~E.}\ \bibnamefont {Karniadakis}},\ }\href {\doibase 10.1137/19M1274067} {\bibfield  {journal} {\bibinfo  {journal} {SIAM Review}\ }\textbf {\bibinfo {volume} {63}},\ \bibinfo {pages} {208} (\bibinfo {year} {2021})}\BibitemShut {NoStop}%
\bibitem [{\citenamefont {Penaud-Polge}, \citenamefont {Velasco-Forero},\ and\ \citenamefont {Angulo}(2022)}]{RN142}%
  \BibitemOpen
  \bibfield  {author} {\bibinfo {author} {\bibfnamefont {V.}~\bibnamefont {Penaud-Polge}}, \bibinfo {author} {\bibfnamefont {S.}~\bibnamefont {Velasco-Forero}}, \ and\ \bibinfo {author} {\bibfnamefont {J.}~\bibnamefont {Angulo}},\ }\href {\doibase 10.1109/Icip46576.2022.9897734} {\bibfield  {journal} {\bibinfo  {journal} {2022 Ieee International Conference on Image Processing, Icip}\ ,\ \bibinfo {pages} {2421}} (\bibinfo {year} {2022})}\BibitemShut {NoStop}%
\bibitem [{\citenamefont {Li}\ \emph {et~al.}(2020{\natexlab{b}})\citenamefont {Li}, \citenamefont {Kovachki}, \citenamefont {Azizzadenesheli}, \citenamefont {Liu}, \citenamefont {Bhattacharya}, \citenamefont {Stuart},\ and\ \citenamefont {Anandkumar}}]{RN174}%
  \BibitemOpen
  \bibfield  {author} {\bibinfo {author} {\bibfnamefont {Z.}~\bibnamefont {Li}}, \bibinfo {author} {\bibfnamefont {N.}~\bibnamefont {Kovachki}}, \bibinfo {author} {\bibfnamefont {K.}~\bibnamefont {Azizzadenesheli}}, \bibinfo {author} {\bibfnamefont {B.}~\bibnamefont {Liu}}, \bibinfo {author} {\bibfnamefont {K.}~\bibnamefont {Bhattacharya}}, \bibinfo {author} {\bibfnamefont {A.}~\bibnamefont {Stuart}}, \ and\ \bibinfo {author} {\bibfnamefont {A.}~\bibnamefont {Anandkumar}},\ }\href {\doibase 10.48550/arXiv.2010.08895} {\enquote {\bibinfo {title} {Fourier neural operator for parametric partial differential equations},}\ } (\bibinfo {year} {2020}{\natexlab{b}})\BibitemShut {NoStop}%
\bibitem [{\citenamefont {Kovachki}\ \emph {et~al.}(2021)\citenamefont {Kovachki}, \citenamefont {Li}, \citenamefont {Liu}, \citenamefont {Azizzadenesheli}, \citenamefont {Bhattacharya}, \citenamefont {Stuart},\ and\ \citenamefont {Anandkumar}}]{kovachki2021neural}%
  \BibitemOpen
  \bibfield  {author} {\bibinfo {author} {\bibfnamefont {N.~B.}\ \bibnamefont {Kovachki}}, \bibinfo {author} {\bibfnamefont {Z.}~\bibnamefont {Li}}, \bibinfo {author} {\bibfnamefont {B.}~\bibnamefont {Liu}}, \bibinfo {author} {\bibfnamefont {K.}~\bibnamefont {Azizzadenesheli}}, \bibinfo {author} {\bibfnamefont {K.}~\bibnamefont {Bhattacharya}}, \bibinfo {author} {\bibfnamefont {A.~M.}\ \bibnamefont {Stuart}}, \ and\ \bibinfo {author} {\bibfnamefont {A.}~\bibnamefont {Anandkumar}},\ }\href@noop {} {\bibfield  {journal} {\bibinfo  {journal} {CoRR}\ }\textbf {\bibinfo {volume} {abs/2108.08481}} (\bibinfo {year} {2021})}\BibitemShut {NoStop}%
\bibitem [{\citenamefont {{Paszke}}\ \emph {et~al.}(2019)\citenamefont {{Paszke}}, \citenamefont {{Gross}}, \citenamefont {{Massa}}, \citenamefont {{Lerer}}, \citenamefont {{Bradbury}}, \citenamefont {{Chanan}}, \citenamefont {{Killeen}}, \citenamefont {{Lin}}, \citenamefont {{Gimelshein}}, \citenamefont {{Antiga}}, \citenamefont {{Desmaison}}, \citenamefont {{K{\"o}pf}}, \citenamefont {{Yang}}, \citenamefont {{DeVito}}, \citenamefont {{Raison}}, \citenamefont {{Tejani}}, \citenamefont {{Chilamkurthy}}, \citenamefont {{Steiner}}, \citenamefont {{Fang}}, \citenamefont {{Bai}},\ and\ \citenamefont {{Chintala}}}]{2019arXiv191201703P}%
  \BibitemOpen
  \bibfield  {author} {\bibinfo {author} {\bibfnamefont {A.}~\bibnamefont {{Paszke}}}, \bibinfo {author} {\bibfnamefont {S.}~\bibnamefont {{Gross}}}, \bibinfo {author} {\bibfnamefont {F.}~\bibnamefont {{Massa}}}, \bibinfo {author} {\bibfnamefont {A.}~\bibnamefont {{Lerer}}}, \bibinfo {author} {\bibfnamefont {J.}~\bibnamefont {{Bradbury}}}, \bibinfo {author} {\bibfnamefont {G.}~\bibnamefont {{Chanan}}}, \bibinfo {author} {\bibfnamefont {T.}~\bibnamefont {{Killeen}}}, \bibinfo {author} {\bibfnamefont {Z.}~\bibnamefont {{Lin}}}, \bibinfo {author} {\bibfnamefont {N.}~\bibnamefont {{Gimelshein}}}, \bibinfo {author} {\bibfnamefont {L.}~\bibnamefont {{Antiga}}}, \bibinfo {author} {\bibfnamefont {A.}~\bibnamefont {{Desmaison}}}, \bibinfo {author} {\bibfnamefont {A.}~\bibnamefont {{K{\"o}pf}}}, \bibinfo {author} {\bibfnamefont {E.}~\bibnamefont {{Yang}}}, \bibinfo {author} {\bibfnamefont {Z.}~\bibnamefont {{DeVito}}}, \bibinfo {author} {\bibfnamefont {M.}~\bibnamefont {{Raison}}}, \bibinfo {author} {\bibfnamefont
  {A.}~\bibnamefont {{Tejani}}}, \bibinfo {author} {\bibfnamefont {S.}~\bibnamefont {{Chilamkurthy}}}, \bibinfo {author} {\bibfnamefont {B.}~\bibnamefont {{Steiner}}}, \bibinfo {author} {\bibfnamefont {L.}~\bibnamefont {{Fang}}}, \bibinfo {author} {\bibfnamefont {J.}~\bibnamefont {{Bai}}}, \ and\ \bibinfo {author} {\bibfnamefont {S.}~\bibnamefont {{Chintala}}},\ }\href {\doibase 10.48550/arXiv.1912.01703} {\bibfield  {journal} {\bibinfo  {journal} {arXiv e-prints}\ ,\ \bibinfo {eid} {arXiv:1912.01703}} (\bibinfo {year} {2019})},\ \Eprint {http://arxiv.org/abs/1912.01703} {arXiv:1912.01703 [cs.LG]} \BibitemShut {NoStop}%
\bibitem [{\citenamefont {Bonev}\ \emph {et~al.}(2023)\citenamefont {Bonev}, \citenamefont {Kurth}, \citenamefont {Hundt}, \citenamefont {Pathak}, \citenamefont {Baust}, \citenamefont {Kashinath},\ and\ \citenamefont {Anandkumar}}]{bonev2023spherical}%
  \BibitemOpen
  \bibfield  {author} {\bibinfo {author} {\bibfnamefont {B.}~\bibnamefont {Bonev}}, \bibinfo {author} {\bibfnamefont {T.}~\bibnamefont {Kurth}}, \bibinfo {author} {\bibfnamefont {C.}~\bibnamefont {Hundt}}, \bibinfo {author} {\bibfnamefont {J.}~\bibnamefont {Pathak}}, \bibinfo {author} {\bibfnamefont {M.}~\bibnamefont {Baust}}, \bibinfo {author} {\bibfnamefont {K.}~\bibnamefont {Kashinath}}, \ and\ \bibinfo {author} {\bibfnamefont {A.}~\bibnamefont {Anandkumar}},\ }in\ \href {https://proceedings.mlr.press/v202/bonev23a.html} {\emph {\bibinfo {booktitle} {Proceedings of the 40th International Conference on Machine Learning}}},\ \bibinfo {series} {Proceedings of Machine Learning Research}, Vol.\ \bibinfo {volume} {202},\ \bibinfo {editor} {edited by\ \bibinfo {editor} {\bibfnamefont {A.}~\bibnamefont {Krause}}, \bibinfo {editor} {\bibfnamefont {E.}~\bibnamefont {Brunskill}}, \bibinfo {editor} {\bibfnamefont {K.}~\bibnamefont {Cho}}, \bibinfo {editor} {\bibfnamefont {B.}~\bibnamefont {Engelhardt}}, \bibinfo
  {editor} {\bibfnamefont {S.}~\bibnamefont {Sabato}}, \ and\ \bibinfo {editor} {\bibfnamefont {J.}~\bibnamefont {Scarlett}}}\ (\bibinfo  {publisher} {PMLR},\ \bibinfo {year} {2023})\ pp.\ \bibinfo {pages} {2806--2823}\BibitemShut {NoStop}%
\bibitem [{\citenamefont {Tang}\ and\ \citenamefont {Lu}(2001)}]{TANG2001149}%
  \BibitemOpen
  \bibfield  {author} {\bibinfo {author} {\bibfnamefont {Y.}~\bibnamefont {Tang}}\ and\ \bibinfo {author} {\bibfnamefont {B.~C.-Y.}\ \bibnamefont {Lu}},\ }\href {\doibase https://doi.org/10.1016/S0378-3812(01)00600-8} {\bibfield  {journal} {\bibinfo  {journal} {Fluid Phase Equilibria}\ }\textbf {\bibinfo {volume} {190}},\ \bibinfo {pages} {149} (\bibinfo {year} {2001})}\BibitemShut {NoStop}%
\bibitem [{\citenamefont {Soares}, \citenamefont {Barreto},\ and\ \citenamefont {Tavares}(2023)}]{Soares2023}%
  \BibitemOpen
  \bibfield  {author} {\bibinfo {author} {\bibfnamefont {E.~d.~A.}\ \bibnamefont {Soares}}, \bibinfo {author} {\bibfnamefont {A.~G.}\ \bibnamefont {Barreto}}, \ and\ \bibinfo {author} {\bibfnamefont {F.~W.}\ \bibnamefont {Tavares}},\ }\href {\doibase 10.1016/j.fluid.2023.113887} {\bibfield  {journal} {\bibinfo  {journal} {Fluid Phase Equilibria}\ ,\ \bibinfo {pages} {113887}} (\bibinfo {year} {2023})}\BibitemShut {NoStop}%
\bibitem [{\citenamefont {Tan}\ and\ \citenamefont {Chen}(2022)}]{RN173}%
  \BibitemOpen
  \bibfield  {author} {\bibinfo {author} {\bibfnamefont {L.}~\bibnamefont {Tan}}\ and\ \bibinfo {author} {\bibfnamefont {L.}~\bibnamefont {Chen}},\ }\href {\doibase 10.48550/arXiv.2202.08942} {\enquote {\bibinfo {title} {Enhanced deeponet for modeling partial differential operators considering multiple input functions},}\ } (\bibinfo {year} {2022})\BibitemShut {NoStop}%
\bibitem [{\citenamefont {Jiang}, \citenamefont {Zhu},\ and\ \citenamefont {Lu}(2024)}]{RN170}%
  \BibitemOpen
  \bibfield  {author} {\bibinfo {author} {\bibfnamefont {Z.}~\bibnamefont {Jiang}}, \bibinfo {author} {\bibfnamefont {M.}~\bibnamefont {Zhu}}, \ and\ \bibinfo {author} {\bibfnamefont {L.}~\bibnamefont {Lu}},\ }\href {\doibase https://doi.org/10.1016/j.ress.2024.110392} {\bibfield  {journal} {\bibinfo  {journal} {Reliability Engineering \& System Safety}\ }\textbf {\bibinfo {volume} {251}},\ \bibinfo {pages} {110392} (\bibinfo {year} {2024})}\BibitemShut {NoStop}%
\bibitem [{\citenamefont {Sammüller}\ and\ \citenamefont {Schmidt}(2024)}]{RN152}%
  \BibitemOpen
  \bibfield  {author} {\bibinfo {author} {\bibfnamefont {F.}~\bibnamefont {Sammüller}}\ and\ \bibinfo {author} {\bibfnamefont {M.}~\bibnamefont {Schmidt}},\ }\href {\doibase ARTN L032601 10.1103/PhysRevE.110.L032601} {\bibfield  {journal} {\bibinfo  {journal} {Physical Review E}\ }\textbf {\bibinfo {volume} {110}} (\bibinfo {year} {2024}),\ ARTN L032601 10.1103/PhysRevE.110.L032601}\BibitemShut {NoStop}%
\bibitem [{\citenamefont {Li}\ \emph {et~al.}(2021{\natexlab{b}})\citenamefont {Li}, \citenamefont {Zheng}, \citenamefont {Kovachki}, \citenamefont {Jin}, \citenamefont {Chen}, \citenamefont {Liu}, \citenamefont {Azizzadenesheli},\ and\ \citenamefont {Anandkumar}}]{RN181}%
  \BibitemOpen
  \bibfield  {author} {\bibinfo {author} {\bibfnamefont {Z.}~\bibnamefont {Li}}, \bibinfo {author} {\bibfnamefont {H.}~\bibnamefont {Zheng}}, \bibinfo {author} {\bibfnamefont {N.}~\bibnamefont {Kovachki}}, \bibinfo {author} {\bibfnamefont {D.}~\bibnamefont {Jin}}, \bibinfo {author} {\bibfnamefont {H.}~\bibnamefont {Chen}}, \bibinfo {author} {\bibfnamefont {B.}~\bibnamefont {Liu}}, \bibinfo {author} {\bibfnamefont {K.}~\bibnamefont {Azizzadenesheli}}, \ and\ \bibinfo {author} {\bibfnamefont {A.}~\bibnamefont {Anandkumar}},\ }\href {\doibase 10.48550/arXiv.2111.03794} {\enquote {\bibinfo {title} {Physics-informed neural operator for learning partial differential equations},}\ } (\bibinfo {year} {2021}{\natexlab{b}})\BibitemShut {NoStop}%
\bibitem [{\citenamefont {Goswami}\ \emph {et~al.}(2023)\citenamefont {Goswami}, \citenamefont {Bora}, \citenamefont {Yu},\ and\ \citenamefont {Karniadakis}}]{RN182}%
  \BibitemOpen
  \bibfield  {author} {\bibinfo {author} {\bibfnamefont {S.}~\bibnamefont {Goswami}}, \bibinfo {author} {\bibfnamefont {A.}~\bibnamefont {Bora}}, \bibinfo {author} {\bibfnamefont {Y.}~\bibnamefont {Yu}}, \ and\ \bibinfo {author} {\bibfnamefont {G.~E.}\ \bibnamefont {Karniadakis}},\ }\enquote {\bibinfo {title} {Physics-informed deep neural operator networks},}\ in\ \href {\doibase 10.1007/978-3-031-36644-4_6} {\emph {\bibinfo {booktitle} {Machine Learning in Modeling and Simulation: Methods and Applications}}},\ \bibinfo {editor} {edited by\ \bibinfo {editor} {\bibfnamefont {T.}~\bibnamefont {Rabczuk}}\ and\ \bibinfo {editor} {\bibfnamefont {K.-J.}\ \bibnamefont {Bathe}}}\ (\bibinfo  {publisher} {Springer International Publishing},\ \bibinfo {address} {Cham},\ \bibinfo {year} {2023})\ pp.\ \bibinfo {pages} {219--254}\BibitemShut {NoStop}%
\end{thebibliography}%

\clearpage
\appendix

\section{Analytical cDFT functional relationships } \label{appendix:c1_HR}
For 1D-\acrshort{HR} fluids confined between two hard walls such that the density profile $\rho(x)$ exists only within the range $x \in (0,L)$, we have an exact functional relation between $\corrf(x)$ and $\rho(x)$\cite{RN158}
\begin{equation}\label{c1_rho_relation}
\begin{aligned}
c_1(x)=&\ln\left[1-\int_{x}^{x+\HRdia}\rho(x')dx'\right]\\
&-\int_{x-\HRdia}^{x}\frac{\rho(x')}{1-\int_{x'}^{x'+\HRdia}\rho(x'')dx''}dx'
\end{aligned}
\end{equation}
where $\HRdia$ is the \acrshort{HR} length, $x=-\HRdia$ and $x=L+\HRdia$ are the positions of the hard walls. For the \acrshort{HR} system under an external potential $V_{ext}(x)$, an analytical for $\rho(x)$ was reported by Vanderlick et al.\cite{RN183}
\begin{equation}\label{hl_solution}
    \rho(x)=h(x)\times l(x)
\end{equation}
with
%\begin{widetext}
\begin{equation}\label{h_funcion}
\begin{aligned}
    h(x)&=\frac
    {\exp \left[ g(x) \right]}
    {{\exp[-\beta V_{loc}(x_0)]}/{h(x_0)}+\int_{x_0}^{x}\exp\left[g(x'')\right]dx''}\\
%\end{aligned}
%\end{equation}
%\begin{equation}\label{g_funcion}
%\begin{aligned}
    g(x)&=-\beta V_{loc}(x)+\int_{x_0}^{x}h(x'-\HRdia)dx' 
\end{aligned}
\end{equation}
%\end{widetext}
and
\begin{equation}\label{l_funcion}
\begin{aligned}
    l(x)=&l(x_0)\exp\left[ \int_{x_0}^{x} h(x')dx'\right]\\
    &-\int_{x_0}^x \exp\left[-\int_{x_0}^{x'} h(x'')dx''\right] l(x'+\HRdia)h(x'+\HRdia)dx'
\end{aligned}
\end{equation}
where $x_0\in[0,L] $ is an arbitrary reference position, and $V_{loc}(x)=V_{ext}(x)-\mu$ is the background potential in the domain $x\in[0,L]$. Eq.\eqref{h_funcion} can be solved successively over the intervals $k\HRdia<x<(k+1)\HRdia, k=0,1,2,\cdots$, beginning with $0<x<\HRdia$, 
\begin{equation}
    h_0(x)=\frac{e^{ -\beta V_{loc}(x)}}
        {1+ \int_{0}^{x} e^{-\beta V_{loc}(x')}dx'}
\end{equation}
Similarly, Eq.\eqref{l_funcion} can be solved successively over intervals $L-(k+1)\HRdia<x<L-k\HRdia, k=0,1,2,\cdots$, beginning with $L-\HRdia<x<L$,
\begin{equation}
    l_0(x)=\exp\left[ -\int_x^L h(x')dx'\right]. 
\end{equation}

In programming, $h(x)$ and $l(x)$ can be too small with 32-bit floating number, leading to high round-off errors in the integration steps. $\ln h(x)$ and $\ln l(x)$ are used to control round-off errors.
The default $\Delta x$ would cause high truncation errors in the numerical integration of $h(x)$ and $l(x)$ with small $|x-x_0|$ in \cref{h_funcion} and \cref{l_funcion}, which is controlled by increasing the sampling resolution 4 times. 
%while truncation errors are managed through the sampling resolution.\JW{Please explain what these statements mean}

\section{External potential and chemical potential sets} \label{appendix:V_ext}
The density profiles $\rho(x)$ and one-body correlation functions $\corrf(x)$ investigated in this work were generated from five parametric forms of the external potential, defined over the domain $0\leq x\leq L$, with the reduced chemical potential in the range $\beta\mu\in (0,4)$. In dimensionless form, these external potentials are given by:

\textbf{Group I}: Simple hard walls
\begin{equation}
\beta V_{ext}(x)=0
\end{equation}

\textbf{Group II}: Van der Waals-like attractive walls
\begin{equation}
\beta V_{ext}(x)= -\epsilon \left[ \left(\frac{\HRdia}{x}\right)^3
           +\left(\frac{\HRdia}{L-x}\right)^3
           \right]
\end{equation}
with $\epsilon \in (0.1,2.2)$.

\textbf{Group III}: Linear potentials (Constant Force)
\begin{equation}
    \beta V_{ext}(x) = m_g x 
\end{equation}
with $m_g \in (0.1,3)$

\textbf{Group IV}: Power-law potentials
\begin{equation}
    \beta V_{ext}(x) = u_0 \left|
           \frac
           {x-L/2}
           {x_0}
           \right|^{a_0}
\end{equation}
with $u_o \in (1,3), x_0\in (1,3), a_0 \in (2,5)$.

\textbf{Group V}: Linear combination of groups 2 and 3 potentials
\begin{equation}
    \beta V_{ext}(x) = m_g x-\epsilon \left[ \left(\frac{\HRdia}{x}\right)^3
           +\left(\frac{\HRdia}{L-x}\right)^3
           \right]
\end{equation}
with $\epsilon \in (0.1,2.2),m_g \in (0.1,3)$.

\section{Activation functions} \label{appendix:activations}
The following four activation functions were tested for training different \acrshort{ML} models: 
I. Rectified linear unit (\acrshort{Relu}):
%\begin{equation}
%    \begin{aligned}
%        \acti(x) &= 0, & x\leq 0 \\
%                  &= x, & x>0 \\
%    \end{aligned}
%\end{equation}
\begin{equation}
    \acti(x) = 
    \begin{cases}
        0, & x \leq 0 \\
        x, & x > 0
    \end{cases}
\end{equation}
II. Scaled exponential linear unit (\acrshort{Selu}) function:
\begin{equation}
\acti(x)=
    \begin{cases}
        \lambda*x, & x>0 \\
         \lambda*\alpha*(e^x-1), & x\leq 0
    \end{cases}
\end{equation}
where $\lambda=1.67326$ and $\alpha=1.05070$ are fixed constants of \acrshort{Selu}, allowing auto internal normalization.

III. Logistic function:
\begin{equation}
        \acti(x) = \frac{1}{1+e^{-x}} 
\end{equation}

IV. Squared Relu (\acrshort{sRelu}):
\begin{equation} \label{eqn:sRelu}
        \acti(x)=
    \begin{cases}
            0, & x\leq 0 \text{ or }  x\geq 2 \\
            x(2-x), & 0<x<2 
    \end{cases}
\end{equation}
%\clearpage
\section{Additional results} \label{appendix:additional}
%\JW{This section includes all Tables and Figures not included in the main text. Discuss each figure/table in detail and refer them in the main text}

\setcounter{table}{0}
\renewcommand{\thetable}{S\arabic{table}}
\setcounter{figure}{0}
\renewcommand{\thefigure}{S\arabic{figure}}

\begin{figure}[h!]
\centering
\vspace{-0.75\baselineskip}
\subfloat[]{
\includegraphics[width=0.45\linewidth]{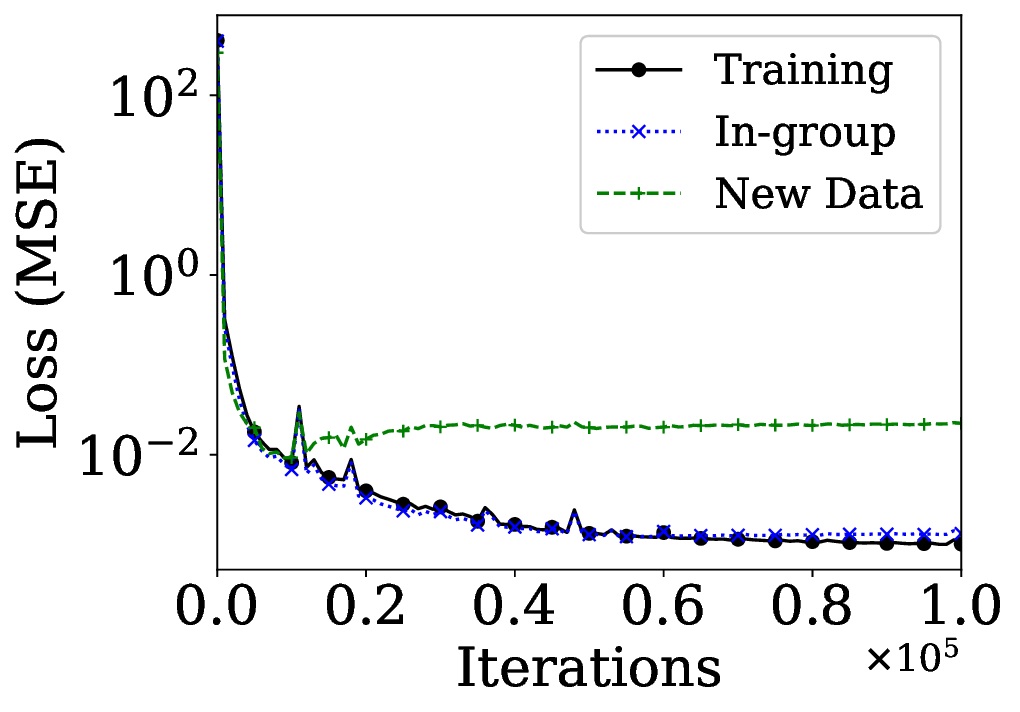} 
    %\caption{DNN-DeepONet}
}
\vspace{-0.75\baselineskip}
\subfloat[]{
\includegraphics[width=0.45\linewidth]{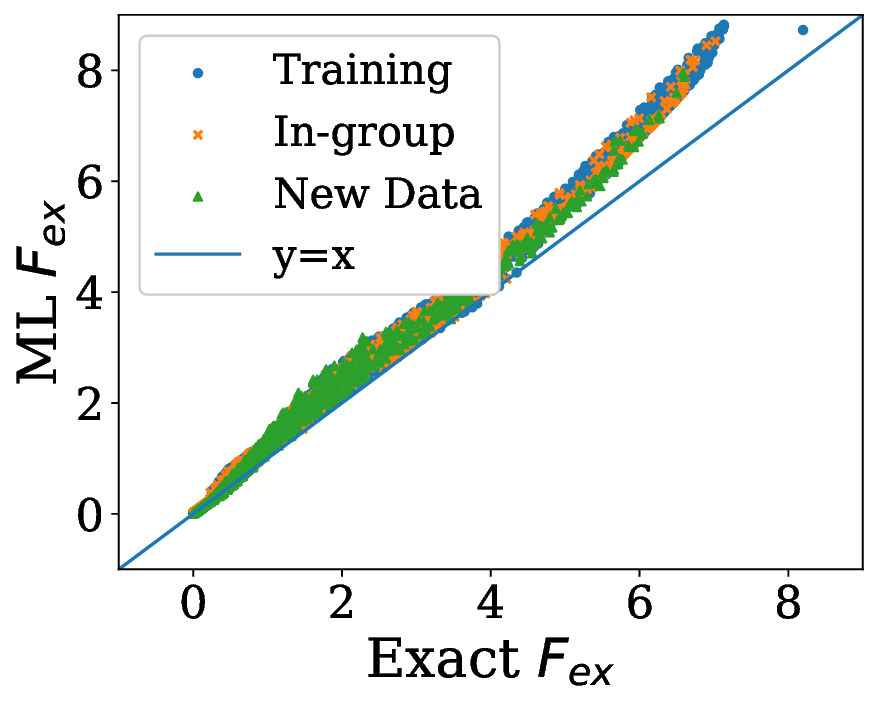} 
    %\caption{CNN-DeepONet}
}

\vspace{-0.75\baselineskip}
\subfloat[]{
\includegraphics[width=0.45\linewidth]{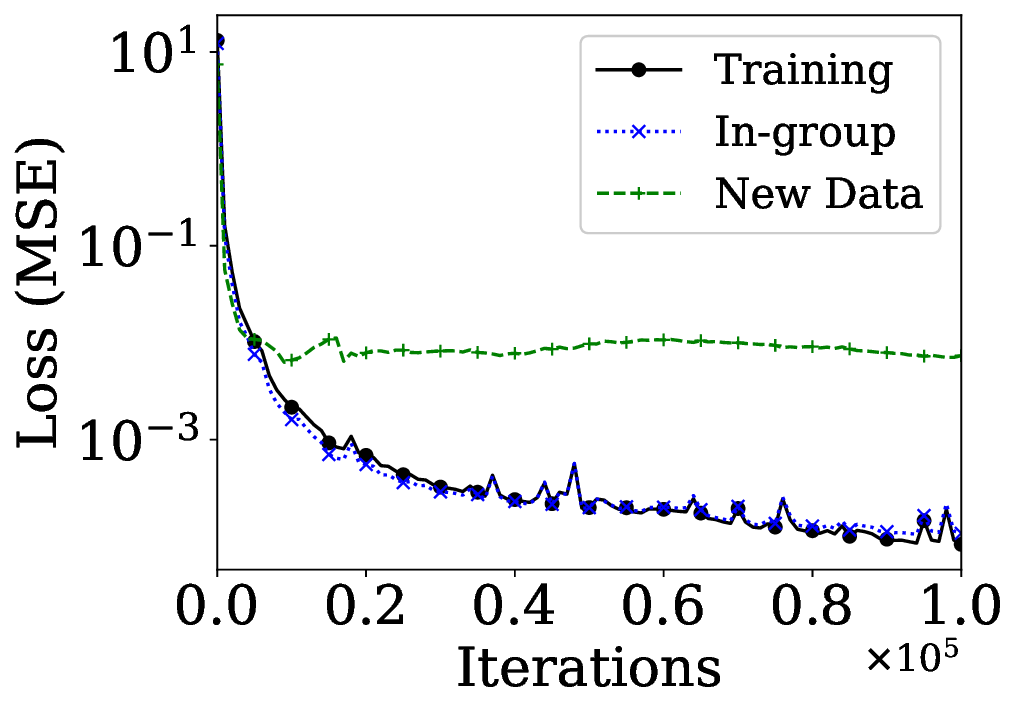} 
    %\caption{MSCNN-DeepONet}
}
\vspace{-0.75\baselineskip}
\subfloat[]{
\includegraphics[width=0.45\linewidth]{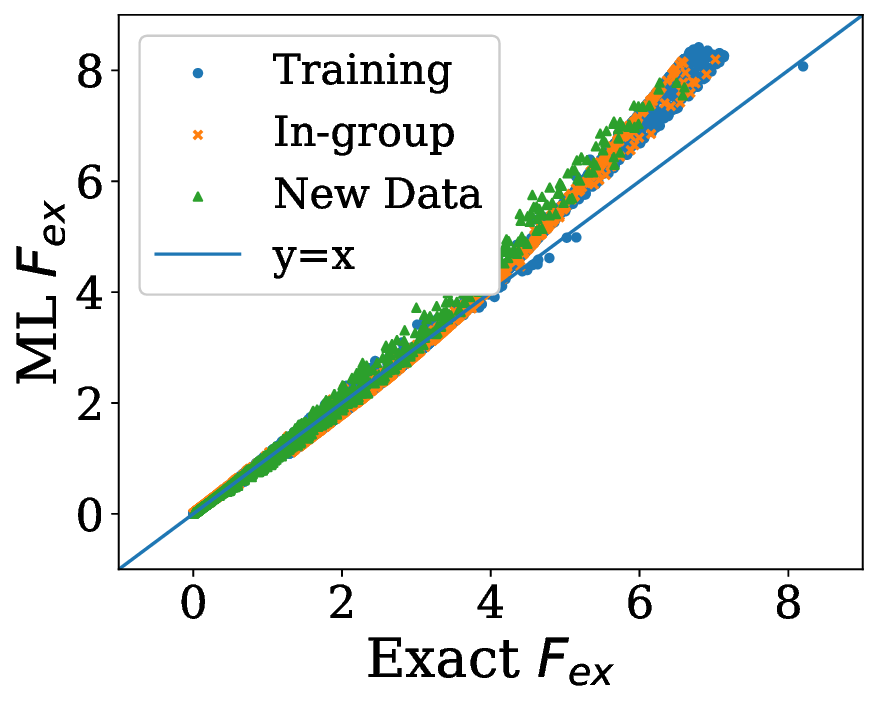} 
    %\caption{GCNN-DeepONet}
}

\vspace{-0.75\baselineskip}
\subfloat[]{
\includegraphics[width=0.45\linewidth]{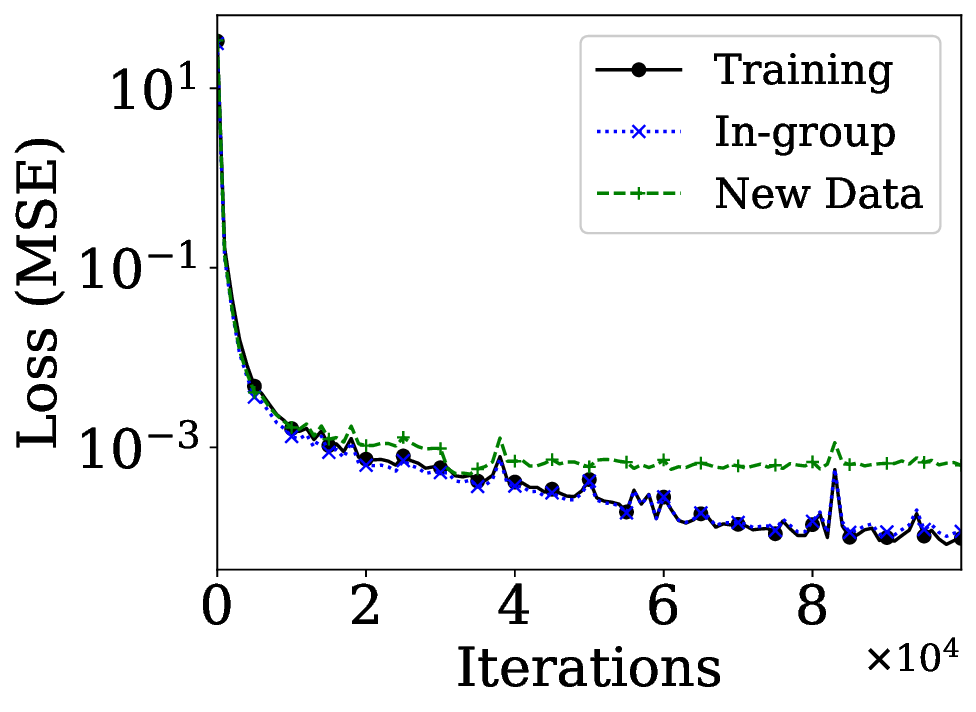} 
    %\caption{DNN-DeepONet}
}
\vspace{-0.75\baselineskip}
\subfloat[]{
\includegraphics[width=0.45\linewidth]{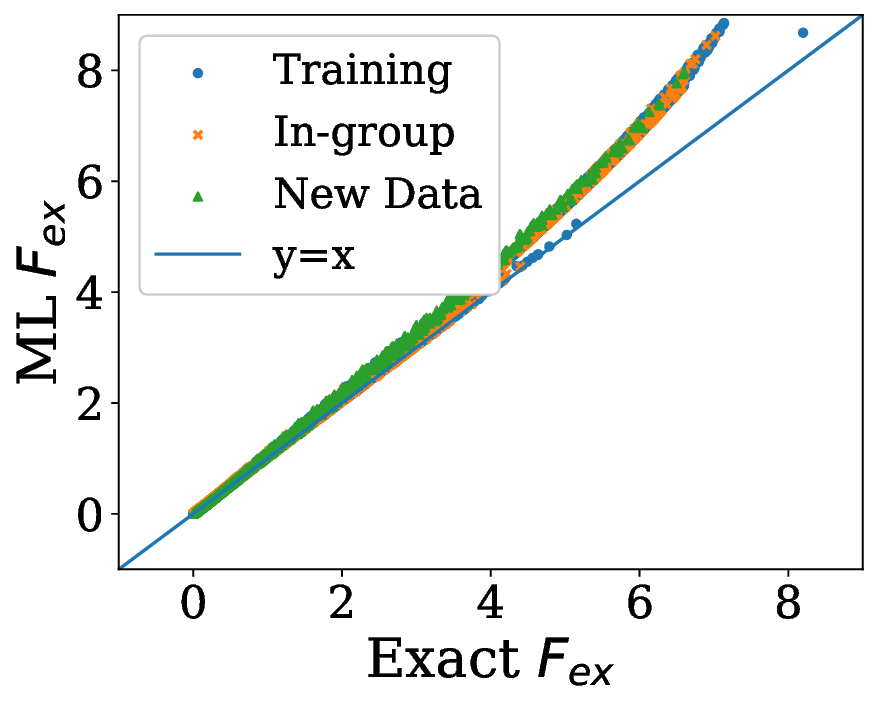} 
    %\caption{CNN-DeepONet}
}
\caption{(ace)Mean Squared error Loss-iteration relationship of training data, in-group and new dataset test data by remaining operator learning methods. Activation function is squared Relu (sRelu) 
(bdf) Excess Free energy prediction of training data, in-group, and new dataset test data by remaining operator learning methods. Activation function is squared Relu (sRelu)
Machine-learning Method: (ab) DK-CNN-DeepONet (cd) DK-RMSCNN-DeepONet (ef) GK-CNN-DeepONet
}
\label{fig:SI:Loss_iter_other}
\end{figure}

\begin{figure*}[ht!]
\centering
\subfloat[]{
\includegraphics[width=0.425\linewidth]{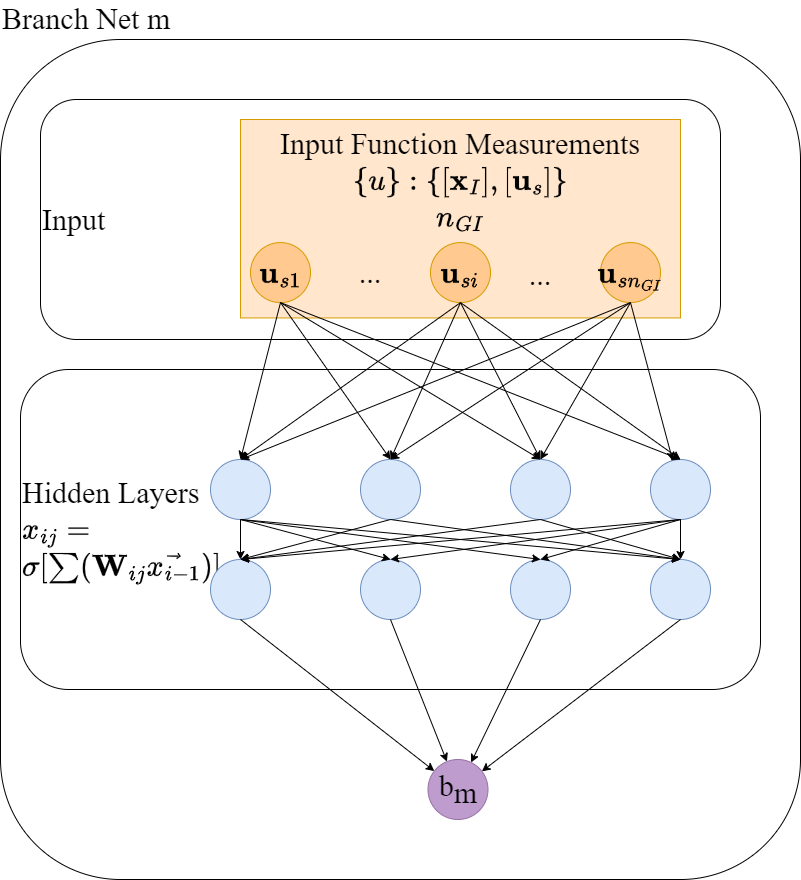}
%\caption{DNN-DeepONet Branch Net- One net}
}
\subfloat[]{
\includegraphics[width=0.525\linewidth]{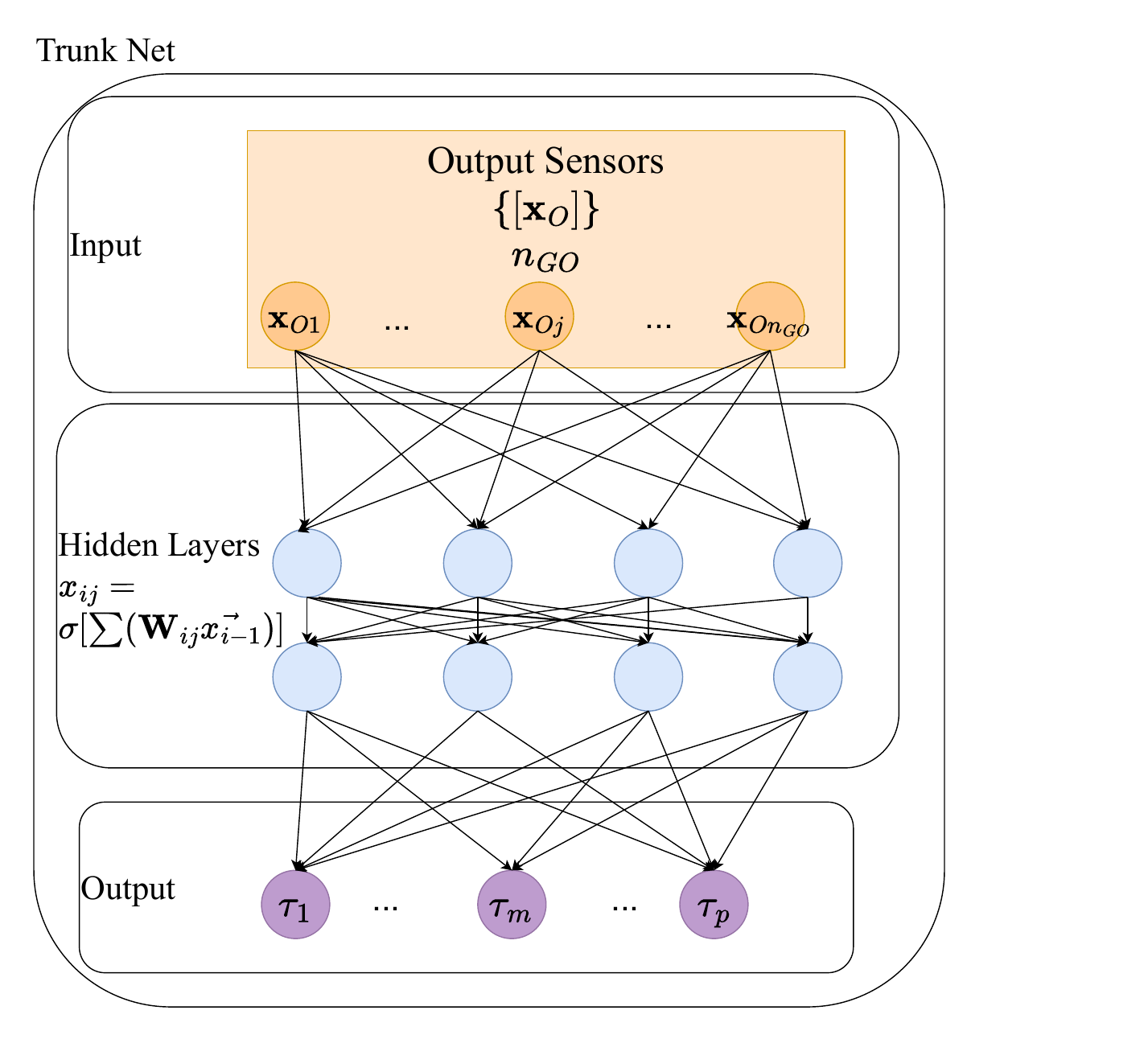}
%\caption{DNN-DeepONet Trunk Net- Stacked Net}
}
\caption{\label{fig:Structures_sub} \acrfull{DNN}- \acrshort{DeepONet} subnets structures, with training dataset shown. (a) The m-th DNN Branch Net in unstack Branch Nets (b) Stacked DNN Trunk Net }
\end{figure*}

\begin{figure*}[ht!]
\includegraphics[height=1.0\textheight]{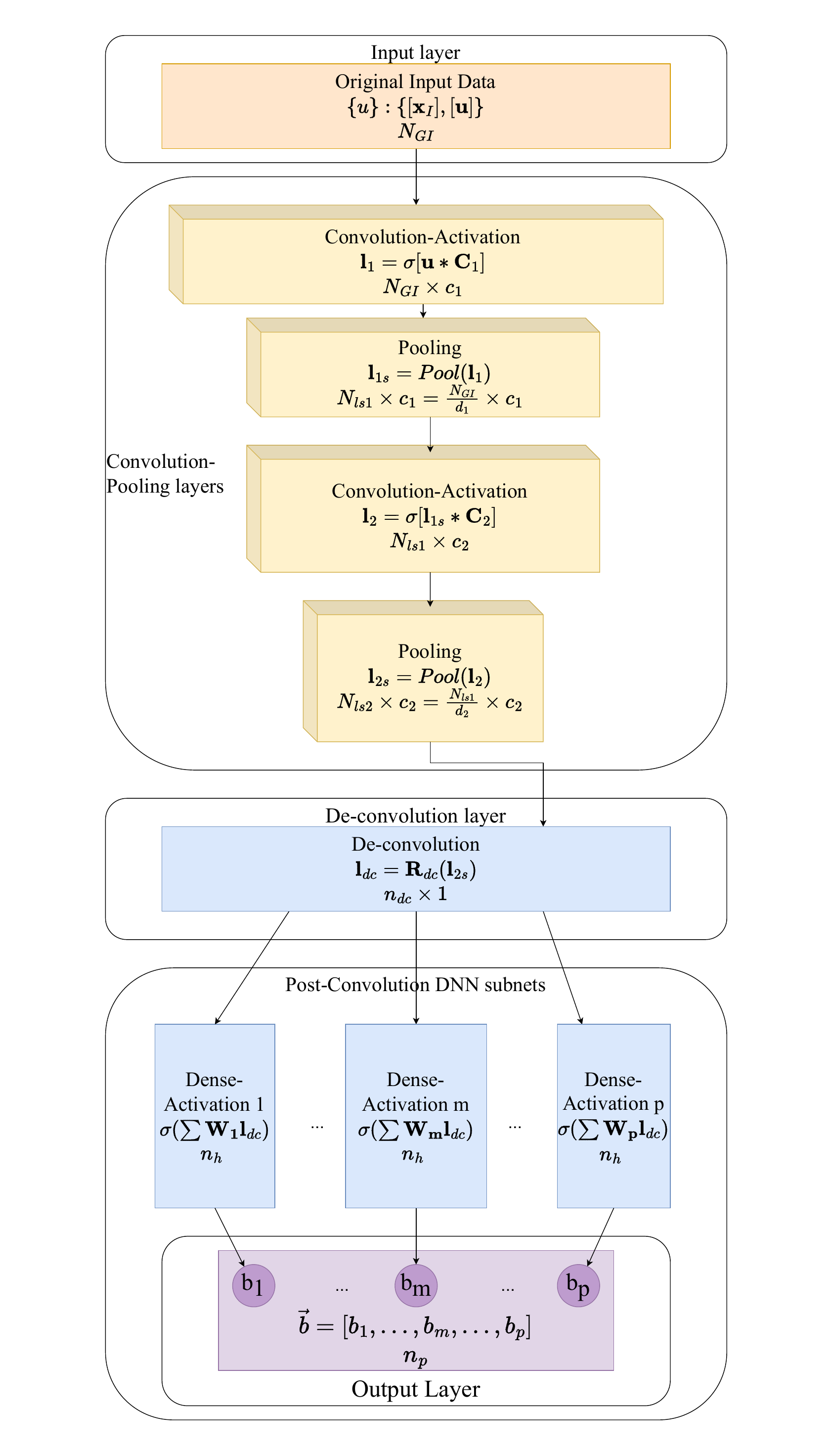}
\caption{\label{fig:CNN_Structures}\acrfull{CNN} Structures for Branch Net  for \acrshort{CNN}-\acrshort{DeepONet}, with operators and datasets in each layer shown. Pool operations are using average pooling operation, $R_dc$ is the transposed convolution operation, $\sigma$ means applying activation function}
\end{figure*}

\begin{figure*}[ht!]
\includegraphics[height=1.0\textheight]{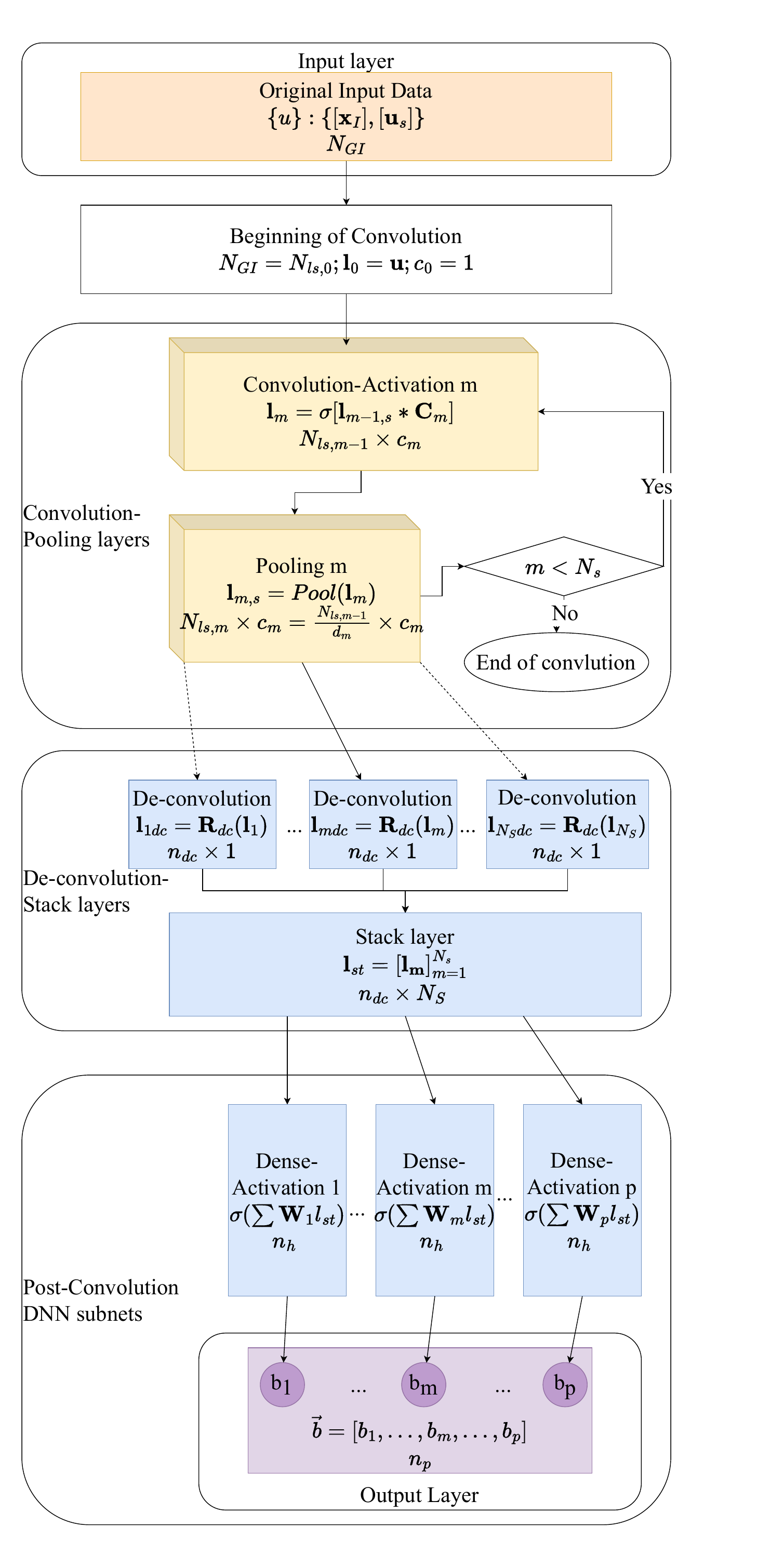}
\caption{\label{fig:MSCNN_Structures} \acrfull{RMSCNN} Structures for Branch Net Structures for \acrshort{RMSCNN}-DeepONet, with operators and datasets in each layer shown. $N_c$ is number of convolution layers, $N_c=5$ for \acrshort{RMSCNN} in this paper.}
\end{figure*}

This section contains additional results and further explanations of the machine-learning methods.  \Cref{tab:SI:Loss_compare_ONs,tab:acti_comp_other} and \Cref{fig:SI:Loss_iter_other} show the additional results for neural operator methods, including \acrshort{DK}-\acrshort{CNN}-\acrshort{DeepONet}, \acrshort{DK}-\acrshort{RMSCNN}-\acrshort{DeepONet} and \acrshort{GK}-\acrshort{CNN}-\acrshort{DeepONet}.
\Cref{fig:Structures_sub,fig:CNN_Structures,fig:MSCNN_Structures} present the structures of the neural operator architectures not included in main text.   

\begin{table}[h!]
\centering
\caption{\label{tab:SI:Loss_compare_ONs} The function MSE loss performance and $\Fex$ Calculation performances of \acrshort{DK}-\acrshort{CNN}-\acrshort{DeepONet}, \acrshort{DK}-\acrshort{RMSCNN}-\acrshort{DeepONet} and \acrshort{GK}-\acrshort{CNN}-\acrshort{DeepONet} using sRelu activation function, Typical \acrshort{DK}-\acrshort{CNN}-\acrshort{DeepONet}s shows poor extrapolation performance and $\Fex$ estimation}
\begin{ruledtabular}\begin{tabular}{c m{5em}<{\centering} m{5em}<{\centering} m{5em}<{\centering}}

  &DK-CNN-DeepONet & DK-RMSCNN-DeepONet &GK-CNN-DeepONet\\
\hline
$\msize$& $ 9.5 \times 10^{5} $ &$ 4.9 \times 10^{5} $    &   $ 5.2 \times 10^{5} $\\
$\corrf-\text{MSE}_T$&$ 8.5 \times 10^{-4} $ & $ 1.02 \times 10^{-5} $ &$ 8.03 \times 10^{-5} $\\ 
$\corrf-\text{MSE}_{IG}$&$ 1.32 \times 10^{-3} $             & $ 7.87 \times 10^{-4} $ &$ 1.07 \times 10^{-4} $\\ 
$\corrf-\text{MSE}_{ND}$&$ 2.25 \times 10^{-3} $ & $ 1.62 \times 10^{-2} $ &$ 1.69 \times 10^{-3} $\\ 
Training time (s)& 2735 & 4223 &3633\\
$\Fex-{R^2}_T$&$0.94$ & $ 0.95 $ &$ 0.97 $\\ 
$\Fex-{R^2}_{IG}$&$0.93 $             & $0.95 $ &$ 0.97$\\ 
$\Fex-{R^2}_{ND}$&$ 0.95 $ & $ 0.93 $ &$0.95 $\\ 
Calculation time (s)& 0.6 & 0.8 &0.8\\
\end{tabular}\end{ruledtabular}
\end{table}
\bigbreak
%\clearpage
\begin{table}[h!]
\centering
\caption{The function MSE loss performance of Typical DK-CNN-DeepONet and GK-CNN-DeepONets using all four different activation function, Typical DK-CNN-DeepONet shows poor extrapolation performance and $F_{ex}$ estimation}
\label{tab:acti_comp_other}
\begin{ruledtabular}\begin{tabular}{c m{5em}<{\centering} m{5em}<{\centering} m{5em}<{\centering} m{5em}<{\centering} } 
Model&   &  Relu&  Selu&  Logistic\\ 
\hline 
\multirow{4}{5em}{\centering DK-CNN-DeepONet}
    & $\corrf-\text{MSE}_T$&  $ 4.31 \times 10^{-4} $&  $ 4.30 \times 10^{-3} $&  $ 2.60 \times 10^{-3} $\\ 
    & $\corrf-\text{MSE}_{IG}$&  $ 4.26 \times 10^{-4} $&  $ 4.36 \times 10^{-3} $&  $ 2.22 \times 10^{-3} $\\ 
    & $\corrf-\text{MSE}_{ND}$&  $ 1.01 \times 10^{-2} $&  $ 4.46 \times 10^{-3} $&  $ 5.87 \times 10^{-3} $\\ 
    & time (s) & 1976 & 2235 & 2543 \\ 
    %\hline 
\multirow{4}{5em}{\centering DK-RMSCNN-DeepONet}
    & $\corrf-\text{MSE}_T$&  $ 1.94 \times 10^{-5} $&  $ 6.54 \times 10^{-4} $&  $ 5.64 \times 10^{-4} $\\ 
    & $\corrf-\text{MSE}_{IG}$& $ 8.15 \times 10^{-5} $& $ 6.33 \times 10^{-4} $& $ 4.98 \times 10^{-4} $\\  
    & $\corrf-\text{MSE}_{ND}$& $ 4.9 \times 10^{-2} $& $ 6.12 \times 10^{-3} $& $ 2.49 \times 10^{-3} $\\ 
    & time (s) & 2123 & 2311 & 2678\\ 
    %\hline 
\multirow{4}{5em}{\centering GK-CNN-DeepONet}
    & $\corrf-\text{MSE}_T$& $ 5.39 \times 10^{-4} $& $ 1.05 \times 10^{-3} $& $ 1.72 \times 10^{-3} $\\ 
    & $\corrf-\text{MSE}_{IG}$& $ 5.5 \times 10^{-4} $& $ 1.00 \times 10^{-3} $&$ 1.34 \times 10^{-3} $\\ 
    & $\corrf-\text{MSE}_{ND}$& $ 5.01 \times 10^{-3} $& $ 4.54 \times 10^{-3} $&$ 6.42 \times 10^{-3} $\\
    & time (s) & 3256 & 3369   &  4376 \\ 
\end{tabular}\end{ruledtabular}
\end{table}

\begin{table}[h!]
\centering
\caption{The detailed hyperparameters range of models used in Neural Scaling law analysis explained in \Cref{sec:Hyperparameters}, grouped by number of layers for each model. $N_{layer}$ showing the number of \acrshort{DNN} layers used in \acrshort{DNN}s and \acrshort{DNN}-\acrshort{DeepONet}, and the number of Fourier layers in \acrshort{FNO}. $N_{d}$ is the number of nodes in each dense layer of \acrshort{DNN}. $d_v$ is the number of channels of each Fourier layer in \acrshort{FNO}. $k$ is the number of modes remained in truncated Fourier transforms of \acrshort{FNO}. The least and most number of trainable parameters (\glssymbol{msize}) are also listed for comarison. Some combination of hyperparameters are filiterd out because of their inferior performances comparing to model with less \glssymbol{msize}, or the \glssymbol{msize} is large enough for the diminishing return on model performance.  }
\label{tab:SI:Hyperparameter}
\begin{ruledtabular}\begin{tabular}{ccccc}
Model Type& $N_{layer}$ & $N_{d}$ or $d_v$ & $k$ &$\msize$ Range\\ \hline
\multirow{4}{5em}{\centering Full-scale DNN }
 &2&$[4,8,32]$&/&4013-53153\\
 &3&$[4,8,32,128]$&/&7253-239009\\
 &4&$[8,32,128,512]$&/&13841-1609505\\
 &5&$[8,32,128,512]$&/&56321-1872161\\
\multirow{4}{5em}{\centering Quasi-local DNN }
 &2&$[4,8,32]$&/&2081-17537\\
 &3&$[4,8,32,128,512]$&/&2101-788993\\
 &4&$[8,32,128,512]$&/&6721-1462273\\
 &5&$[8,32,128]$&/&7293-131969\\
\multirow{4}{5em}{\centering DNN-DeepONet}
&2&$[4,8,32]$&/&2081-82433\\
&3&$[4,8,32,128,512]$&/&4265-788993\\
&4&$[8,32,128,512]$&/&4337-1051649\\
&5&$[8,32,128,512]$&/&4409-1314305\\
\multirow{3}{5em}{\centering FNO}
&2&$[4,8,16,64,128]$&$[8,32,128]$&521-4393217\\
&3&$[4,8,16,64,128]$&$[8,32,128]$&727-6556353\\
&4&$[16,64,128]$&$[32,128]$&2469-8719489\\
\end{tabular}\end{ruledtabular}
\end{table}

\clearpage
%\subsection{Figures of Network structures in this figure}

\end{document}